\documentclass{aa}  

\usepackage{natbib}
\bibpunct{(}{)}{;}{a}{}{,}
\usepackage{graphicx}
\usepackage{txfonts}
\begin{document}

\title{A LOFAR study of non-merging massive galaxy clusters}

\author{F. Savini\inst{1}
          \and
        A. Bonafede\inst{1,2,3}
        \and
        M. Br{\"u}ggen\inst{1}
        \and 
        D. Rafferty\inst{1}
        \and
        T. Shimwell\inst{4,5}
        \and
        A. Botteon\inst{2,3}
        \and
        G. Brunetti\inst{2}
        \and
        H. Intema\inst{4}
         \and
        A. Wilber\inst{1}
        \and
       R. Cassano\inst{2}
         \and
         F. Vazza\inst{3}
        \and
         R. van Weeren\inst{4}
        \and
       V. Cuciti\inst{2,3}
         \and
        F. De Gasperin\inst{1}
                \and
       H. R{\"o}ttgering$^{4}$   
                \and
        M. Sommer\inst{6}
           \and
       L. B\^irzan\inst{1}
       \and
       A. Drabent\inst{7}
      }

   \institute{
   Hamburger Sternwarte, Universit\"at Hamburg, Gojenbergsweg 112, 21029, Hamburg, Germany.
       \and
   INAF IRA, via P. Gobetti 101, 40129 Bologna, Italy.
       \and
   Dipartimento di Fisica e Astronomia, Universit\'a di Bologna, via P. Gobetti 93/2, 40129 Bologna, Italy.
  \and 
 Leiden Observatory, Leiden University, PO Box 9513, 2300 RA Leiden, The Netherlands.
 \and
ASTRON, the Netherlands Institute for Radio Astronomy, Postbus 2, 7990 AA, Dwingeloo, The Netherlands.
\and
Argelander-Institut f\"ur Astronomie, Auf dem Hügel 71, D-53121, Bonn, Germany.
\and
Th\"uringer Landessternwarte, Sternwarte 5, D-07778 Tautenburg, Germany.\\ \\
\email{federica.savini@hs.uni-hamburg.de}
 }

   \date{Received ...; accepted ...}

% \abstract{}{}{}{}{} 
% 5 {} token are mandatory
   \abstract{Centrally located diffuse radio emission has been observed in both merging and non-merging galaxy clusters. Depending on their morphology and size, we distinguish between giant radio haloes, which occur predominantly in merging clusters, and mini haloes, which are found in non-merging, cool-core clusters. 
In recent years, cluster-scale radio emission has also been observed in clusters with no sign of major mergers, showing that our knowledge of the mechanisms that lead to particle acceleration in the intra-cluster medium (ICM) is still incomplete. Low-frequency sensitive observations are required to assess whether the emission discovered in these few cases is common in galaxy clusters or not. With this aim, we carried out a campaign of observations with the LOw Frequency ARay (LOFAR) in the frequency range 120 - 168 MHz of nine massive clusters selected from the \textit{Planck} SZ catalogue, which had no sign of major mergers. In this paper, we discuss the results of the observations that have led to the largest cluster sample studied within the LOFAR Two-metre Sky Survey, and we present \textit{Chandra} X-ray data used to investigate the dynamical state of the clusters, verifying that the clusters are currently not undergoing major mergers, and to search for traces of minor or off-axis mergers. We discover large-scale steep-spectrum emission around mini haloes in the cool-core clusters PSZ1G139.61+24 and RXJ1720.1+2638, which is not observed around the mini halo in the non-cool-core cluster A1413. We also discover a new 570 kpc-halo in the non-cool-core cluster RXCJ0142.0+2131. We derived new upper limits to
the radio power for clusters in which no diffuse radio emission was found,  and we discuss the implication of our results to constrain the cosmic-ray energy budget in the ICM. We conclude that radio emission in non-merging massive clusters is not common at the sensitivity level reached by our observations and that no clear connection with the cluster dynamical state is observed. Our results might indicate that the sloshing of a dense cool core could trigger particle acceleration on larger scales and generate steep-spectrum radio emission.} 

%We confirm the presence of the giant radio halo previously found in A2261, and argue that the giant radio halo claimed in A2390 is likely contaminated by a central radio galaxy. The case of A478 is uncertain with no diffuse emission found despite the claim of mini halo present in literature. No diffuse radio emission is found in the clusters A1576 and A1423. 

%radio sources, such as ultra-steep-spectrum halos around the mini halos in the clusters PSZ1G139.61+24 and RXJ1720.1+2638, and new head-tail radio galaxies in the cluster environment of A1423 and A1413. We find a 600-kpc central double radio galaxy in the massive cool-core cluster A2390 and confirm the presence of a mini halo in the cluster A1413. Furthermore, we discover a radio halo in the non-cool-core cluster RXCJ0142.0+2131, and we confirm the presence of a radio halo in A2261. We place new upper limits on radio halo power for the clusters A1423 and A1576. Finally, we discuss the implications of our results regarding the connection between the cluster dynamical state and the formation of diffuse radio emission.} 
%We then conclude investigating the connection
%...investigate if there is a connection between halos and mini halos. 
%mass-selected sample of galaxy clusters from the PSZ catalogue.


   \maketitle
%
%________________________________________________________________

\section{Introduction}

Diffuse radio emission in galaxy clusters is caused by relativistic electrons that emit synchrotron radiation in the intra-cluster magnetic field (see \citealp{Feretti2012} and references therein). 
This emission has a low surface brightness ($\sim$ 0.1 - 1 $\, \mu$Jy \, arcsec$^{-2}$ at $\nu \sim$ 1.4 GHz) and, depending on its morphology, location, and size (hundreds of kpc up to few Mpc), it is classified as radio relic, radio halo, or mini halo. In this paper, we are only concerned with radio haloes and mini haloes, which are sources that are located at the centres of galaxy clusters. Both type of sources are commonly characterised by a steep radio spectrum with a spectral index of \footnote{The spectrum is defined by $S(\nu) \propto \nu^{\alpha}$.} $\alpha < -1$.\\

Giant radio haloes ($\sim$ Mpc-scale) have predominantly been found in merging clusters (e.g. \citealp{Buote2001}, \citealp{Cassa2010}, \citealp{Cuci2015}) that typically do not have a cool core, i.e. a core characterised by a peaked X-ray surface brightness, high gas density, and significant drop in temperature within the inner $\sim 100$ kpc. Haloes indicate the presence of a cluster-wide particle acceleration mechanism, such as in situ re-acceleration (see \citealp{bru2014} for a review).\\

Mini haloes ($<$ 500 kpc-scale) are detected in cool-core clusters that have not been disrupted or disturbed by major mergers. Often, the brightest cluster galaxy (BCG) resides at the centre of the mini halo and, when the BCG is radio loud, its radio lobes are embedded in the mini halo. The origin of the relativistic cosmic-ray electrons (CRe) of mini haloes is still unclear. The main contending theories are either leptonic (re-acceleration) models and hadronic models. Leptonic models involve the re-acceleration of a lower energy population of CRe by turbulent motions in the intra-cluster medium (ICM; \citep{Gitti2002}). The turbulence can be caused by the sloshing of the low-entropy gas falling inside the dark matter potential well of the cluster. The sloshing can be instigated, for example, by a minor merger with a lower mass galaxy cluster or group. When the gas meets the higher entropy ICM, a cold front, i.e. a discontinuity in the X-ray emissivity, is formed \citep{ZuHone2013}. Hadronic models involve the injection of secondary electrons produced continuously by inelastic collisions of relativistic cosmic-ray protons (CRp) with the cluster thermal proton population \citep{Pf2004}. The CRp are expected to be present in the ICM, therefore some level of synchrotron emission from secondary electrons is expected in galaxy clusters, especially in non-merging clusters with a dense cool core, where the rate of hadronic interactions should be the highest.\\

%The electron population may be supplied by active galaxies in the cluster environment, so called active galactic nuclei (AGN) injection (e.g. \citealp{Pin2017}).

%\textbf{MB: As we expect there to be some level of relativistic protons in the ICM (which have long life-times), there must be some level of synchrotron emission from secondary electrons. This emission is best found by observing deep in non-merging, cool-core galaxy clusters.}\\
%and they may also probe the physics of gas heating in cool-core clusters via turbulence/shocks or cosmic ray driven instabilities (e.g. Pfrommer \& Ensslin, 2004).\\

\begin{table*}
 \centering
  \caption{Observation details of the targets. The letter P indicates the clusters observed as part of dedicated proposals.}
\begin{tabular}{c c c c c c c c}
  \hline
Name &  RA     &   DEC & LoTSS obs.& LoTSS obs. & Radio flux &  \textit{Chandra} obs. & \textit{Chandra} clean\\ 
& \scriptsize{(h m s, J2000)}                   &        \scriptsize{($^{\circ}$ $'$ $''$, J2000)} &  ID & date & calibrator & ID & exp. time \scriptsize{(ks)}\\ 

\hline
\hline

RXCJ0142.0+2131 &  01 42 02.6 & +21 31 19.0  &  LC9\_011 (P) & 21-02-18 & 3C196 & 10440 & 19.9\\
A478 & 04 13 20.7 & +10 28 35.0 &  LC8\_006 (P) & 7/21-11-17 & 3C196 & 1669, 602, 6928 & 139.3 \\
&&&&&&6929, 7217, 7128 &\\
&&&&&&7222, 7231, 7232 &\\
&&&&&&7233, 7234, 7235 &\\
PSZ1G139.61+24 &  06 22 04.6 &  +74 40 51.6 &LC8\_022 & 27-07-17 &  3C295 & 15139, 15297 & 23.1\\
A1413 & 11 55 18.9 & +23 24 31.0 & LC9\_020 (P) & 25-01-18 & 3C295 & 537, 1661, 5002 & 127.8\\
&&&&&&5003, 7696 &\\
A1423 & 11 57 22.5 & +33 39 18.0 & LC8\_022 & 21-09-17 & 3C196 & 528, 11724 & 33.3 \\
A1576 &  12 37 59.0 & +63 11 26.0  & LT5\_007  & 04-04-16 & 3C196 & 7938, 15127 & 28.5 \\
RXJ1720.1+2638 & 17 20 10.1 & +26 37 29.5 & LC7\_004 (P) & 25-01-17 & 3C295 & 1453, 3224, 4361 & 42.3\\
A2261 & 17 22 27.1 & +32 08 02.0 &LC6\_015   & 27-07-17& 3C295 & 550, 5007 & 30.6\\
A2390 & 21 53 34.6  & +17 40 11.0  & LC9\_030  & 18-12-17 & 3C196 & 500, 501, 4193 & 99.1 \\
\hline
\end{tabular}
\label{obs}
\end{table*}

\begin{table*}
 \centering
  \caption{Properties and literature information of the selected sample of non-merging massive clusters. Col. 1: Name of the cluster; Col. 2: redshift; Col. 3: angular to physical scale conversion at the cluster redshift; Col. 4:  mass within the radius enclosing a mean density of 500 times the critical density \citep{Planck2014}; Col. 5: core entropy from X-ray data \citep{Giaci2017}; Col. 6: presence of a cool core (CC) from X-ray data \citep{Giaci2017}; Col. 7: presence of radio diffuse emission as present in the literature (UL = upper limit; MH = mini halo; cMH = candidate mini halo; H = halo); Col. 8: observations present in the literature; Col. 9: radio references: (1) \citealp{Kale2013}; (2) \citealp{Giaci2014};  (3) \citealp{Giaci2017}; (4) \citealp{Govo2009}  (5) \citealp{Ventu2008};  (6) \citealp{Giaci2014b};  (7) \citealp{Sommer2017}. }
\begin{tabular}{c c c c c c c c c }
  \hline
 
Name & $z$  & Scale &  $M_{500}$ & $K_0$ & X-ray & Radio & Telescope \& frequency & Ref. \\ 
        &   &   \scriptsize{(kpc/$''$)} & \scriptsize{ ($10^{14} M_{\odot}$)}&  \scriptsize{(keV cm$^2$)} & & & \\
\hline
\hline

RXCJ0142.0+2131  &  0.280 & 4.28 & 6.07 & 131 & nCC &UL & GMRT: 235/610  MHz & 1\\
A478 & 0.088& 1.66 & 7.06 & 8&CC& MH& VLA: 1.4 GHz & 2\\
PSZ1G139.61+24 &   0.270 &4.17 & 7.09&10 & CC& cMH & GMRT: 610 MHz & 3\\
A1413 & 0.143 & 2.53 & 5.98  &64& nCC & cMH  & VLA: 1.4 GHz & 4\\
A1423 & 0.214 & 3.50& 6.09 & 27&CC& UL  &GMRT: 610 MHz & 5\\
A1576 &    0.302 & 4.51 &   5.98& 186& nCC & UL   & GMRT: 610 MHz & 1\\
RXJ1720.1+2638  &0.164&  2.84& 6.34 & 21&CC& MH  & GMRT: 317/617 MHz, 1.28 GHz & 6\\
& & & & & & & VLA: 1.5, 4.9, 8.4 GHz & 6\\
A2261 &0.224 &3.63 & 7.39& 61& nCC & H & VLA: 1.4 GHz & 7\\
A2390  & 0.234 & 3.75 &  9.48 & 15&CC& H & VLA: 1.4 GHz & 7\\
\hline
\end{tabular}
\label{info1}
\end{table*}

Recent observations with the Giant Metrewave Radio Telescope (GMRT) and the Very Large Array (VLA) have revealed the presence of radio haloes in a few clusters that are not undergoing major mergers, and that - in some cases - host a cool core (i.e. CL1821+643, \citealp{Bona2014}; A2261, A2390, \citealp{Sommer2017}; A2142, \citealp{Ventu2017}), challenging the notion that radio haloes only form in major mergers. It has been proposed that these sources might be connected to the occurrence of minor/off-axis mergers, although it remains unclear how minor mergers could initiate continuum emission on  megaparsec scales. Models (e.g. \citealp{Cassa2006}) predict that minor mergers can generate ultra-steep spectrum (USS) emission ($\alpha < -1.5$) due to the smaller amount of energy available compared to major mergers. However, flatter spectrum sources such as that in CL1821+643, where $\alpha_{323}^{1665} \sim$ -1.1, are difficult to explain in this scenario. Finally, we note that signatures of minor-merging activities and gas-sloshing mechanisms have been detected in clusters containing mini haloes (e.g. \citealp{Gitti2007}; \citealp{Giaci2014b}, \citealp{Savini2018b}). To assess whether the emission discovered in these few cases is common in galaxy clusters or not if looking at low radio frequencies, we selected and studied a sample of nine non-merging galaxy clusters. The observations were carried out with the LOw Frequency ARray (LOFAR; \citealp{VH2013}) with the aim of studying large-scale radio emission to shed light on the non-thermal phenomena in galaxy clusters and the connection with cluster dynamics. Sensitive low-frequency observations allow us to detect steep-spectrum emission that cannot be observed at higher frequencies and are fundamental in the case of slightly disturbed clusters.\\ 

%shed light on the non-thermal phenomena in galaxy clusters and the connection between radio emission from the ICM and cluster dynamics

In case of the non-detection of large-scale diffuse sources, we provide new upper limits on the radio power of haloes. These constraints demonstrate the importance of low-frequency observations to determine the cosmic-ray pressure and energy budget in clusters.\\

%that are more stringent than those derived from FERMI-LAT and GMRT data , are important to determine the cosmic ray pressure and the cosmic ray energy budget in clusters.\\ These limits,  ...."

In this paper, we assume a flat, $\Lambda$CDM cosmology with matter density $\Omega_M = 0.3$ and Hubble constant $H_0 = 67.8$ km s$^{-1}$ Mpc$^{-1}$ \citep{Planck2016}. All our images are in the J2000 coordinate system.

%are required to fully understand the physical mechanisms

\begin{figure}
        \includegraphics[width=\columnwidth]{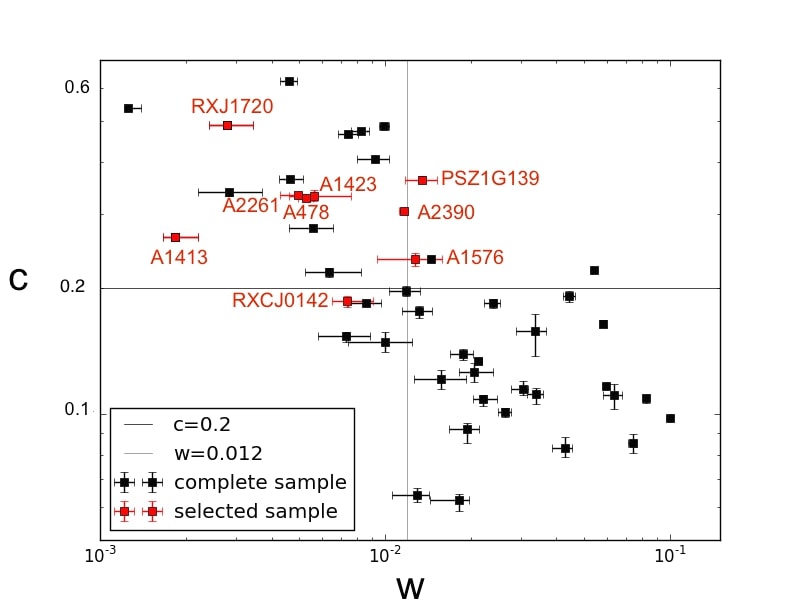}
        \caption{ Diagram of the X-ray morphological indicators based on \textit{Chandra} observations for the galaxy clusters, the emission centroid shift, $w$, and the concentration parameter, $c$, of the mass-selected cluster sample in \citet{Cuci2015} and \citet{Cassa2016}.  Red squares indicate the sources of our sample. Following \citet{Cassa2010}, we adopted the values $w \le 0.012$ and $c \ge 0.2$ as an indication of the distinction between merging/non-merging clusters. Merging clusters lie in the bottom right region of the plot, whilst non-merging clusters in the top left region. }
       \label{plot1}
\end{figure}

\section{The sample}
%: RXCJ0142.0+2131, A478, PSZ1G139.61+24.20, A1413, A1423, A1576, RXJ1720.1+2638, A2261, and A2390
In order to investigate the presence of diffuse emission in non-merging systems at low radio frequencies, we selected a sample of galaxy clusters (listed in Tab. \ref{obs}) on the basis of their cluster-scale X-ray morphology and lack of evidence of a recent major merger. \citet{Cuci2015} and \citet{Cassa2016} studied a sample of clusters selected from the \textit{Planck} Sunyaev–Zel'dovich catalogue (PSZ; \citealt{Planck2014}) with a mass of $M_{500} \ge 6 \times 10^{14} M_{ \odot}$ in the redshift range\footnote{
The sample in \citet{Cuci2015} consists of clusters with available radio and X-ray data. The sample in \citet{Cassa2016}, although covering a smaller range in redshift, also includes clusters with available X-ray data but without radio observations. In total, a completeness in mass greater than 80\% is achieved.} $0.08 < z < 0.33$ and  $0.2 < z < 0.33$, respectively. For this study, we selected the clusters whose dynamical status has been classified as non-merging from \textit{Chandra} X-ray observations. In particular, we focus on two morphological indicators: the concentration parameter and centroid shift, which have been shown to be the most sensitive to the cluster dynamical state  \citet{Lov2017}, We briefly describe these two parameters, as follows:
%by using three morphological indicators derived from X-ray observations:
%taking into account both the completeness in mass and in the radio information. 
\begin{itemize}

%\item \textbf{the power ratio, which is the multipole decomposition of the projected two-dimensional mass distribution inside a given aperture, centred on the cluster X-ray centroid. The lowest ratio of multipoles that provides a good indication of substructure is $P_3/P_0$ (e.g. \citealp{Bo2010})};
%parameter which indicates the presence of multiple peaks in the X-ray distribution providing a clear substructure measure (e.g. %Buote \& Tsai, 1995; Jeltema et al., 2005; Ventimiglia et al., 2008; Buote, 2001; 

\item the concentration parameter, $c$, is defined as the ratio of the X-ray surface brightness within 100 kpc over the X-ray surface brightness within 500 kpc. It helps to select clusters with a compact core, i.e. clusters whose core has not been disrupted by a merger \citep{Santos2008},

\begin{equation}
c = \frac{S(< 100 kpc)}{S(< 500 kpc)} ;
\end{equation}

%the peak over the ambient X-ray surface brightness. 
%It allows to distinguish clusters with a compact core that has not been disrupted by a merger from clusters with a spread distribution of the gas in the core (e.g. \citealp{Santos2008}).

\item and the emission centroid shift, $w$, defined as the standard deviation of the projected separation $\Delta$ between the peak and the cluster X-ray centroid computed within N circles of increasing radius R (e.g. \citealp{Bo2010}),

\begin{equation}
w = \frac{1}{R} \times \sqrt[]{\frac{ \Sigma^{N}_{i=0} (\Delta_i - \langle \Delta \rangle )^2 }{N - 1}}  .
\end{equation}

%Mohr et al., 1993; Poole et al., 2006; O'Hara et al., 2006; Ventimiglia et al., 2008; Maughan et al., 2008; 

 \end{itemize}

High values of $w$ indicate a dynamically disturbed system, while high values of $c$ indicate a peaked core that is typical of non-merging systems. We selected the clusters labelled as ``relaxed'' in \citet{Cassa2016} and \citet{Cuci2015}. In addition, we added the constraint that clusters must be easily observed with LOFAR, i.e. with declination greater than $10^{\circ}$. The selected clusters, which are shown as red symbols in Fig.~\ref{plot1} (together with the clusters belonging to the complete sample), lie in the top left region of the~$c-w$ plot (non-merging clusters). We note that many of these clusters lie near the merging/non-merging boundary (defined with the values $w \sim 0.012$ and $c \sim 0.2$ in \citet{Cassa2010},  where $w$ is derived at 500 kpc.). These clusters are more likely to be slightly disturbed systems, thus through their observations we can search for low-frequency steep-spectrum emission powered by minor/off-axis mergers.

 %Recently, \citet{Lov2017} have shown that $c$ and $w$ are the X-ray morphological indicators that are the most sensitive to the cluster dynamical state, hence, in the following, we will only consider $c$ and $w$. 
%to classify clusters as non-merging systems as those that satisfy: $P_3/P_0 \le 1.2 \cdot 10^{-7}$, $c \ge 0.2$, and $w \le 0.012$,

\section{Data reduction}
\label{sec:radio}

\subsection{Radio: LOFAR observations}

Five of the targets were observed as part of the deep imaging survey LoTSS (LOFAR Two-metre Sky Survey; \citealp{Shim2017}). The remaining four were observed as part of dedicated proposals (LC7\_004, LC8\_006, LC9\_011, LC9\_020), adopting the LoTSS observing set-up \footnote{We note that the clusters observed for the proposals lie at the centre of the pointings and the clusters observed within LoTSS lie at a maximum distance from the phase centre of the pointing of $1^\circ$, where the primary beam response is $> 0.8$. This allows us to reach a consistent sensitivity to diffuse emission for the whole sample.}. The observations were carried out at 120 - 168 MHz using the high band antenna (HBA) with a total on-source time of 8 h preceded and followed by a flux density calibrator observation for 10 minutes. More details can be found in Tab. \ref{obs}. 
The calibration and imaging procedure is based on the facet calibration scheme presented in \citet{vanWeeren2016}. A complete outline of the procedure can be found in \citet{Savini2018a}. The main steps are as follows:

\begin{itemize}
\item Preliminary pre-processing, which was performed by the Radio Observatory (ASTRON) and has been applied to the data
\item Direction-independent calibration, which was obtained by executing the Prefactor pipeline\footnote{https://github.com/lofar-astron/prefactor}, following the strategy outlined in De Gasperin et al. (2018) %\citet{Dega2018}
\item Direction-dependent calibration using the FACTOR pipeline\footnote{https://github.com/lofar-astron/factor}
\end{itemize}

Each field of view was divided into a discrete number of facets that are separately calibrated. The calibrator for each facet was selected with a minimum flux density specified by the user in a range between 0.3 Jy and 0.6 Jy.  The facets are usually processed in order of calibrator flux density, before processing those bordering the cluster facet. The cluster facet is the last to be processed, so that it could benefit from the calibration of the preceding facets. The facet images are stitched together to form a mosaic and the mosaicked final image is corrected for the primary beam. In line with other LOFAR HBA studies, we adopted a systematic calibration error of $15\%$ on all the measured flux densities (\citealp{Shi2016}, \citealp{vanWeeren2016b}, \citealp{Savini2018a}). 
Radio imaging was performed through the {\tt WSClean} package \citep{Off2014} implemented in FACTOR, varying the robust values of the Briggs weighting \citep{Briggs} and \textit{uv}-taper to obtain different resolutions and increase the sensitivity to diffuse emission. To obtain specific images, such as spectral index maps, compact-source-subtracted images, and halo-injected images, we used the Common Astronomy Software Applications (CASA, version 4.5.2; \citealp{Mc2007}) with the multi-scale option of the \texttt{clean} task and took  the \textit{w}-projection parameter into account.

\subsection{Radio: GMRT observation}

We processed archival GMRT observations at 610~MHz of the clusters A478 and RXJ1720.1+2638 using the Source Peeling and Atmospheric Modeling (SPAM) pipeline (see \citealp{Int2017} for details) to perform a detailed spectral analysis with the LOFAR observations. A478 and RXJ1720.1+2638 were observed, respectively, on October 10, 2011 (under project code 21\_017) and July 24, 2011 (under project code 11MOA01). Visibilities at 610~MHz were recorded in one polarisation (RR) over a bandwidth of 32~MHz, as part of a dual-frequency observation. The on-source time was 3.4 h for A478 and 5.2~h for RXJ1720.1+2638. The primary calibrator used for flux and bandpass calibrations in both observations was 3C\,48. We adopted the same flux standard as for LOFAR \citep{SH2012}. A $T_{\rm sys}$ gain correction of $0.963$ for A478 and $0.981$ for RXJ1720.1+2638 was derived using the all-sky map at 408~MHz by \citet{Haslam1995}; this $T_{\rm sys}$ gain correction was subsequently applied. The pipeline removed $56\%$ of the data of A478 due to Radio Frequency Interferences (RFI) and various telescope issues and the pipline removed $48\%$ for RXJ1720.1+2638. The pipeline output visibilities were imported into CASA for final imaging, using the multi-scale option of the \texttt{clean} task. Our highest fidelity images reach a sensitivity of $65 \, \mu$Jy\,beam$^{-1}$ with a $4.8'' \times 3.9''$ beam and $45 \, \mu$Jy\,beam$^{-1}$ with a $5.0'' \times 4.8''$ beam, for A478 and RXJ1720.1+2638 respectively. We adopted a 10\% scale error on all flux density measurements \citep{Chandra2004}.

%with a second band centered on 235~MHz

\subsection{X-ray: \textit{Chandra} observation}

To investigate the connection between the thermal and non-thermal components in the ICM, we reprocessed \textit{Chandra} X-ray observations for each cluster in the sample. The ID and clean exposure time of each observation can be found in Tab. \ref{obs}. We carried out a standard data reduction using CIAO v4.9 and \textit{Chandra} CALDB v4.7.3 to produce the exposure-corrected images in the $0.5-2.0$ keV band shown in the paper (see \citet{Botteon2017} for the procedure outline).\\ 

The \textit{Chandra} X-ray density and temperature profiles of the clusters we selected were already obtained and combined by \citet{Giaci2017} to derive the specific entropy at the cluster centre, $K_0$. We used the values of $K_0$ for the clusters in our sample to distinguish between cool-core or non-cool-core clusters following \citet{Giaci2017}: clusters with low central entropies ($K_0 < 30 - 50$ keV cm$^2$) are expected to host a cool core. The values relevant for our sample are reported in Tab. \ref{info1}, where we specify whether the clusters host a cool core or not, according to the above classification.

\section{Results}
\label{res}

\begin{figure*}
 \includegraphics[width=1\textwidth]
  {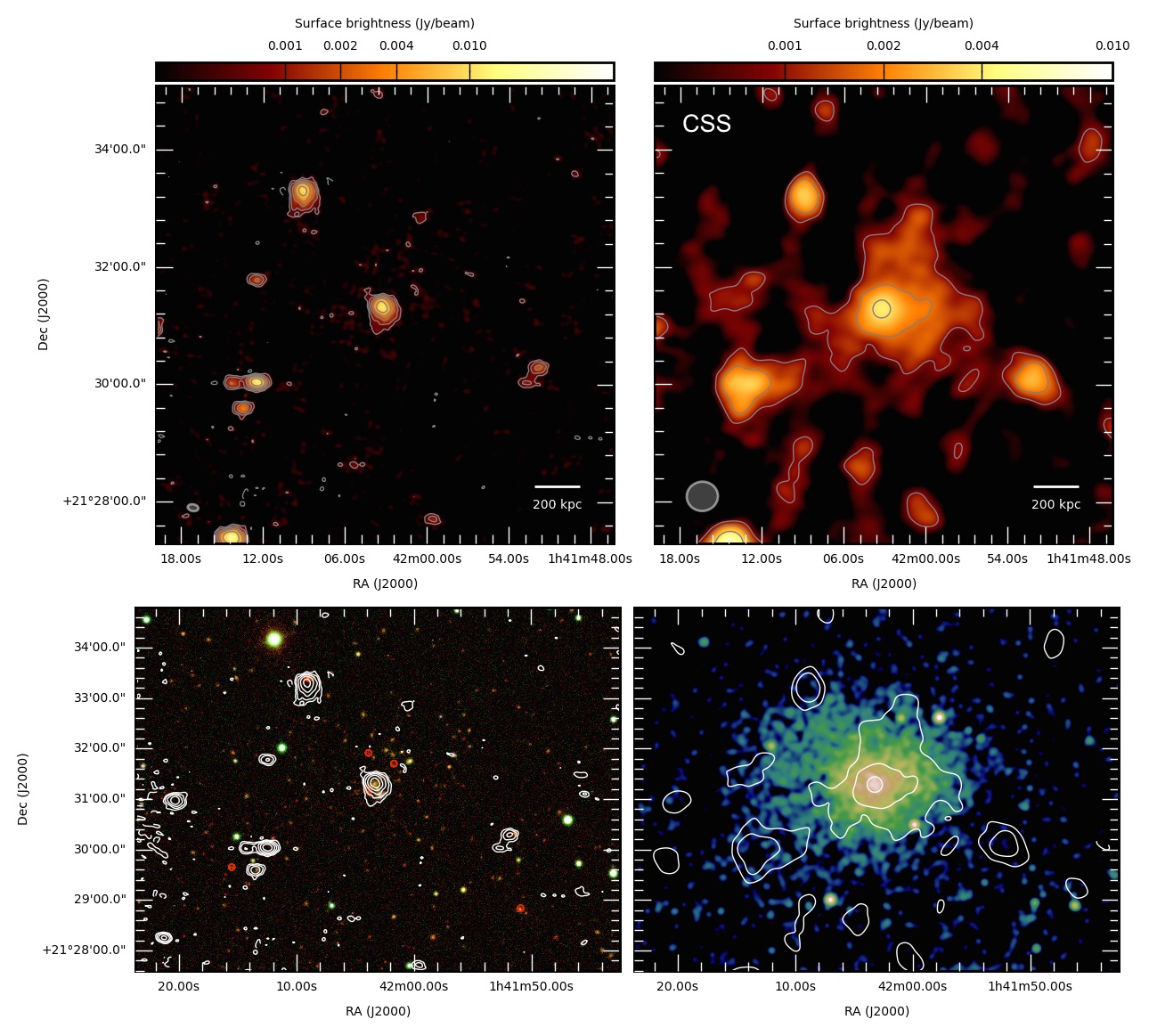}
   \caption{ \, \textit{RXCJ0142.0+2131} \,
   {\bf Top left panel:} 
   High-resolution 144 MHz LOFAR image of RXCJ0142.0+2131. The contour levels start at $3\sigma$ where $\sigma$ = 150 $\mu$Jy\,beam$^{-1}$, and are spaced by a factor of two. The negative contour level at $-3\sigma$ is overlaid with a dashed line. The beam is $11'' \times 7''$ and is shown in grey in the bottom left corner of the image. 
  {\bf Top right panel:} 
   Low-resolution 144 MHz LOFAR image of RXCJ0142.0+2131. The contour levels start at $3\sigma$ where $\sigma$ = 300 $\mu$Jy\,beam$^{-1}$, and are spaced by a factor of two. The negative contour level at $-3\sigma$ is overlaid with a dashed line. The beam is $26'' \times 24''$ and is shown in grey in the bottom left corner of the image. This image was obtained after central source subtraction (CSS) with a taper of $15''$ and Briggs weighting (robust = 0).
 {\bf Bottom left panel:} Optical Pan-STARRS image of RXCJ0142.0+2131 with the high-resolution ($11'' \times 7''$) 144 MHz LOFAR contours overlaid. The red circles indicate the cluster-member galaxies with available spectroscopic redshift.   {\bf Bottom right panel:} \textit{Chandra} X-ray image  of RXCJ0142.0+2131 smoothed on a scale of 5$''$ with the low-resolution ($26'' \times 24''$) 144 MHz LOFAR contours overlaid.}
  \label{rxc}
\end{figure*}

\begin{figure*}
 \includegraphics[width=1\textwidth]
  {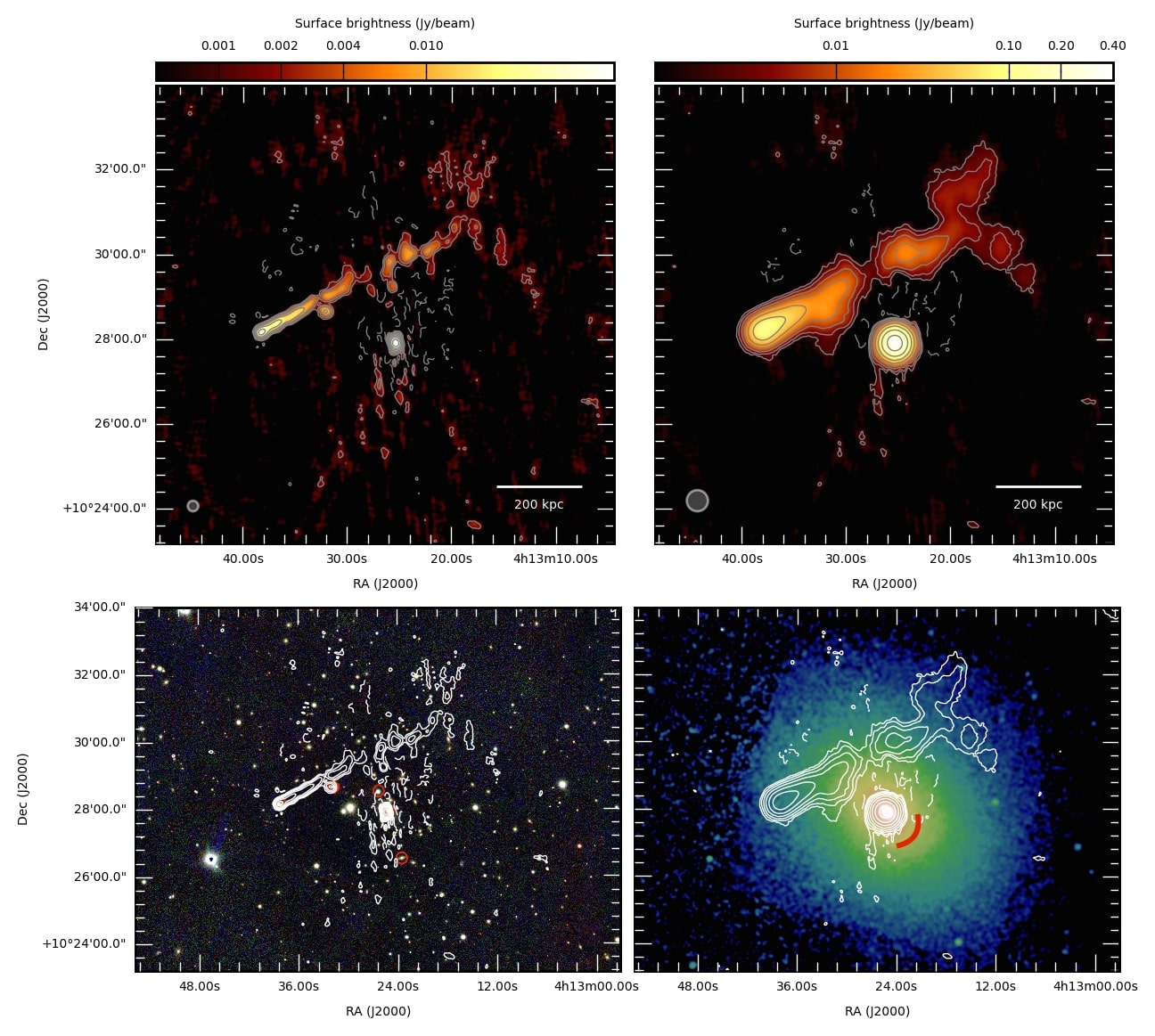}
   \caption{\, \textit{A478} \,   {\bf Top left panel:} 
   High-resolution 144 MHz LOFAR image of A478. The contour levels start at $3\sigma$ where $\sigma$ = 460 $\mu$Jy\,beam$^{-1}$ and are spaced by a factor of two. The negative contour level at $-3\sigma$ is overlaid with a dashed line. The beam is $10'' \times 10''$ and is shown in grey in the bottom left corner of the image. 
   {\bf Top right panel:}  
   Low-resolution 144 MHz LOFAR image of A478. The contour levels start at $3\sigma$, where $\sigma$ = 620 $\mu$Jy\,beam$^{-1}$, and are spaced by a factor of two. The negative contour level at $-3\sigma$ is overlaid with a dashed line. The beam is $30'' \times 30''$ and is shown in grey in the bottom left corner of the image.  
   {\bf Bottom left panel:}  Optical Pan-STARRS image of A478 with the high-resolution ($10'' \times 10''$) 144 MHz LOFAR contours overlaid. The red circles indicate the cluster-member galaxies with available spectroscopic redshift.    
     {\bf Bottom right panel:} \textit{Chandra} X-ray image  of A478 smoothed on a scale of 5$''$ with the low-resolution ($30'' \times 30''$) 144 MHz LOFAR contours overlaid. The red arc indicates the position of the cold front found by \citet{Mark2003}.}
  \label{a478}
\end{figure*}

\begin{figure*}
 \includegraphics[width=0.5\textwidth]
  {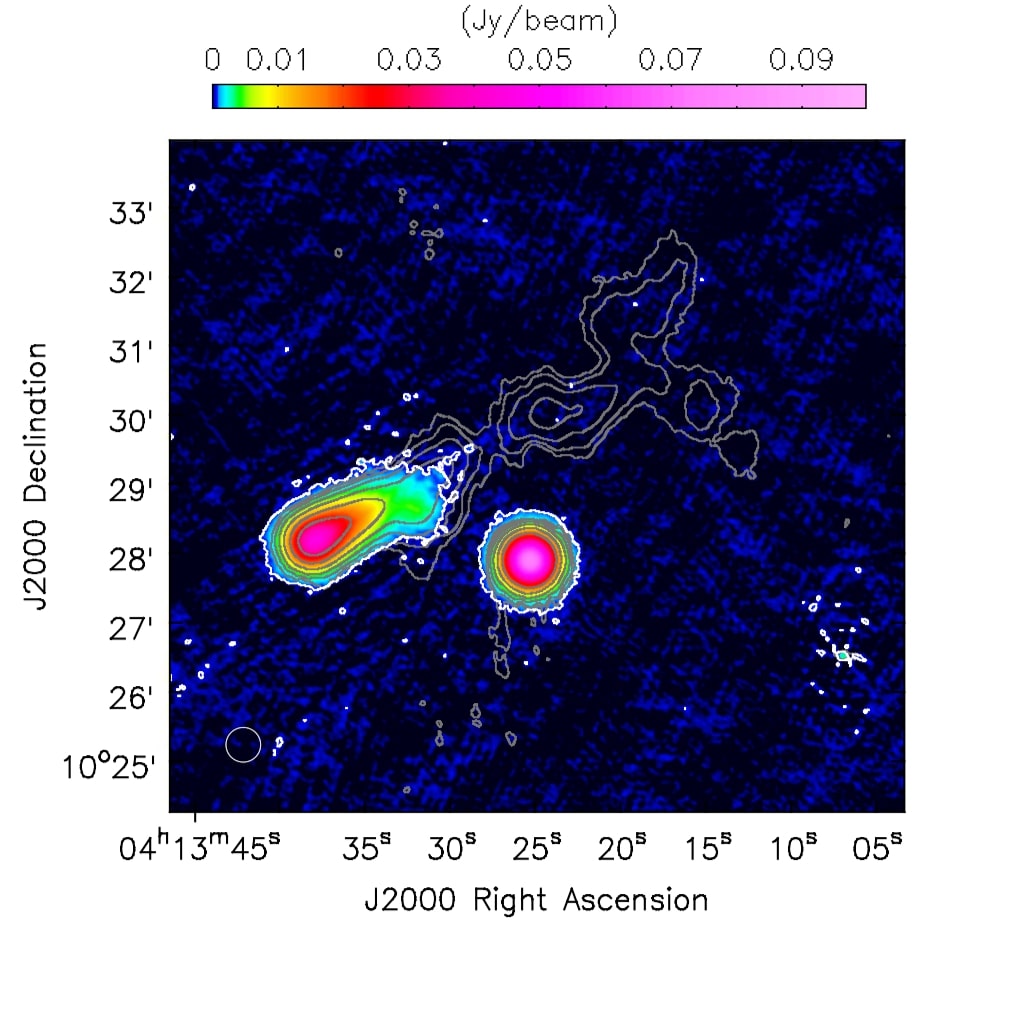}
 \includegraphics[width=0.5\textwidth]
  {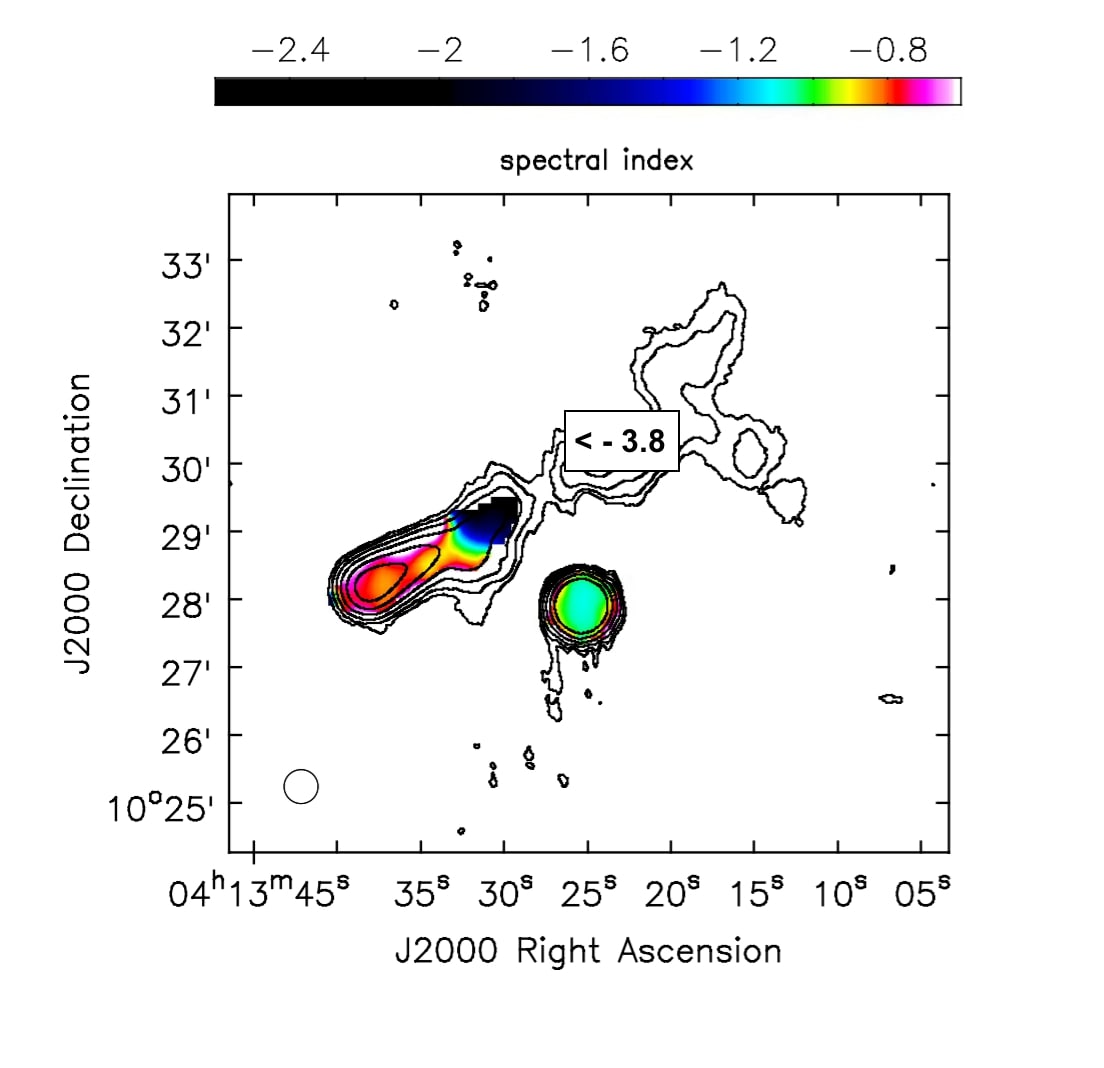}
   \caption{  Contour levels from the low-resolution LOFAR image of A478 in Fig. \ref{a478} are overlaid in grey and black.
   {\bf Left panel:} 
   610 MHz GMRT image of A478 with its $3\sigma$ contour level in white where $\sigma$ = 90 $\mu$Jy\,beam$^{-1}$. The beam is $30'' \times 30''$ for both GMRT and LOFAR, and is shown in the bottom left corner of the image. 
  {\bf Right panel:} Spectral index map between the 610 MHz GMRT and 144 MHz LOFAR images of A478. Pixels below 3$\sigma$ are blanked. The error map is shown in Fig. \ref{spix-err1}.}
  \label{a478-spix}
\end{figure*}

\begin{figure*}
 \includegraphics[width=1\textwidth]
  {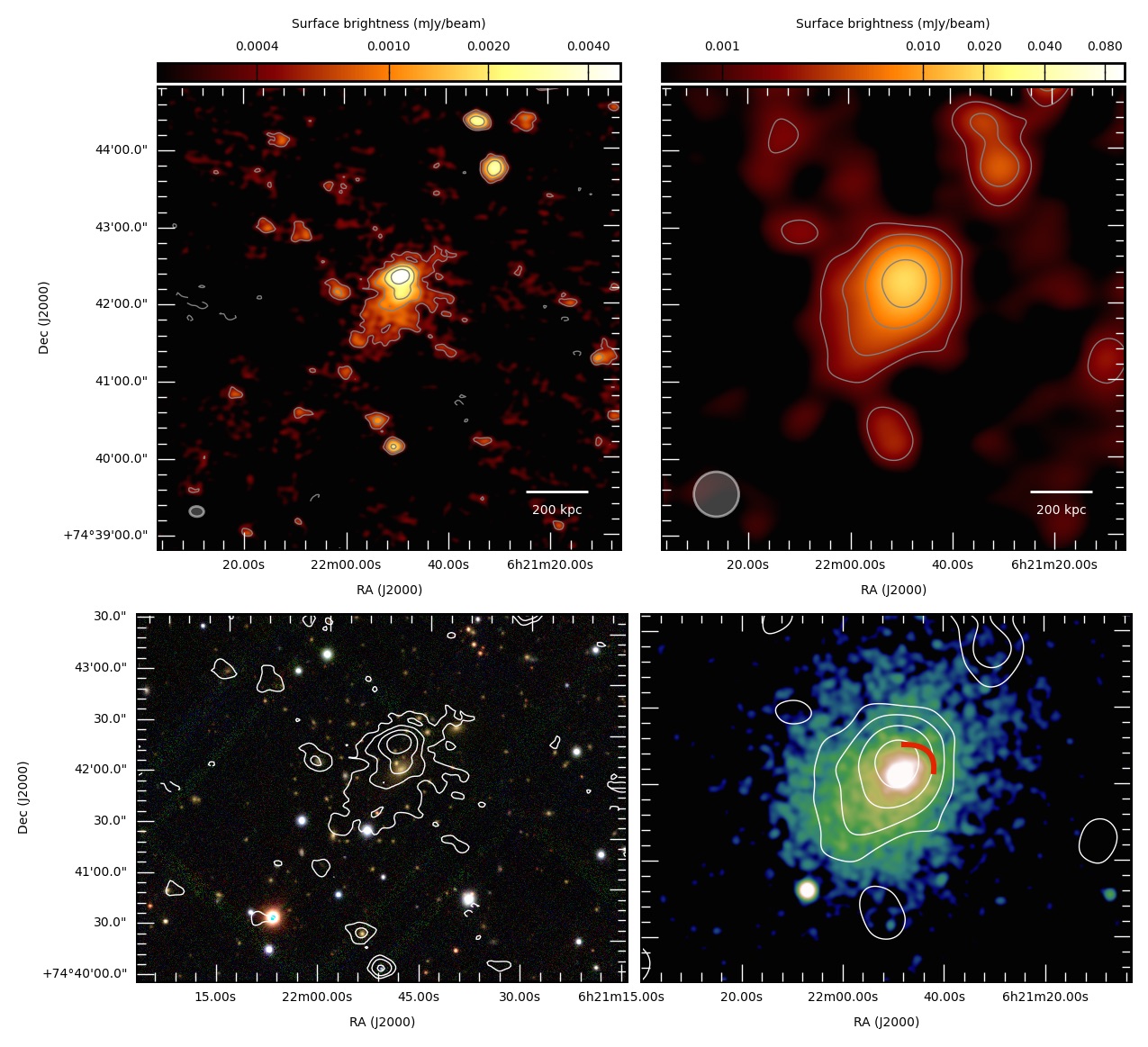}
   \caption{ \, \textit{PSZ1G139.61+24} \, 
   {\bf Top left panel:} 
 High-resolution 144 MHz LOFAR image of PSZ1G139.61+24. The contour levels start at $3\sigma$, where $\sigma$ = 150 $\mu$Jy\,beam$^{-1}$, and are spaced by a factor of two. The negative contour level at $-3\sigma$ is overlaid with a dashed line. The beam is $11'' \times 8''$ and is shown in grey in the bottom left corner of the image. 
  {\bf Top right panel:} 
Low-resolution 144 MHz LOFAR image of PSZ1G139.61+24. The contour levels start at $3\sigma$, where $\sigma$ = 500 $\mu$Jy\,beam$^{-1}$, and are spaced by a factor of two. The negative contour level at $-3\sigma$ is overlaid with a dashed line. The beam is $35'' \times 35''$ and is shown in grey in the bottom left corner of the image. 
  {\bf Bottom left panel:} 
  Optical Pan-STARRS image of PSZ1G139.61+24 with the high-resolution ($11'' \times 8''$) 144 MHz LOFAR contours overlaid.  
   {\bf Bottom right panel:} 
   \textit{Chandra} X-ray image of PSZ1G139.61+24 smoothed on a scale of 6$''$ with the low-resolution ($35'' \times 35''$) 144 MHz LOFAR contours overlaid. The red arc indicates the position of the cold front found by \citet{Savini2018b}.}
  \label{psz}
\end{figure*}

\begin{figure*}
 \includegraphics[width=1\textwidth]
  {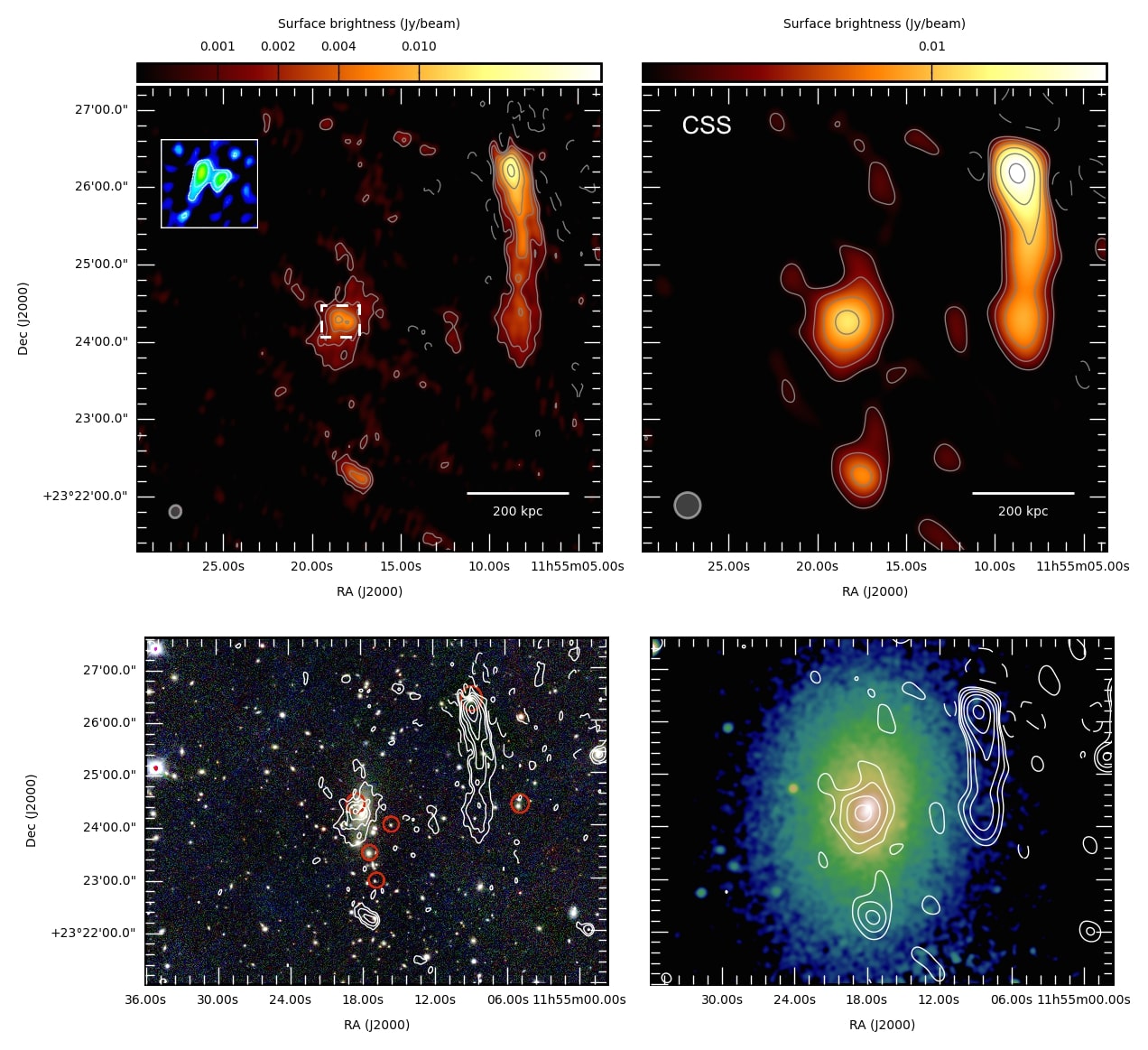}
   \caption{ \, \textit{A1413} \, 
   {\bf Top left panel:} 
 High-resolution 144 MHz LOFAR image of A1413.The contour levels start at $3\sigma$, where $\sigma$ = 270 $\mu$Jy\,beam$^{-1}$, and are spaced by a factor of two. The negative contour level at $-3\sigma$ is overlaid with a dashed line. The beam is $10'' \times 9''$ and is shown in grey in the bottom left corner of the image. The insert box shows the \textit{uv}-cut high-resolution image of the central sources that are subtracted to obtained the image in top right panel.
  {\bf Top right panel:} 
Low-resolution 144 MHz LOFAR image of A1413. The contour levels start at $3\sigma$, where $\sigma$ = 450 $\mu$Jy\,beam$^{-1}$, and are spaced by a factor of two. The negative contour level at $-3\sigma$ is overlaid with a dashed line. The beam is $20'' \times 20''$ and is shown in grey in the bottom left corner of the image. This image was obtained after the CSS with a taper of $20''$ and Briggs weighting (robust = 0).
  {\bf Bottom left panel:} 
  Optical SDSS image of A1413 with the high-resolution ($10'' \times 9''$) 144 MHz LOFAR contours overlaid.  The red circles indicate the cluster-member galaxies with available spectroscopic redshift.     
   {\bf Bottom right panel:} 
   \textit{Chandra} X-ray image  of A1413 smoothed on a scale of 5$''$ with the low-resolution ($20'' \times 20''$) 144 MHz LOFAR contours overlaid.}
  \label{a1413}
\end{figure*}

\begin{figure*}
 \includegraphics[width=1\textwidth]
  {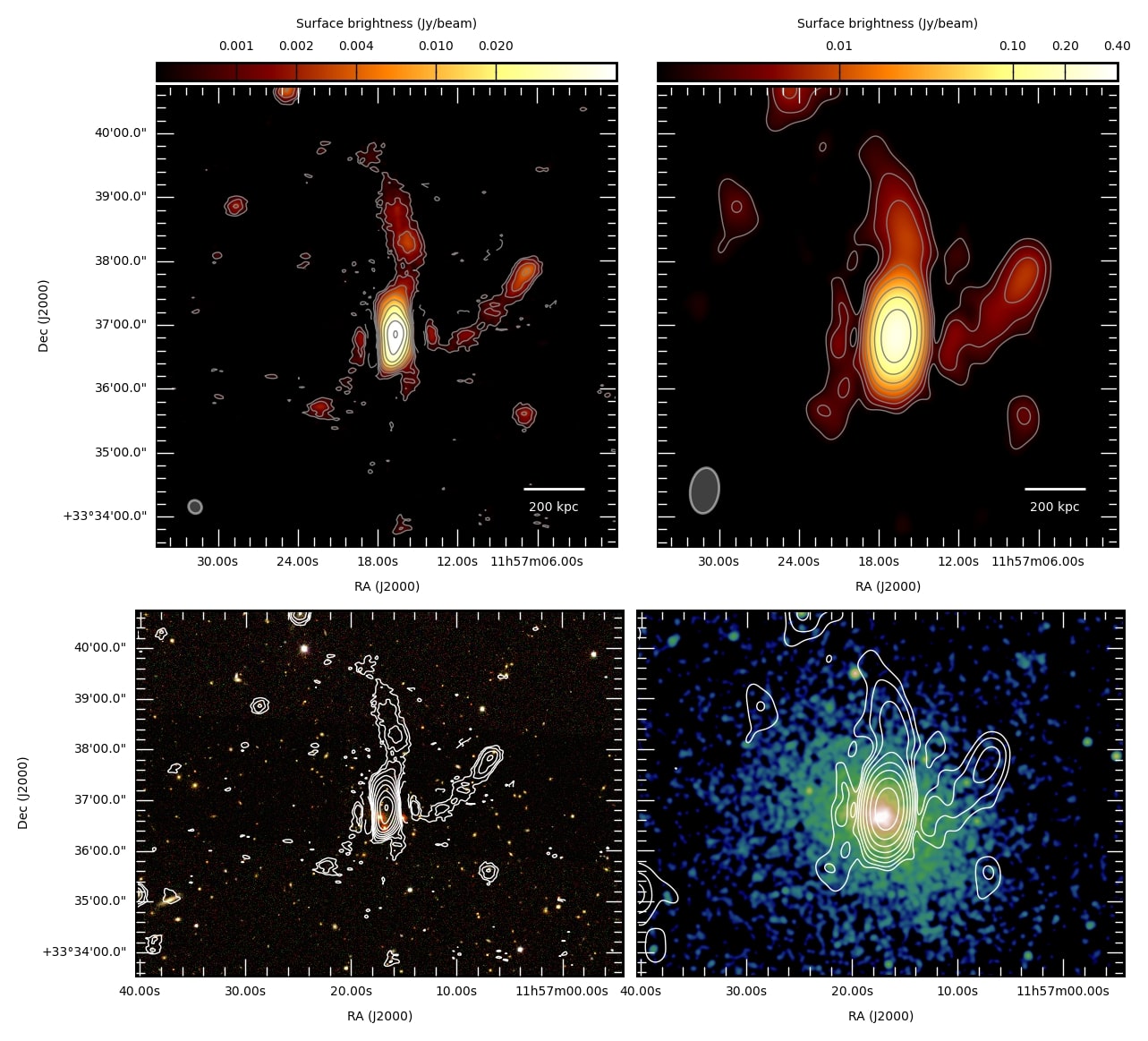}
   \caption{ \, \textit{A1423} \, 
   {\bf Top left panel:} 
 High-resolution 144 MHz LOFAR image of A1423. The contour levels start at $3\sigma$, where $\sigma$ = 170 $\mu$Jy\,beam$^{-1}$, and are spaced by a factor of two. The negative contour level at $-3\sigma$ is overlaid with a dashed line. The beam is $13'' \times 12''$ and is shown in grey in the bottom left corner of the image. 
  {\bf Top right panel:} 
Low-resolution 144 MHz LOFAR image of A1423. The contour levels start at $3\sigma$, where $\sigma$ = 420 $\mu$Jy\,beam$^{-1}$, and are spaced by a factor of two. The negative contour level at $-3\sigma$ is overlaid with a dashed line. The beam is $43'' \times 27''$ and is shown in grey in the bottom left corner of the image. 
  {\bf Bottom left panel:} 
  Optical Pan-STARRS image of A1423 with the high-resolution ($43'' \times 27''$) 144 MHz LOFAR contours overlaid.   The red circles indicate the cluster-member galaxies with available spectroscopic redshift.    
   {\bf Bottom right panel:} 
   \textit{Chandra} X-ray image  of A1423 smoothed on a scale of 5$''$ with the low-resolution ($43'' \times 27''$) 144 MHz LOFAR contours overlaid.}
  \label{a1423}
\end{figure*}

\begin{figure*}
 \includegraphics[width=1\textwidth]
  {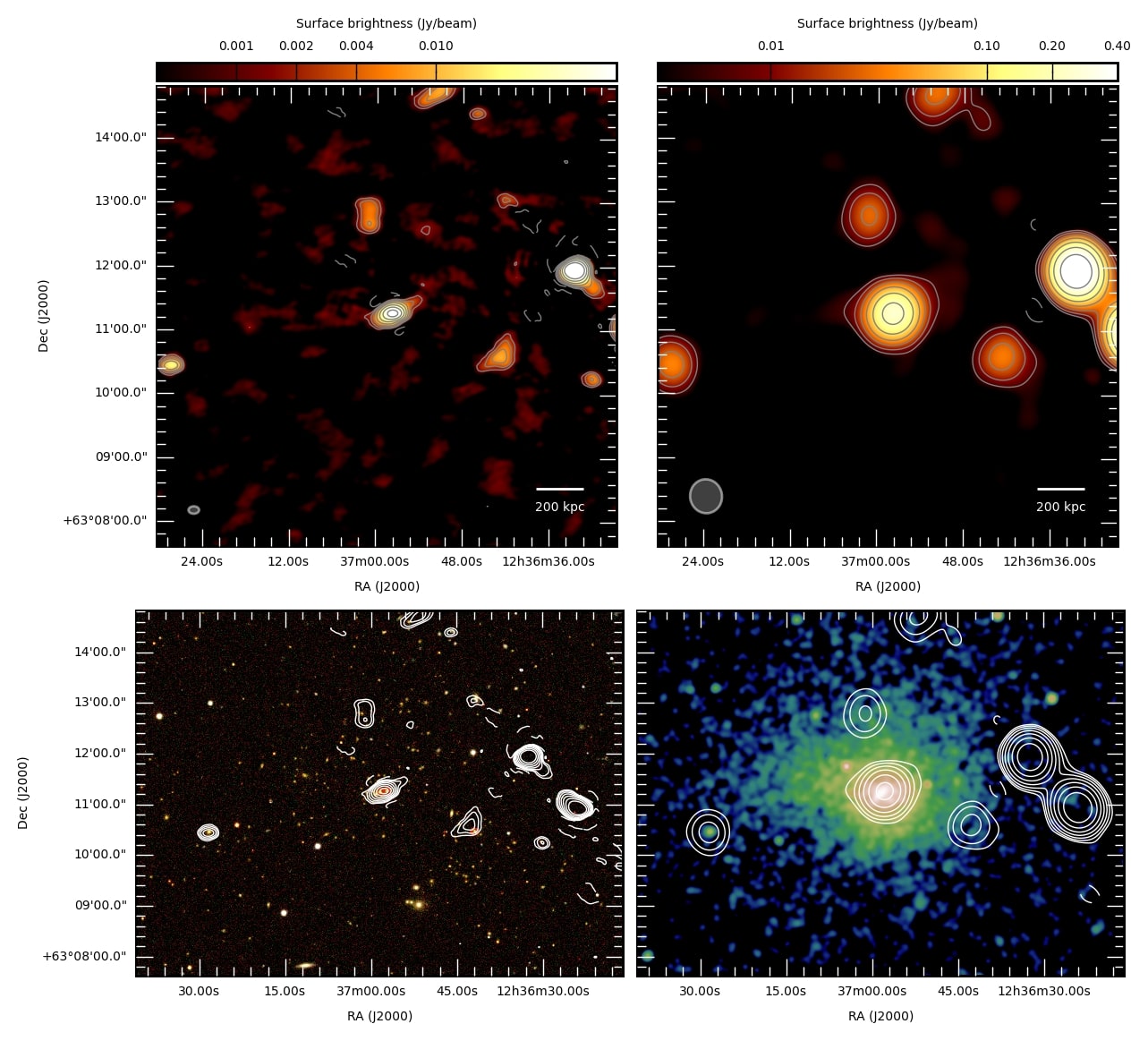}
   \caption{ \, \textit{A1576} \, 
   {\bf Top left panel:} 
   High-resolution 144 MHz LOFAR image of A1576. The contour levels start at $3\sigma$, where $\sigma$ = 500 $\mu$Jy\,beam$^{-1}$, and are spaced by a factor of two. The negative contour level at $-3\sigma$ is overlaid with a dashed line. The beam is $10'' \times 7''$ and is shown in grey in the bottom left corner of the image. 
   {\bf Top right panel:}  
   Low-resolution 144 MHz LOFAR image of A1576. The contour levels start at $3\sigma$, where $\sigma$ = 150 $\sigma$ = 2 mJy\,beam$^{-1}$, and are spaced by a factor of two. The negative contour level at $-3\sigma$ is overlaid with a dashed line. The beam is $32'' \times 30''$ and is shown in grey in the bottom left corner of the image.  
   {\bf Bottom left panel:}  Optical Pan-STARRS image of A1576 with the high-resolution ($10'' \times 7''$) 144 MHz LOFAR contours overlaid.  The red circles indicate the cluster-member galaxies with available spectroscopic redshift.      
     {\bf Bottom right panel:} \textit{Chandra} X-ray image  of A1576 smoothed on a scale of 5$''$ with the low-resolution ($32'' \times 30''$) 144 MHz LOFAR contours overlaid.}
  \label{a1576}
\end{figure*}

\begin{figure*}
 \includegraphics[width=1\textwidth]
  {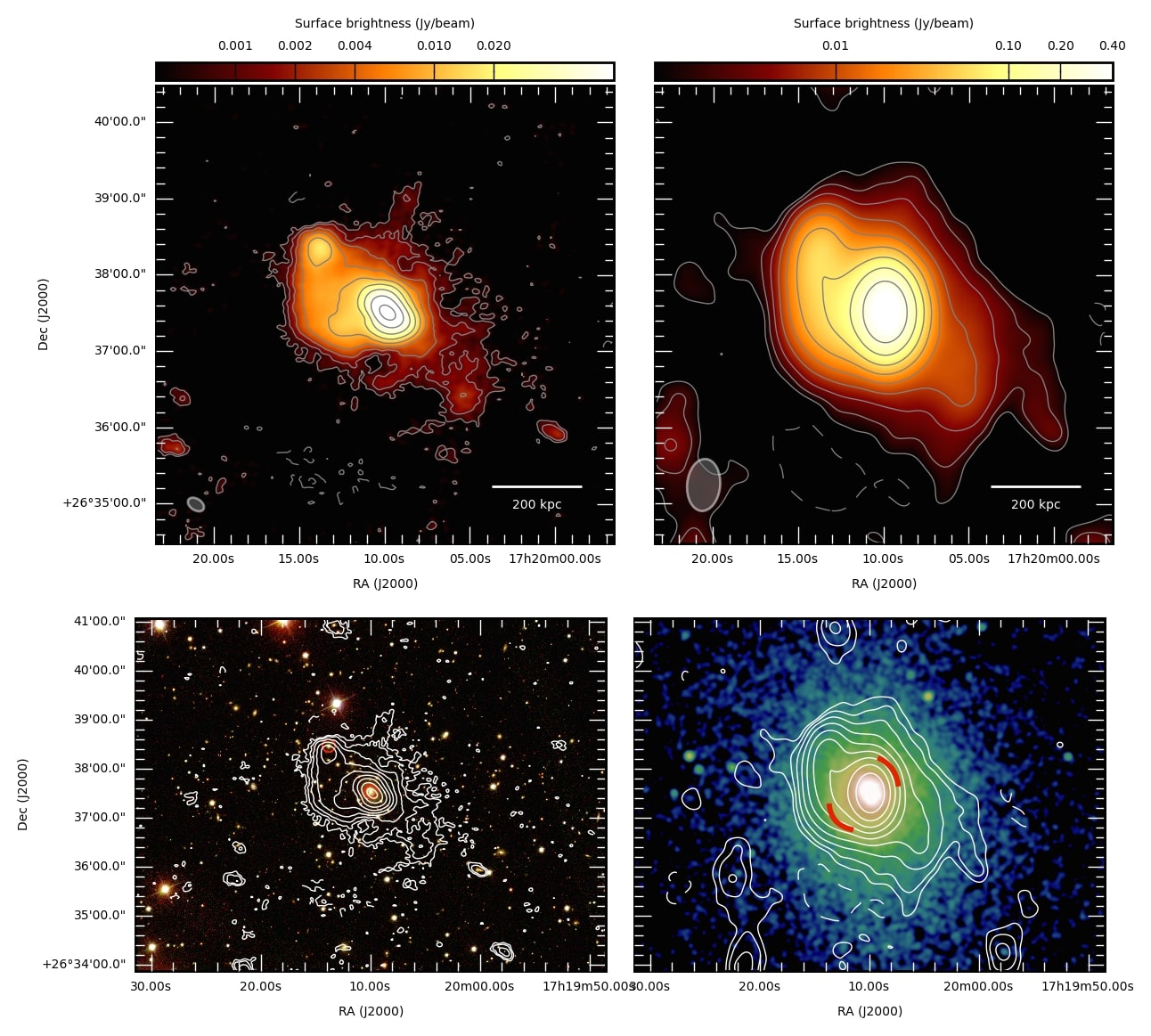}
   \caption{ \, \textit{RXJ1720.1+2638} \, 
   {\bf Top left panel:} 
   High-resolution 144 MHz LOFAR image of RXJ1720.1+2638. The contour levels start at $3\sigma$, where $\sigma$ = 200 $\mu$Jy\,beam$^{-1}$, and are spaced by a factor of two. The negative contour level at $-3\sigma$ is overlaid with a dashed line. The beam is $14'' \times 9''$ and is shown in grey in the bottom left corner of the image. 
  {\bf Top right panel:} 
   Low-resolution 144 MHz LOFAR image of RXJ1720.1+2638. The contour levels start at $3\sigma$, where $\sigma$ = 330 $\mu$Jy\,beam$^{-1}$, and are spaced by a factor of two. The negative contour level at $-3\sigma$ is overlaid with a dashed line. The beam is $41'' \times 26''$ and is shown in grey in the bottom left corner of the image. 
 {\bf Bottom left panel:} Optical Pan-STARRS image of RXJ1720.1+2638 with the high-resolution ($14'' \times 9''$) 144 MHz LOFAR contours overlaid.   The red circles indicate the cluster-member galaxies with available spectroscopic redshift.    
    {\bf Bottom right panel:} \textit{Chandra} X-ray image  of RXJ1720.1+2638 smoothed on a scale of 5$''$ with the low-resolution ($41'' \times 26''$) 144 MHz LOFAR contours overlaid. The red arcs indicate the position of the two cold fronts found by \citet{Giaci2014b}.}
  \label{rxj1720}
\end{figure*}

\begin{figure*}
 \includegraphics[width=0.5\textwidth]
  {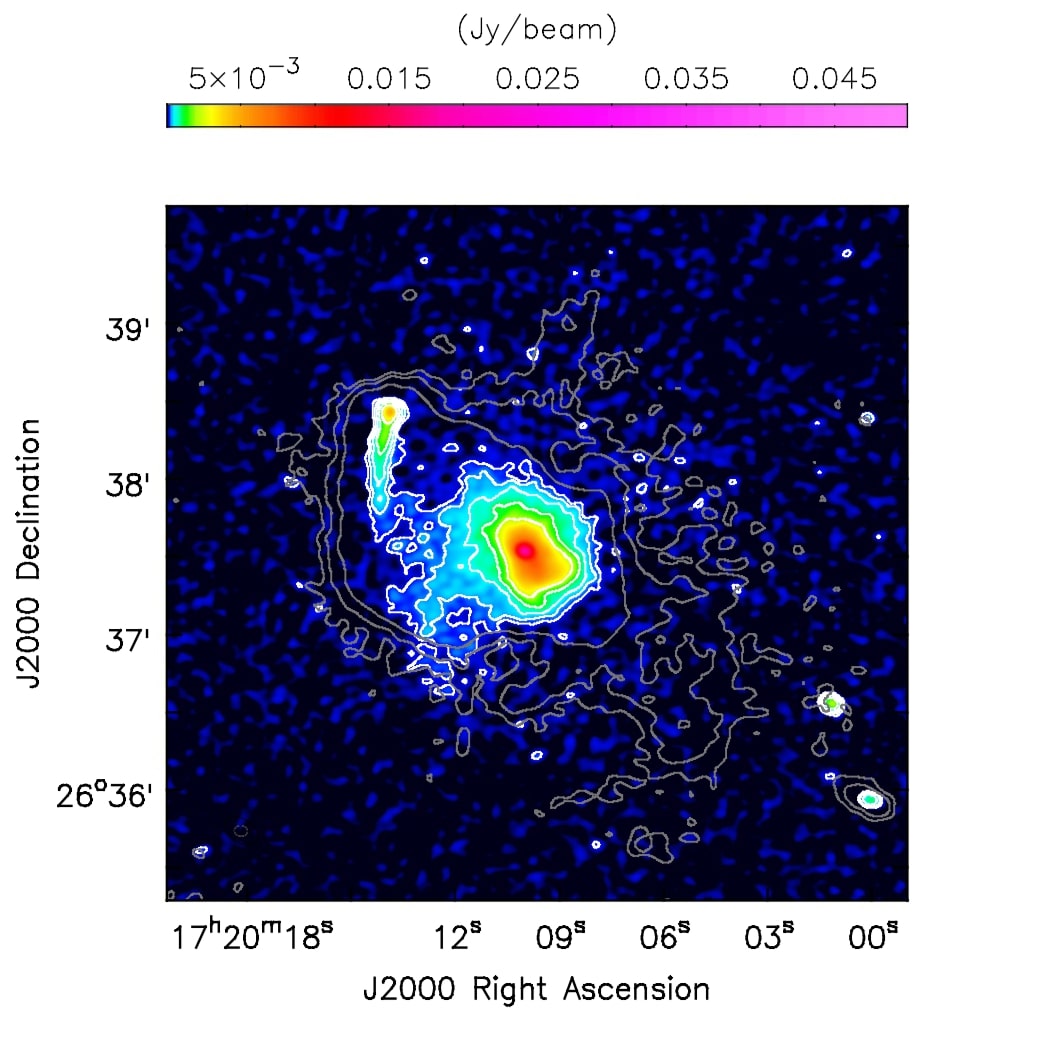}
 \includegraphics[width=0.45\textwidth]
  {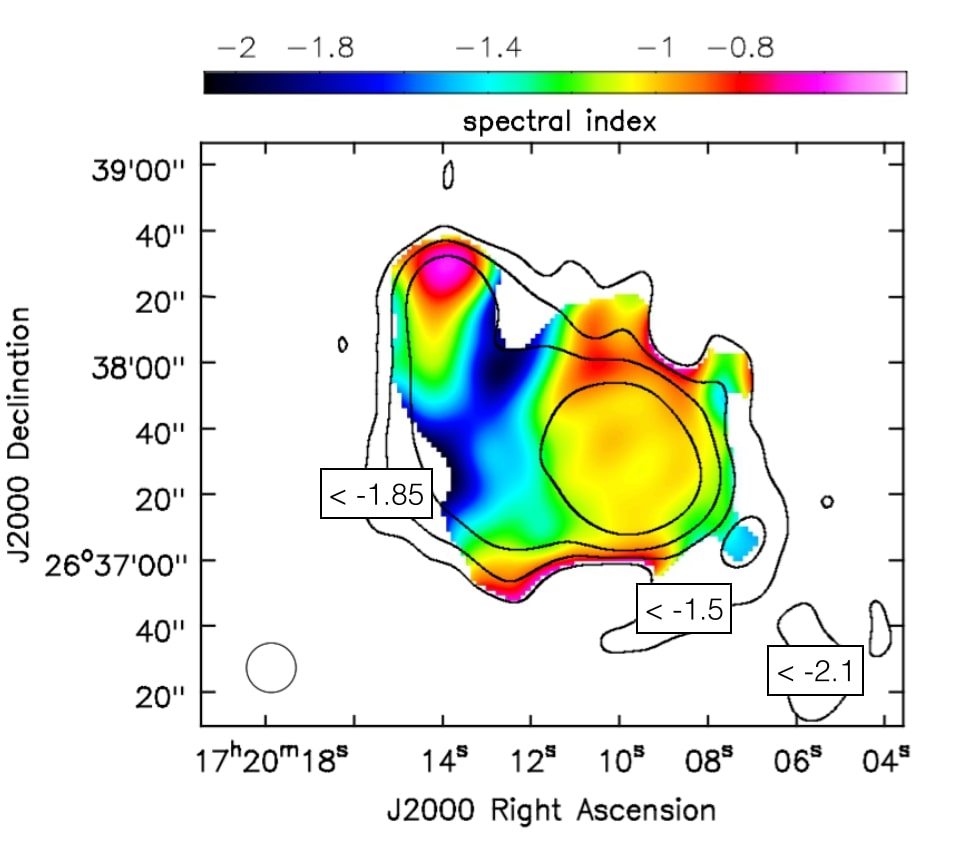}
   \caption{ 
   {\bf Left panel:} 
   610 MHz GMRT image of RXJ1720.1+2638 with its contour levels in white starting at $3\sigma$, where $\sigma$ = 45 $\mu$Jy\,beam$^{-1}$ spaced by a factor of two. The first three contour levels from the high-resolution LOFAR image of Fig. \ref{rxj1720} are overlaid in grey. The beam are $5'' \times 5''$ and $14'' \times 9''$ for GMRT and LOFAR, respectively, and are shown in the bottom left corner of the image. 
  {\bf Right panel:} Spectral index map between the 610 MHz GMRT and 144 MHz LOFAR images of RXJ1720.1+2638. The contour levels, overlaid in black, are from the LOFAR image obtained for the spectral analysis (uniform weighting, \textit{uv}-cut, see main text), not shown in the paper. Pixels below 3$\sigma$ are blanked. The error map is shown in Fig. \ref{spix-err2}.}
  \label{rxj-spix}
\end{figure*}

\begin{figure*}
 \includegraphics[width=1\textwidth]
  {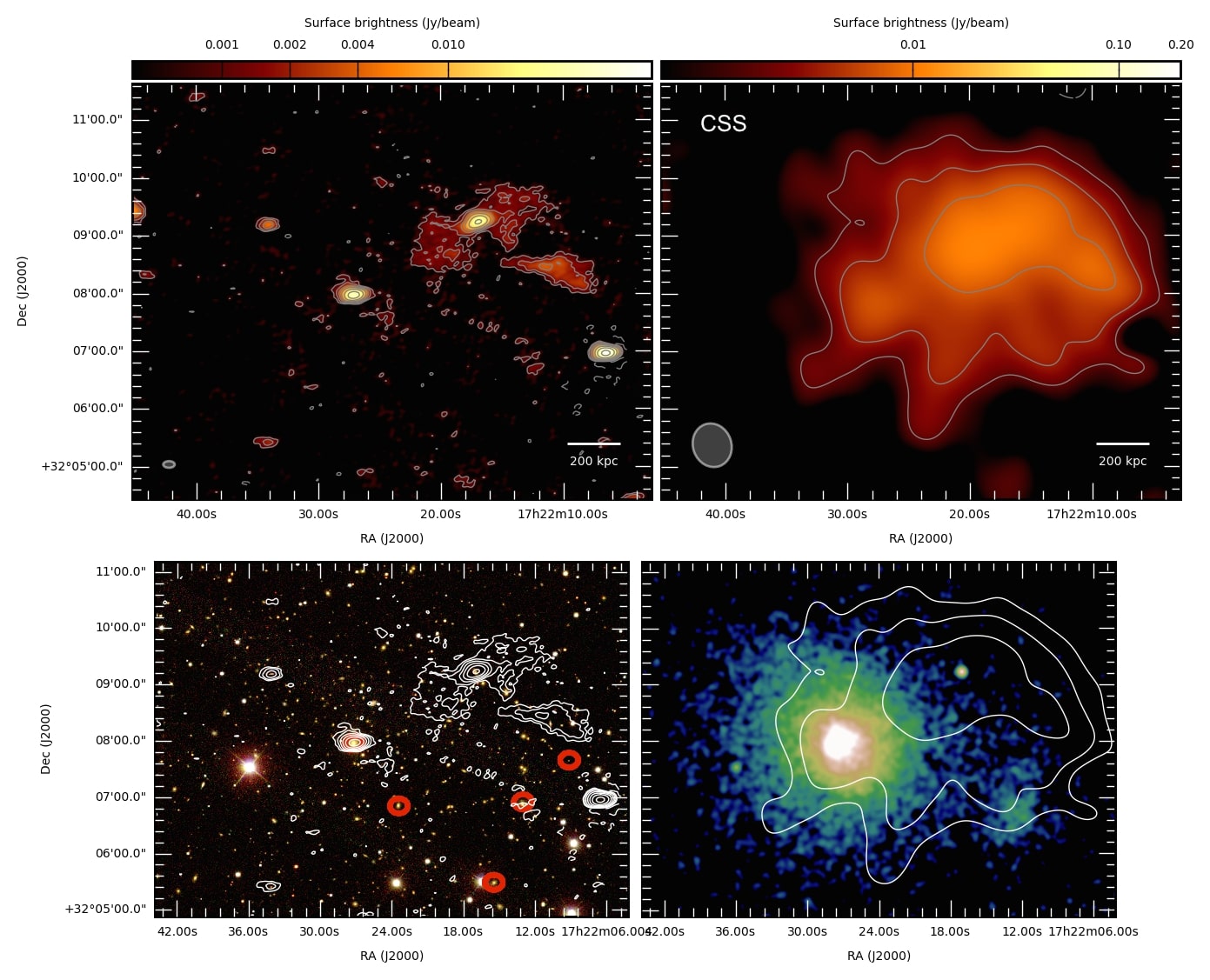}
   \caption{ \, \textit{A2261} \,
   {\bf Top left panel:} 
   High-resolution 144 MHz LOFAR image of A2261. The contour levels start at $3\sigma$, where $\sigma$ = 280 $\mu$Jy\,beam$^{-1}$, and are spaced by a factor of two.  The negative contour level at $-3\sigma$ is overlaid with a dashed line. The beam is $12'' \times 7''$ and is shown in grey in the bottom left corner of the image. 
   {\bf Top right panel:}  
   Low-resolution 144 MHz LOFAR image of A2261. The contour levels start at $3\sigma$, where $\sigma$ = 600 $\mu$Jy\,beam$^{-1}$, and are spaced by a factor of two.  The negative contour level at $-3\sigma$ is overlaid with a dashed line. The beam is $46'' \times 40''$ and is shown in grey in the bottom left corner of the image. This image was obtained after CSS with a taper of $35''$ and Briggs weighting (robust = 0).
   {\bf Bottom left panel:}  Optical SDSS image of A2261 with the high-resolution ($12'' \times 7''$) 144 MHz LOFAR contours overlaid.  The red circles indicate the cluster-member galaxies with available spectroscopic redshift.      {\bf Bottom right panel:} \textit{Chandra} X-ray image  of A2261 smoothed on a scale of 5$''$ with the low-resolution ($46'' \times 40''$) 144 MHz LOFAR contours overlaid.}
  \label{a2261}
\end{figure*}

\begin{figure*}
 \includegraphics[width=1\textwidth]
  {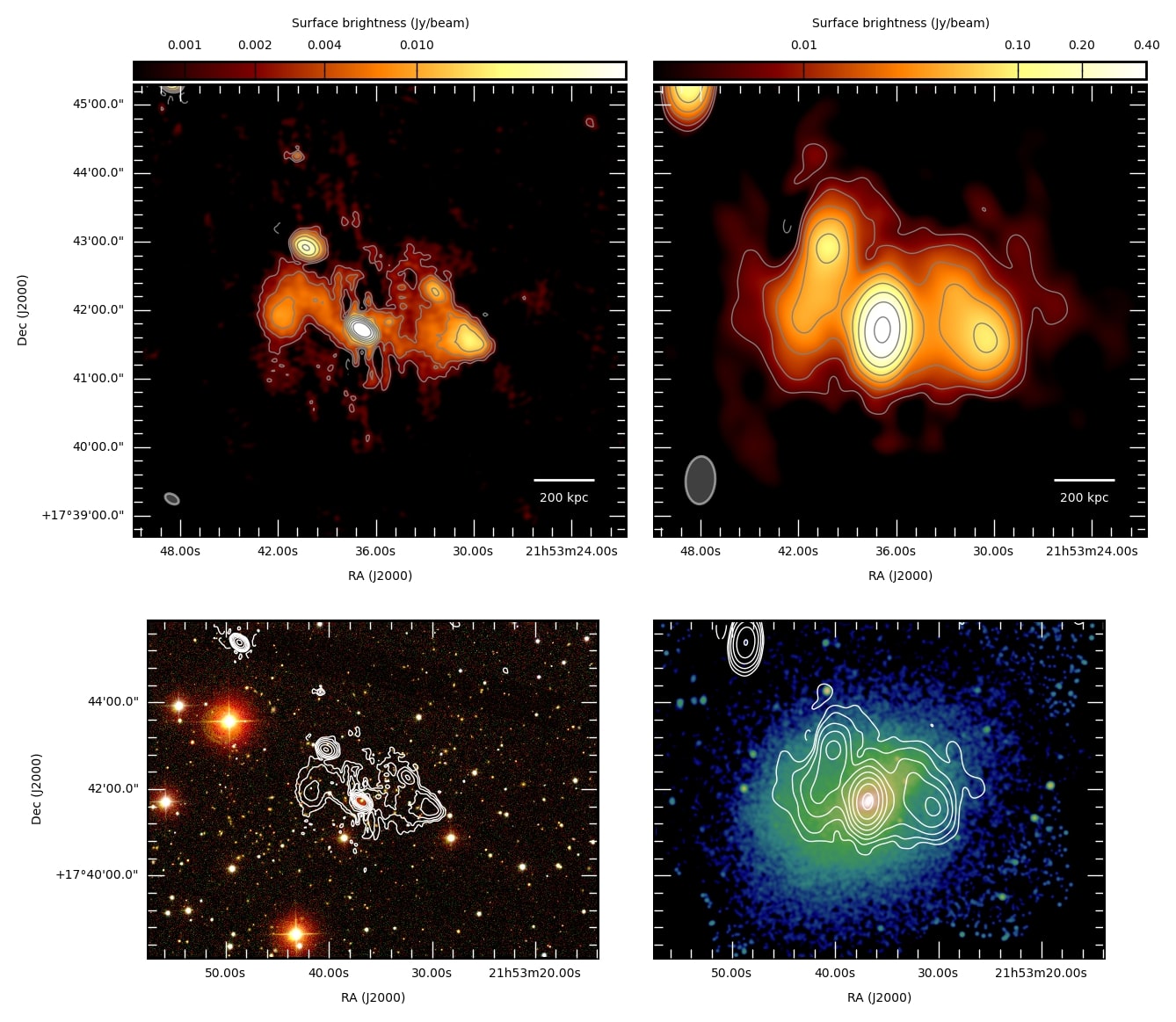}
   \caption{ \, \textit{A2390} \,
   {\bf Top left panel:} 
   High-resolution 144 MHz LOFAR image of A2390. The contour levels start at $3\sigma$, where $\sigma$ = 400 $\mu$Jy\,beam$^{-1}$, and are spaced by a factor of two.  The negative contour level at $-3\sigma$ is overlaid with a dashed line. The beam is $13'' \times 8''$ and is shown in grey in the bottom left corner of the image. 
  {\bf Top right panel:} 
   Low-resolution 144 MHz LOFAR image of A2390. The contour levels start at $3\sigma$, where $\sigma$ = 1.2 mJy\,beam$^{-1}$, and are spaced by a factor of two.  The negative contour level at $-3\sigma$ is overlaid with a dashed line. The beam is $42'' \times 26''$ and is shown in grey in the bottom left corner of the image. 
 {\bf Bottom left panel:} Optical Pan-STARRS image of A2390 with the high-resolution ($13'' \times 8''$) 144 MHz LOFAR contours overlaid. 
    {\bf Bottom right panel:} \textit{Chandra} X-ray image  of A2390 smoothed on a scale of 5$''$ with the low-resolution ($42'' \times 26''$) 144 MHz LOFAR contours overlaid.}
  \label{a2390}
\end{figure*}

In the following, each cluster is described in a separate subsection. The information from the literature is summarised in Tab.~\ref{info1}, while the LOFAR results are summarised in Tab.~\ref{info2}. In Fig.~\ref{rxc} - \ref{a2390}, we present the images obtained at low frequencies in our campaign of LOFAR observations: a high- and a low-resolution image at the central frequency of 144 MHz for each cluster and overlays with the X-ray images taken by \textit{Chandra} and optical images taken within the Sloan Digital Sky Survey (SDSS) or the Panoramic Survey Telescope and Rapid Response System (Pan-STARRS). Our high-resolution LOFAR images with a larger field of view, which show the quality of the facet calibration and the presence of additional sources in the cluster periphery, can be found in the Appendix. The size of the diffuse emission is given by $D_{\rm radio} = \sqrt{D_{\rm min} \times D_{\rm max}}$, where $D_{\rm min}$ and $D_{\rm max}$ are the minimum and maximum diameter of the 3$\sigma$ surface brightness isocontours (e.g. \citealt{Cassa2008}; \citealt{Giaci2017}), as measured from the low-resolution LOFAR images.\\

In some cases, to reveal diffuse emission, it is necessary to subtract the contribution of compact sources in the cluster centre. Therefore, we image the sources at high resolution applying a cut in the \textit{uv}-range, subtract them from the \textit{uv}-data, and re-image the data sets at lower resolution (see \citet{Wilber2018} for details on the subtraction procedure). The value of the \textit{uv}-cut varies from target to target, depending on the extension of the sources that we attempt to subtract. The error on the total flux density is computed as $\sqrt{\sigma_{\rm cal}^2 + \sigma_{\rm sub}^2}$, where $\sigma_{\rm cal}$ is equal to $15\%$ of the measured flux density and $\sigma_{\rm sub}$ is the subtraction error. The latter is derived by varying the range of the {\it uv}-cut that is used for modelling the compact sources. We can then quantify the error on the total flux density of the diffuse emission after subtracting different models.\\

When no hint of cluster-scale radio emission is detected, we provide upper limits on the radio power through the mock halo injection procedure (e.g. \citealt{Ventu2008}, \citealt{Bona2017}). We created a mock radio halo with a central brightness $I_0$ and an $e$-folding radius $r_{\rm e}$ derived from the total radio power at 1.4 GHz ($P_{1.4}$) and halo size ($R_{\rm H}$) following the known power-mass correlation found by \citet{Cassa2013} for haloes. We used the relation $R_{\rm H}/r_{\rm e} = 2.6$ found by \citet{Bona2017}, who compared the values of $R_{\rm H}$ and $r_{\rm e}$ of the clusters; these values are both in the samples of \citet{Cassa2007} and \citet{Murgia2009}. We then carried out a Fourier transform of the mock source into the {\it uv}-data of the LOFAR observation, which is then imaged taking into account the \textit{w}-projection parameter. We chose a relatively empty region near the cluster centre, void of bright sources or artefacts, to host the injected flux density. We decreased the total flux density of the mock halo until it could not be detected in the LOFAR image, i.e. when the surface brightness above $2 \sigma$ has a maximum linear size $< 3 r_{\rm e}$ following \citet{Bona2017} and \citet{Wilber2018}. Assuming the typical spectral index value for the haloes ($\alpha = -1.3$), we then rescaled the  total flux density of the mock halo to 1.4 GHz and computed the limit to the total radio power at 1.4 GHz, which can be compared with values present in the literature.
 
%(i.e. the radius at which the brightness drops to $I_0/e$)

 %has been estimated as in \citet{Giaci2017} and \citet{Cassa2007} as $R_{\rm radio} = \sqrt{R_{\rm min} R_{\rm max}}$,  where $R_{\rm min}$ and $R_{\rm max}$ are the minimum and maximum radii of the 3$\sigma$ surface brightness isocontour in the low-resolution LOFAR image. 

%The negative contour level at $-3\sigma$ is overlaid with a dashed line. 

%The contour levels, overlaid in black, are from the LOFAR image obtained for the spectral analysis (uniform weighting, \textit{uv}-cut, see main text), not shown in the paper. 

\begin{figure*}
 \includegraphics[width=0.55\textwidth]
  {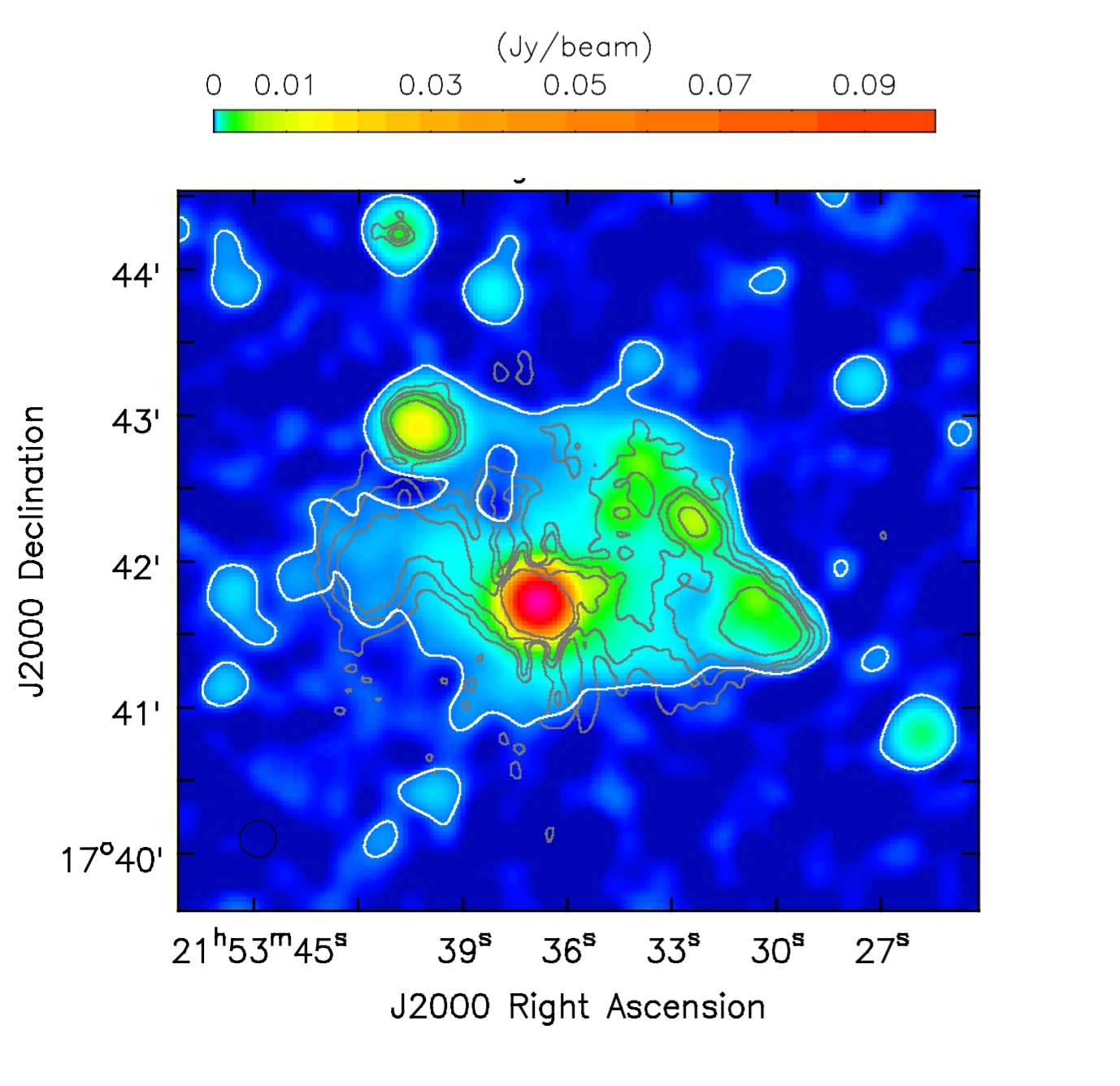}
 \includegraphics[width=0.45\textwidth]
  {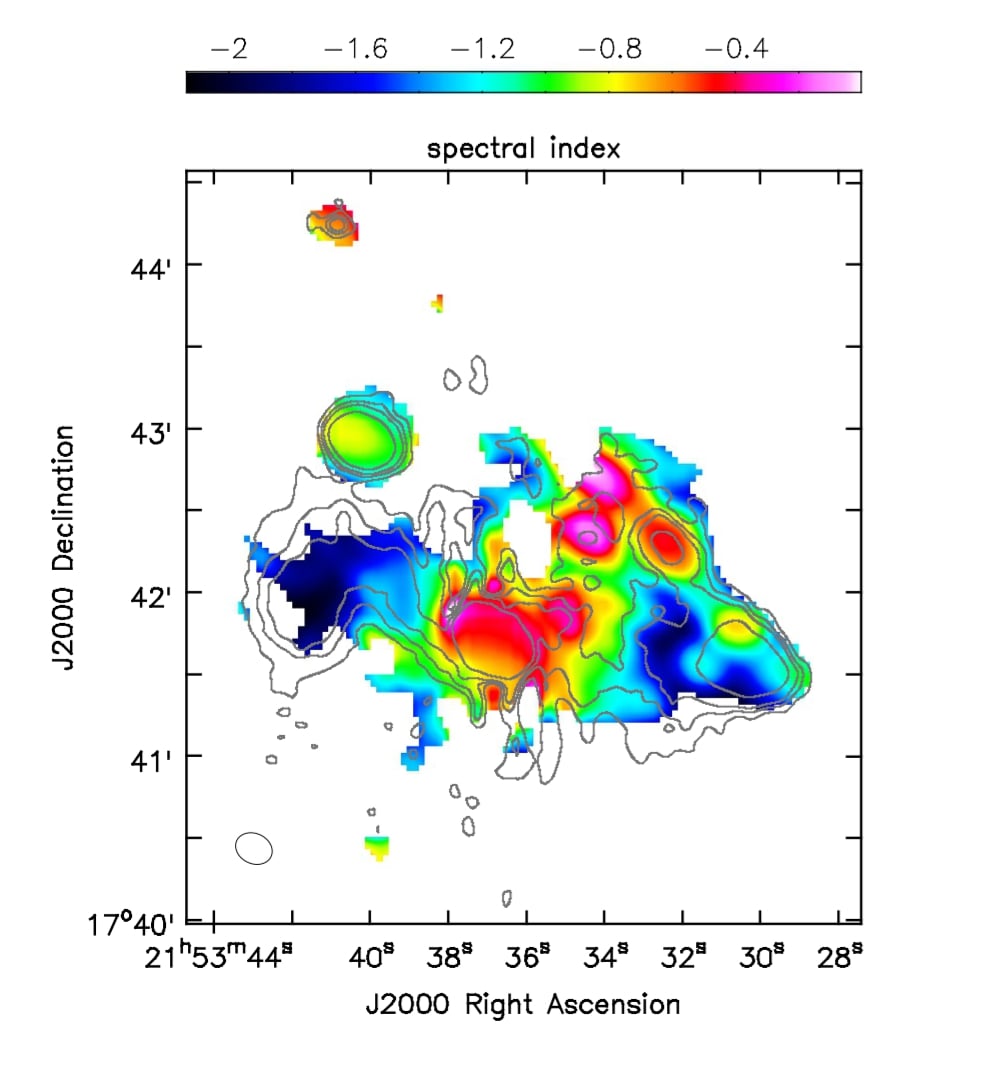}
   \caption{The contour levels, overlaid in grey, are from the LOFAR high-resolution image in Fig.\ref{a2390} with $(4, 8, 12, 24)\, \times \, \sigma$ where $\sigma$ = 400 $\mu$Jy\,beam$^{-1}$.
   {\bf Left panel:} 
   1.5 GHz VLA image of A2390 with a beam of $15'' \times 15''$ and an rms noise of 26 $\mu$Jy\,beam$^{-1}$. The first 3$\sigma$ contour is shown in white. {\bf Right panel:} 
   Spectral index map between the 1.5 GHz VLA and 144 MHz LOFAR images of A2390. Pixels below 3$\sigma$ are blanked. The error map is shown in Fig. \ref{spix-err3}.}
  \label{a2390-spix}
\end{figure*}

\subsection{RXCJ0142.0+2131} 

The non-cool-core cluster RXCJ0142.0+2131 is known to host a radio source at the cluster centre, that is coincident with the peak of the X-ray emission. Several bright radio sources in the field are visible in the image of the cluster \citep{Kale2013}. The X-ray distribution has a relatively disturbed morphology with no strong features. We note that this cluster exhibits the smallest value of $c$ in the sample.\\

%he X-ray luminosity is less than $10^{45}$ erg\,s$^{-1}$, and th
%These properties are similar to those of systems with no radio halos \citep{Cassa2010}.
%An amorphous distribution of X-ray emission is seen in the \textit{Chandra} image with no strong central peak (Fig. 1). 
%This cluster has a velocity dispersion of 1278±134 km s−1; however, no substructure in the velocity distribution (Barr et al. 2005). 
%Based on the mass-to-light ratios of galaxies and α-element abundance ratios, 
Based on optical data, the occurrence of a possible merger event was inferred \cite{Barr2006}, considering the presence of the second BCG (to the north-east) at z=0.283 located at $\sim 650$ kpc from the cluster centre and the X-ray morphology elongated in the same direction \citep{Linden2014}. \\

Since no diffuse radio emission has been detected in this cluster, an upper limit on the radio power was derived by \citet{Kale2013} using GMRT observations at 610 MHz and 235 MHz. The upper limit rescaled at 1.4 GHz is $P_{1.4} \le 0.45 \times 10^{24}$ W\, Hz$^{-1}$ \citep{Cassa2013}.\\

%An upper limit on the radio flux was obtained using GMRT data in : at 610 MHz $88.2\pm 0.05$ Jy\,beam$^{-1}$ with a beam of $9.5'' \times 7.1''$ and at 235 MHz $54.2\pm 1.3$ Jy\,beam$^{-1}$ with a beam of $27.3'' \times 13.5''$. 

The high-resolution LOFAR image we obtained (left panel of Fig. \ref{rxc}) is consistent with the 610 MHz GMRT image in \citet{Kale2015} with a central radio source and few other radio sources to the north-east and south-west. 
We subtracted the contribution of the sources detected at high resolution using only the longer baselines with a \textit{uv}-cut of $>1750 \lambda$ (that corresponds to 500 kpc) and re-imaged the data at lower resolution to search for diffuse emission. We were then able to see the presence of centrally located diffuse emission with $D_{\rm radio} \sim 570$ kpc, as shown in the right panel of Fig. \ref{rxc}. The total flux density is $32 \pm 6$ mJy corresponding to a total radio power at 144 MHz of $(8.6 \pm 1.6) \times 10^{24}$ W\, Hz$^{-1}$. Using the upper limit at 1.4 GHz placed by \citet{Kale2013}, we can estimate the spectral index to be $\alpha^{1400}_{144} < -1.3$. We note that this is a conservative estimate, since the upper limit is computed considering a 1 Mpc halo, while the halo we detect is on a scale of less than 600 kpc. We can use the power-luminosity plot in Fig. 3 in \citet{Bru2007}, where upper limits for 500 kpc-haloes are also derived for the typical GMRT observations of the radio halo survey. At the redshift of this cluster, a limit evaluated on a size of 500 kpc would be 1.6 - 1.8 times deeper than that on a size of 1 Mpc. Hence, we derive a more stringent limit on the spectrum of $\alpha^{1400}_{144}< -1.6$, which would lead to an USS halo classification.\\

The size, although smaller than that of giant haloes, and the estimated spectral index value are consistent with the properties of a radio halo, likely an USS halo. We note that the power measured with LOFAR and extrapolated at 1.4 GHz lies below the known power-mass correlation of \citet{Cassa2013} (see Fig. \ref{corr}).
        
%Other two cases of ``underluminous'' radio halos have been recently discovered in the clusters A1451 and Zwcl0634.1+4750 \citep{Cuci2018}. Similarly to the radio halo in RXC0142.0+2131, these sources are smaller with respect to the classical giant radio halos, and are associated with clusters that are not extremely disturbed. However, their spectrum does not seem to be steep, as the halo in RXC0142.0+2131 is likely to be.

\subsection{A478}%Giaci2014

A478 is a cool-core cluster with a relaxed X-ray morphology and a spherically symmetric temperature distribution on large scales.
A number of X-ray substructures at the cluster centre, such as small cavities with sizes of few kpc, are found by \citet{Sun2003}. \citet{Giaci2014} presented
VLA radio images at 1.4 GHz  and reveal the presence of a central double-lobe radio galaxy with a size of $\sim$ 13 kpc. The hosting galaxy is the BCG, as can be seen in the optical overlay. %BCG (04h13m25.3s, +10$^{\circ}$27$'$55$''$)
Diffuse, low-brightness radio emission encompassing the central active galactic nucleus (AGN) and extending on a scale of $\sim$ 300 kpc is also detected and classified as a mini halo. The low-resolution image at 1.4 GHz (Fig. 1c in \citealt{Giaci2014}) shows that the mini halo blends with the tail of a cluster-member head-tail radio galaxy located to the north-east of the cluster centre. The tail extends for $\sim$ 200 kpc and encompasses an unresolved source, which is also a cluster member. A second unresolved source to the north with no optical identification is thought to be a background galaxy.\\

%is likely interacting with the ICM, creating the small cavities.
%Two small X-ray cavities, located within the central 15 kpc, are partially filled by the radio galaxy lobes at 1.4 GHz. With a size of only ∼ 4 kpc, these are among the smallest cavities found in cluster cores.
\citet{Mark2003} reported the presence of a cold front in the \textit{Chandra} image at $\sim$ 60 kpc to the south-west of the cluster, which we show in the X-ray image of Fig. \ref{a478}. Its position seems to indicate that the radio emission is confined by the front, as expected in simulations of mini haloes \citep{ZuHone2013}.\\

The calibration of the LOFAR data set was difficult owing to the presence of a 3C source (3C109; 04h13m40.37s, +11$^{\circ}$12$'$13.8$''$) with a flux density of $\sim$ 21 Jy located at $0.7^{\circ}$ from the target. We created a sky model of this double-lobe radio galaxy using a high-resolution VLA image at 4.8 GHz to improve its calibration. However, the image sensitivity is limited by the dynamic range, therefore we consider the caveats relative to the presence of negative holes around the central source, and the higher noise of the target facet (460 $\mu$Jy\,beam$^{-1}$) compared to that obtained for the other clusters. Nevertheless, the final image of A478 at 144 MHz (Fig.~\ref{a478}) confirms the presence of the central AGN; the head-tail radio galaxy to the north-east with a tail is overall much longer than previously seen, extending for more than 650 kpc. Interestingly, the tail appears to be divided into two regions. This is seen in the case of A1033 \citet{Dega2017}, in which the authors have proposed the  gentle re-energisation of electrons as the mechanism that justifies the brightness of the second part of the radio tail, which is only visible at low frequencies.

%Such a scenario can be explained by two different bursts of the AGN, also considering that the east part of the tail can not be seen at 1.4 GHz and is likely to be old steep lobe emission.

%, or by CRe re-acceleration powered by the interaction between a perturbed ICM and the plasma of the radio tail, as proposed by  for the cluster .\\

With LOFAR, we do not see hints of centrally located diffuse emission. As a further argument, we injected mock mini haloes with different sizes and flux densities in the data set, both at the cluster centre and in a void close-by region, to rule out the possibility that the calibration artefacts are responsible for the absence of the mini halo (see Fig. \ref{inj} in the Appendix section). We used the minimum and maximum values for $r_e$ , i.e. 25 and 100 kpc, and for $I_0$ at 1.4 GHz, i.e. $13$ and $1$ $\mu$Jy\,beam$^{-1}$, respectively, found by \citet{Murgia2009} for a sample of mini haloes. Considering the flux density reported in \citet{Giaci2017} for the mini halo in A478, the recovered flux density of the mock mini halo indicates that the limit on the spectral index of the source would be $\alpha > -1$, which is unusually flat for a mini halo. %One possibility is that the emission that has been previously classified as mini halo might merely be related to the central AGN and the diffuse tail of the radio galaxy to the north-east. However, we can not completely exclude the presen
%To get an estimate of the spectral index we are sensitive to, we consider the conservative $2\sigma$ contour level of the LOFAR image (where $\sigma$ = 620 $\mu$Jy\,beam$^{-1}$) and the $3\sigma$ contour level (where $\sigma$ = 50 $\mu$Jy\,beam$^{-1}$) of the 1.4 GHz VLA image of Fig. 1c in \citet{Giaci2014}, which are imaged with the same beam of $30'' \times 30''$. We obtain that the dimmest part of the diffuse emission should have a spectrum flatter than $-0.9$, which would be unusually flat for a mini halo, or re-acceleration models in general.

%The tail emission appears to be blended with the central source as seen in the low-resolution image, however no additional diffuse emission is found in the central region.

%consist of two components. Seed electron from the central AGN lobe + Gently Re-energization ?
%Interestingly, the tail emission is not continuous...

We processed an archival GMRT observation at 610 MHz and obtain the low-resolution image shown in the left panel of Fig.~\ref{a478-spix}. Neither diffuse emission nor the second portion of the tail is visible. We re-imaged the LOFAR and GMRT data sets with a Gaussian taper of 30$''$ and with the same pixel size, baseline range (200 - 40000$\lambda$), and uniform weighting scheme to minimise the differences in the {\it uv}-coverage of the two observations, and we obtain the spectral index map that is shown in the right panel of Fig.~\ref{a478-spix}. The central emission has a spectral index of $\sim -1.1$ and the head-tail radio galaxy has the typical trend of an active radio galaxy with a flat core ($\alpha_{610}^{144} \sim -0.6$) and a steepening along the tail up to $\alpha_{610}^{144} \sim -3$. We can provide an upper limit on the spectrum of the second portion of the tail, using the mean flux density from the LOFAR image and the rms noise from the GMRT image. This results in a spectral index $< -3.8$.

\subsection{PSZ1G139.61+24 }

The cool-core cluster PSZ1G139.61+24 has been studied in detail in a dedicated paper by \citet{Savini2018b}. \citet{Giaci2017} have reported the detection of a tentative mini halo with an overall source size of $\sim$ 100 kpc located at the cluster centre, and the LOFAR image presented in \citet{Savini2018b} reveals new diffuse emission extending for more than 500 kpc. The radio source associated with the cluster consists of a central bright component surrounded by halo-like emission extending beyond the cool core. The two components become apparent in a spectral analysis performed between 144 MHz LOFAR and 610 MHz GMRT images that reveals a central component with $\alpha^{610}_{144} \sim$ -1.3, whilst the large-scale faint emission exhibits an USS with an upper limit of $\alpha^{610}_{144} <$ -1.7. Although the cluster core has a low entropy, typical of non-merging cool-core systems, the X-ray analysis shows that the cluster is slightly disturbed and hosts a cold front (shown in the bottom panel of Fig. \ref{psz}) suggesting the presence of gas sloshing. \citet{Savini2018b} have argued that the large-scale radio emission outside the core is produced by a minor merger that powers electron re-acceleration without disrupting the cool core.

\subsection{A1413} %govoni2009

The 1.4 GHz VLA image in \citet{Govo2009} reveals the presence of diffuse emission at the centre of the cool-core cluster A1413, extending on a scale of $\sim$ 220 kpc with a total flux density of $1.9 \pm 0.7$ mJy. At the sensitivity limit of the FIRST survey \citep{first}, the central optical galaxy does not contain a compact radio source (Fig. 2 in \citealp{Govo2009}), and the authors suggested that the source might be a candidate mini halo, speculating about a scenario in which the central galaxy has switched off while the mini halo continues to emit. \citet{Govo2009} also reported an offset between the emission peak of this radio source and both the central galaxy and the X-ray emission peak.\\
 The X-ray morphology is slightly elongated in the north-south direction and shows a bright core with a moderate entropy value ($K_0 = 64 \pm 8$, \citealt{Giaci2017}) that indicates the absence of a cool core. We note that this cluster is the only (candidate) mini halo found in a non-cool-core cluster in the sample studied in \citet{Giaci2017}.\\

With the LOFAR image in Fig. \ref{a1413}, we confirm the presence of centrally located diffuse emission. A head-tail radio galaxy, not mentioned or shown in \citet{Govo2009}, can be seen at the west of the cluster. The optical counter part is a galaxy member at $z = 0.144$. We obtain a high-resolution image using only the longer baselines with a {\it uv}-cut of $> 2580 \lambda$ (corresponding to $\sim$ 200 kpc) to model the central compact sources (shown in the box within the top left panel of Fig. \ref{a1413}). We note that two sources are actually visible in the central region, one of which is co-located with the X-ray centre, hence no offset is present. We then subtract the compact sources, imaging the central radio diffuse emission that is shown in the low-resolution image of Fig. \ref{a1413}; we classify this source as a mini halo. The size of the diffuse emission is $\sim$ 210 kpc with a total flux density of $40 \pm 7$ mJy at 144 MHz. The implied spectral index value is $\alpha^{1400}_{144} \sim$ -1.3. We note that the spectral index might be steeper, considering that the flux reported by \citet{Govo2009} is likely overestimated, since the authors subtracted only the contribution of one of the central sources, i.e. that visible at 1.4 GHz with FIRST.

%D = 2.53*3600*(0.022*0.012)**0.5 = 150 kpc

\subsection{A1423} %ventu2008
\label{a1423-sec}

No hint of diffuse radio emission has ever been observed in the cool-core cluster A1423 and no radio images are available in the literature.  An upper limit on the radio power was derived by \cite{Ventu2008} using a GMRT observation at 610 MHz. The upper limit rescaled at 1.4 GHz is $P_{1.4} \leq 0.38 \times 10^{24}$ W\, Hz$^{-1}$ \citep{Cassa2013}.\\ 

Our high-resolution LOFAR image of A1423 (Fig. \ref{a1423}) shows a bright central radio source ($\sim$ 0.38 Jy) that is likely connected to the BCG visible in the optical image. This radio source is elongated in the north-south direction with a tail extending to the north for $\sim$ 400 kpc, which might be remnant emission connected to the central source. A second tailed source is found to the north-west of the cluster centre and might be connected to an optical galaxy of unknown redshift. Subtracting the bright and extended central source was not possible since it would leave residuals that cannot be distinguished from diffuse emission. The low-resolution LOFAR observation does not show additional emission. Moreover, no spatial correlation between radio and X-ray emission is found since the X-ray morphology appears disturbed along the east-west axis, whilst the radio emission is elongated along the north-south axis.\\

We computed a new upper limit on the radio power by injecting a mock halo in the data set (see Sec. \ref{res}). The integrated flux density of the mock halo computed within a region centred on the injected halo with a radius equal to $R_{\rm H} = 436$ kpc is $28$ mJy at 144 MHz, which corresponds to a total radio power at 144 MHz of $P_{144} = 4.1 \times 10^{24}$ W\, Hz$^{-1}$. Assuming the typical spectral index value used in the power-mass correlation for haloes ($\alpha = -1.3$), we derive a new upper limit to the radio power at 1.4 GHz, which we plot in Fig.~\ref{corr}. The new upper limit is $P_{1.4} < 0.20 \times 10^{24}$ W\, Hz$^{-1}$, i.e. almost a factor of 2 deeper than that derived in literature.\\

%\textbf{TIM: how are you estimating the flux that is measured? I guess by subtracting the image with the injected halo from the other without the injected halo? If so is this a fair test because you get such good constraints on the flux without the halo but if the halo were real you wouldnt have that. Are we assuming perfect subtraction?}

%We created a mock radio halo with $P_{1.4} = 1.55 \times 10^{24}$ W\, Hz$^{-1}$ and a radius of $R_{\rm H} = 436$ kpc, knowing the mass of the cluster to be $M_{500} \sim 6.09 \times 10^{14} M_{\odot}$, such that $I_{0,1.4}$ =  0.22 $\mu$Jy/arcsec$^2$ and $r_{\rm e} = 168$ kpc. We rescaled the surface brightness to 144 MHz creating a set of mock halos with different $I_{0,144}$. These sources are Fourier transformed into the {\it uv}-data of the LOFAR observation, which is then imaged with a robust parameter of -0.25 and an outer taper of 8$’’$. 

\subsection{A1576} %kale2013

As in the case of A1423, no hint of diffuse radio emission has so far been observed in non-cool-core cluster A1576. A central radio galaxy with indications of a jet and three optical counterparts, which create a multiple core system, is reported by \citet{Kale2013}.
%and it is likely stage of merger (Dahle et al. 2002). almost coincident with the peak in the X-ray emission. 
Two other radio sources are visible in the field to the north (co-located with an optical source with unknown redshift) and to the south-west (likely connected to a cluster-member galaxy). The X-ray morphology is elongated in the east-west direction and does not show a strong central peak. Moreover, based on weak lensing analysis, \citet{Dahle2002} inferred significant dynamical activity. An upper limit on the radio power was derived by \citet{Ventu2008} using a GMRT observation at 610 MHz. The upper limit rescaled at 1.4 GHz is $P_{1.4} \le 0.64 \times 10^{24}$ W\, Hz$^{-1}$ \citep{Cassa2013}.\\

\begin{figure}
 \includegraphics[width=0.5\textwidth]
  {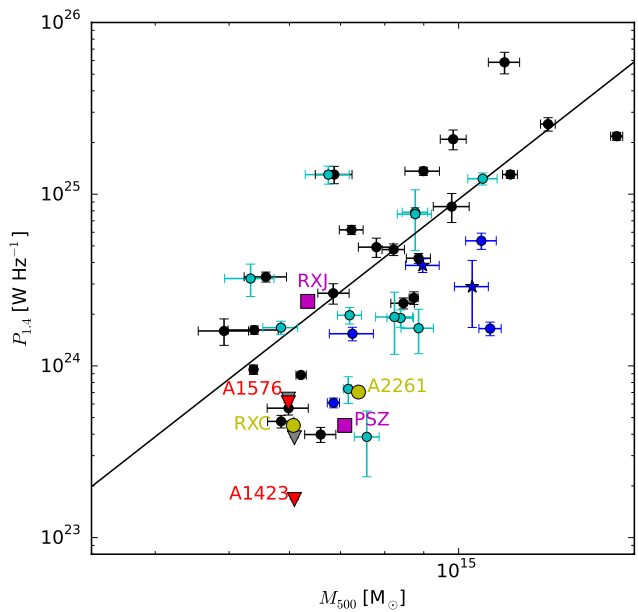}
   \caption{Radio power at 1.4 GHz vs. cluster mass $M_{500}$ for a sample of clusters with radio halo. The plot is reproduced from \citet{MA2016}. Haloes with flux density measured at 1.4 GHz are indicated by black circles and their related fit is shown as a black line. Haloes with
flux density measured at frequencies other than 1.4 GHz are indicated by cyan circles, ultra-steep haloes by blue circles, and ultra-steep haloes with flux density measured at frequencies other than 1.4
GHz by blue stars. 
The upper limit of radio halo power at 1.4 GHz in A1423 and A1576 are indicated by triangles, as derived in the literature in grey and the new limits obtained with LOFAR in red. The limit derived for A1576 is almost coincident with the value in literature, while that for A1423 is almost a factor of 2 smaller. We also indicate the power of the haloes in A2261 and RXCJ0142.0+2131 with yellow circles, and the steep-spectrum sources in PSZ1G139.6+24 and RXJ1720.1+2638 with magenta squares.} 
  \label{corr}
\end{figure}

During the LOFAR observation of this cluster, the ionosphere was very active and we reached a noise of 500 $\mu$Jy\,beam$^{-1}$ (Fig. \ref{a1576}). At this sensitivity level, no radio diffuse emission is detected at 144 MHz either after subtracting the contribution of the compact sources. Hence, we compute a new upper limit on the radio power by injecting a mock halo in the data set (see Sec. \ref{res}). The integrated flux density of the mock halo computed within a region centred on the injected halo with a radius of $R_{\rm H} = 429$ kpc is $37$ mJy at 144 MHz, which corresponds to a total radio power at 144 MHz of $P_{144} = 11.9 \times 10^{24}$ W\, Hz$^{-1}$. Assuming the typical spectral index value used in the power-mass correlation for haloes ($\alpha = -1.3$), we derived the radio power that we plot in Fig.~\ref{corr}. The new upper limit on the radio power at 1.4 GHz is $P_{1.4} < 0.62 \times 10^{24}$ W\, Hz$^{-1}$, i.e. comparable with that derived in literature.

 %P=1.44599e24, flux=4.48 mJy, RH=428 kpc, re=164.9 kpc

%The LOFAR observation at 144 MHz confirms the non-detection of diffuse radio emission, as can be seen also in the low resolution image (Fig \ref{a1576}). Therefore, we proceeded in computing a new deeper upper limit on the radio power. We created a mock radio halo with $P_{1.4} =  1.45 \times 10^{24}$ W\, Hz$^{-1}$ and a radius of $R_{\rm H} = 428$ kpc, knowing the mass of the cluster to be $M_{500} \sim 5.98 \times 10^{14} M_{\odot}$, such that $I_{0,1.4} =  ... \mu$Jy/arcsec$^2$ and $r_{\rm e} = 165$ kpc. We rescaled the surface brightness to 144 MHz creating a set of mock halos with different $I_{0,144}$. These sources are Fourier transformed into the {\it uv}-data of the LOFAR observation, which is then imaged with a robust parameter of -0.25 and an outer taper of 8$’’$. 

%The cluster does not host a cool core.

%The clusters A1576 shows relatively disturbed morphologies in X-rays, but have low X-ray luminosities (LX <1045 erg s-1). These properties are similar to those of other merging systems with low X-ray luminosities that have been found to be radio-quiet (without radio halos/relics) (Cassano et al. 2010; Russell et al. 2012). 

\subsection{RXJ1720.1+2638}  %giaci2014

The cool-core cluster RXJ1720.1+2638 has been studied in detail by \citet{Giaci2014b} %where the radio source associated with the cluster is analysed through observations with the GMRT at 317, 617, and 1280 MHz and with the VLA at 1.5, 4.9, and 8.4 GHz. 
using VLA and GMRT observations.
The central source is classified as a mini halo consisting of a bright central component with a size of $\sim$ 160 kpc, and a fainter spiral-shaped tail of emission extending towards the south for more than $200$ kpc. Two cold fronts detected in the \textit{Chandra} X-ray image of the cluster appear to confine the mini halo.\\

The LOFAR images, shown in Fig.~\ref{rxj1720}, reveal a new diffuse component extending beyond the cold fronts, not visible at higher frequencies. The emission extends towards the south-west with an overall size of $\sim$ 600 kpc. We reprocessed the GMRT observation at 610 MHz and obtained the high-resolution image shown in the left panel of Fig.~\ref{rxj-spix}. The head-tail radio galaxy to the north-east of the cluster is clearly visible and a connection between the central diffuse emission and the tail is already seen at this frequency. We re-imaged the LOFAR and GMRT data sets with a Gaussian taper of 20$''$ and with the same pixel size, baseline range (200 - 40000$\lambda$), and uniform weighting scheme to minimise the differences in the {\it uv}-coverage of the two observations and obtained a spectral index map that is shown in the right panel of Fig.~\ref{rxj-spix}. The mini halo appears to have a constant spectrum with a spectral index of $\alpha^{610}_{144} \sim -1$, whilst the head-tail radio galaxy has the typical trend of an active radio galaxy with a flat core and a steepening along the tail. The emission connecting the tail and the mini halo is steep, ranging between $\alpha^{610}_{144} \sim$ -1.4 and $\alpha^{610}_{144} \sim$ -1.8. The cluster-scale diffuse emission, which cannot be seen at 610 MHz, is ultra steep, and we can only provide upper limits, as shown in Fig.~\ref{rxj-spix}, using the mean flux density from the LOFAR image and the rms noise from the GMRT image. The radio emission in RXJ1720.1+2638 resembles that of PSZ1G139.61+24 \citep{Savini2018b} with an inner, flatter component in the form of the already-known mini halo, and an outer part with a steeper-spectrum halo-like emission on larger scales, with $\alpha^{610}_{144} <$ -1.5. The two inner cold fronts (reported in the X-ray image of Fig. \ref{rxj1720}) appear to separate the two components. Interestingly, the cluster  A2142 \citep{Ventu2017} also shows a two-component emission, and the presence of cold fronts both in the inner and outer region. However, because of the scarcity of X-ray counts in the outer cluster regions, we are not able to search for cold fronts on a larger scale.

\subsection{A2261}

A2261 is a non-cool-core cluster with a central radio source coincident with the BCG. %(2MASX J17222717+3207571).
The cluster has a relaxed morphology in X-rays in the central 500 kpc-radius region, while showing a diffuse patch of X-ray emission towards the west of the cluster; this is likely an infalling group, which suggests a possible minor merger \citep{Linden2014}. \citet{Sommer2017} found a centrally located diffuse component on a scale of $\sim$ 1 Mpc  using VLA observations at 1.4 GHz and classified the large-scale source as a radio halo.\\

The high-resolution LOFAR image (left panel of Fig. \ref{a2261}) shows the presence of a central compact radio source and a radio galaxy located at $\sim$ 540 kpc towards the north-west of the cluster centre; this radio galaxy has a faint, diffuse emission, symmetric with respect to its core, which could be the two remnant radio lobes. The core is identified with an X-ray source and an optical galaxy whose redshift is unknown. At the cluster redshift, the radio galaxy would extend up to $\sim$ 590 kpc. A patch of diffuse radio emission with no clear origin, which might be a relic or a dying AGN, is seen to the west of the cluster centre. After modelling these compact sources using only the longer baselines with a \textit{uv}-cut of $>750 \lambda$ (that corresponds to 1 Mpc), we subtracted their contribution, revealing the presence of diffuse emission extending up to 1.2 Mpc. In general, the \textit{uv}-subtraction method is not indicated for extended sources and in this case the remaining contribution of the diffuse lobes of the radio galaxy can be clearly seen. This affects the morphology of the diffuse emission that shows the brightness peak and an extension at the location of the radio galaxy and the unknown patch of diffuse emission. We note that the image obtained by \citet{Sommer2017} (Fig. 2) shows a different morphology of the large-scale diffuse source, which is likely due to the result of the \textit{uv}-subtraction at a different frequency. For instance, the fact that the radio galaxy lobes are better subtracted might be related to their probable steep spectrum that contributes at high frequencies to a much smaller extent compared to what is observed at low frequencies. However, the radio source as seen at 144 MHz extends on a larger scale than the single radio sources, around the cluster centre, and also towards west and south-west, where the patch of X-ray emission is located, which might indicate a minor merger. This could also explain the offset between radio and X-ray emission peaks, also visible in the 1.4 GHz image. 
We obtained images using different uv-cut for subtraction corresponding to 800 kpc, 1 Mpc, and 1.2 Mpc. We show in Fig. \ref{uvsub} in the Appendix, models, and resulting images. Although the \textit{uv}-subtraction is not entirely reliable as for the case of a compact point source, we can state that diffuse emission that is not related to the radio galaxy is clearly visible. Hence, the LOFAR image at 144 MHz confirms the presence of a radio halo, as suggested by the VLA observation at 1.4 GHz, although its morphology and radio power are hard to determine. We estimate the contribution of the flux densities of the radio galaxy and unknown patch of emission, re-imaging the LOFAR \textit{uv}-subtracted data set with the same taper and weighting scheme used for the halo (Fig.11 in the paper), keeping, in addition, the \textit{uv}-cut of $750 \lambda$ used for modelling the single sources. Subtracting this value of $20$ mJy from the total flux density of the large-scale diffuse emission that is $\sim$ 185 mJy; the radio halo has $\sim$ 165 mJy corresponding to a total radio power at 144 MHz of $\sim 26 \times 10^{24}$ W\, Hz$^{-1}$.\\

%The total flux density is $155 \pm 30$ mJy corresponding to a total radio power at 144 MHz of $(24 \pm 5) \times 10^{24}$ W\, Hz$^{-1}$. The error is large, since we need to take into account the flux density contamination of the radio galaxy, which was difficult to subtract. \footnote{Here, the error on the source subtraction is computed as follows. The total flux density measured after subtraction in the region of the radio galaxy is $\sim$ 60 mJy, while the total flux density of the whole radio source is $\sim$ 185 mJy. The halo emission therefore must be in the range $125 - 185$ mJy, considering the two extreme cases where the residual $\sim$ 60 mJy consist entirely of radio galaxy emission or halo emission, respectively. Hence we estimate the radio halo to have a mean flux density of $155 \pm 30$ mJy.}\\

Considering the total flux density at 240 MHz, 610 MHz, and 1.4 GHz reported in Tab. 3 in \citet{Sommer2017}, we can provide an approximate estimate of the spectral index of the halo including LOFAR measurements to be $\alpha = -1.7 \pm 0.3$, which makes the radio halo in A2261 a candidate USS halo.

%, which is consistent within the errors of the spectral index reported by the authors ($\alpha = -1.2^{+0.23}_{-0.5}$).

%The total flux density (core+diffuse) of the radio galaxy at 610 MHz is 57 ± 6 mJy and 23.3 mJy at 1.4 GHz (NVSS J172216+320910).
%The implied spectral index is 1.08. 
%The compact X-ray source, CXOGBAJ172217.0+220913 (Gilmour et al. 2009), and an uncatalogued galaxy in the SDSS r-band image is co-incident with the compact radio core (Fig. 2). 
%The field of this cluster shows the presence of several radio galaxies (Fig. A.8).
%The low frequency data is not deep enough for an accurate determination of the flux of the extended emission in A2261 the data are largely consistent with what would be expected from a radio halo of moderate spectral slope.

\begin{table*}
 \centering
  \caption{Derived physical parameters of the radio diffuse emission of our target clusters as seen by LOFAR. Col.1: Name of the cluster; Col. 2: classification of the diffuse emission as based on observations at 144 MHz (UL = upper limit; MH = mini halo; cMH = candidate mini halo; H = halo; USSH = USS halo; Col.3: size of the diffuse emission as measured from the low-resolution LOFAR images; Col.4: Total flux density at 144 MHz; Col.5: Total radio power at 144 MHz; Col. 6: Notes on the final image.
  }
\begin{tabular}{c c c c c c}
  \hline
Name & LOFAR class. & $D_{\rm radio}$ & $S^{\rm diff}_{144}$ & $P^{\rm diff}_{144}$ & Notes\\ 
        & & \scriptsize{(kpc)} &   \scriptsize{(mJy)} & \scriptsize{ ($\times 10^{24}$ W\, Hz$^{-1}$)} & \\
\hline
\hline
RXCJ0142.0+2131  & H & 570 & $32 \pm 6$ & $8.6 \pm 1.6$ & new H discovered\\
A478   & uncertain & - & - & - & no diffuse emission\\
PSZ1G139.61+24   & MH+USSH  & 550 & $30 \pm$ 4& $7 \pm 1$ & MH: 3$\times 10^{24}$ W\, Hz$^{-1}$, USSH: 4$\times 10^{24}$ W\, Hz$^{-1}$ \\
A1413   & MH & 210 & $40 \pm 7$ & $2.3 \pm 0.4$ & MH confirmed\\
A1423   &  UL& -& $23$ & \textbf{$3.3$} & mock H injection; $P_{1.4} < 0.17\times 10^{24}$ W\, Hz$^{-1}$ ($\alpha=-1.3$)\\
A1576  & UL & -& $37$ & $11.9$ & mock H injection; $P_{1.4} < 0.62\times 10^{24}$ W\, Hz$^{-1}$ ($\alpha=-1.3$)\\
RXJ1720.1+2638   & MH+USSH & 600& $820 \pm 123$& $72\pm 11$ & MH: 64$\times 10^{24}$ W\, Hz$^{-1}$, USSH: 8$\times 10^{24}$ W\, Hz$^{-1}$\\
A2261  & H & 1200 & $\sim 165$ &  $\sim 26$ & H confirmed\\
A2390    & uncertain & 1100 & - & - & 600 kpc-central double radio galaxy\\
\hline
\end{tabular}
\label{info2}
\end{table*}

\subsection{A2390}

The cool-core cluster A2390 is particularly interesting for its large mass ($\sim 10^{15} M_{\odot}$), which sets it apart from the other clusters in the sample (see top left panel of Fig. \ref{plot}). A2390 has been observed at 1.4 GHz with the VLA. The data were analysed by \citet{Bacchi2003}, who classified the detected emission as a mini halo with an integrated flux density of $63 \pm 3$ mJy and an extension of $\sim$ 550 kpc. Using deeper Jansky VLA observations in the 1-2 GHz frequency band and re-analysing the VLA observation in \citet{Bacchi2003}, \citet{Sommer2017} discovered diffuse radio emission on a larger scale extending for $\sim$ 800 kpc in the form of a radio halo, after subtracting the compact source contribution. A spectral in-band analysis was also performed, finding a steep spectrum of $\alpha^{\rm 1GHz}_{\rm 2GHz} = -1.60 \pm 0.17$.\\

Our high-resolution LOFAR image shows that the central radio source is a double radio galaxy with the lobes extending in the east-west axis for $\sim$ 600 kpc. The morphology indicates the presence of a bright core and fading lobes. Radio galaxies of such a size are not common at the centre of clusters, where the ICM usually prevents the expansion of the lobes to such large scales. The absence of emission in the north-south axis around the galaxy core had already been pointed out in previous works by \citet{Bacchi2003} and \citet{Sommer2017}. We note that there are some imaging artefacts due to the imperfect calibration of the galactic core that is brighter than the diffuse lobes and causes the image sensitivity to be limited by the dynamic range. The low-resolution LOFAR image shows the presence of diffuse emission on a larger scale with a total extension of $1.2$ Mpc. 
%It was not possible to subtract the contribution of the radio galaxy with its diffuse lobes to search for a radio halo. 
The lowest contour of the low-resolution image might still be interpreted as cluster-scale diffuse emission or old lobes emission. It is not possible to disentangle this emission from that of the fading lobes of the radio galaxy, neither it is possible to accurately subtract its core to search for the presence of an underlying radio halo. Hence, we can only conclude that the radio emission is mostly dominated by the contribution of the radio galaxy whose morphology is clearly visible with LOFAR.\\

In the left panel of Fig.~\ref{a2390-spix}, we show the image of the VLA data set that was calibrated in  \citet{Sommer2017}. We  re-imaged this data set to match\footnote{In this case we did not apply a \textit{uv}-cut and the uniform weighting parameter (as we did for the spectral analysis of A2261) because this would prevent us from recovering the morphology of the diffuse lobes. However, we note that the shortest baselines of the VLA and LOFAR observations are able to detect large-scale emission up to few arcminutes, larger than the emission in A2390.} the LOFAR imaging parameters, such as cell size and resolution, and obtain the spectral index map shown in the right panel of Fig. \ref{a2390-spix}. The core appears to have a flat spectrum with $\alpha^{1400}_{144} \sim$ -0.5, which is a typical value for the core of an active radio galaxy, whilst the lobes are much steeper with $\alpha^{1400}_{144}$ ranging between $\sim$ -1.3 and $\sim$ -2. 
We can speculate on the type and evolutionary phase of the radio galaxy. It could be an FR-II radio galaxy that has recently restarted (which would explain the flat core), still showing old lobes and relic hotspots from a previous activity cycle (e.g. \citealp{Shu2015}, \citealp{Brienza2016}). Interestingly, \citet{Aug2006} classified the radio galaxy of A2390 as an FR-II with a flat-spectrum core and a compact twin-jet structure in a north–south direction on a sub-arcsec scale, as seen in the 1.7 - 43 GHz frequency range with very long baseline interferometry (VLBI) observations. They also note that the orientation of the jets is misaligned with respect to the ionisation cones and dust disc of the host galaxy on larger scales. They suggest that the misalignment might be due to a precession of the central super massive black hole, and that the radio source might be an example of a bubble being blown into the ICM at its early stage ($10^3 - 10^4$ yr duration). This is in line with our interpretation that the east-west jets are originated by a previous AGN active phase. The AGN might then be experiencing a second episode of activity with the jets growing in the north-south direction after a precession. With this scenario we expect the presence of X-ray cavities and bubbles. Four inner cavities are found by \citet{Son2015} and coincide with the location of the inner jets (east-west and north-south). Further cavities on large scales at the location of the old lobes are also expected and could be searched in future studies.

%More data at different frequencies from the lobes is needed to reveal more details, and understand the connection between the AGN and the ICM. 

%AB: LR image should show a larger region 

%A2390: $1400\pm$ & $250\pm$ 
%RXJ1720
%815 mJy MH
%105 mJy USSH

% The power values include a k-correction with an averaged spectral index of $\alpha = -1.3$, as in.

\section{Discussion}
%Is the vaule of w on smaller scales similar to the value of w on 500 kpc for merging clusters?

\begin{figure*}
        \includegraphics[width=\columnwidth]{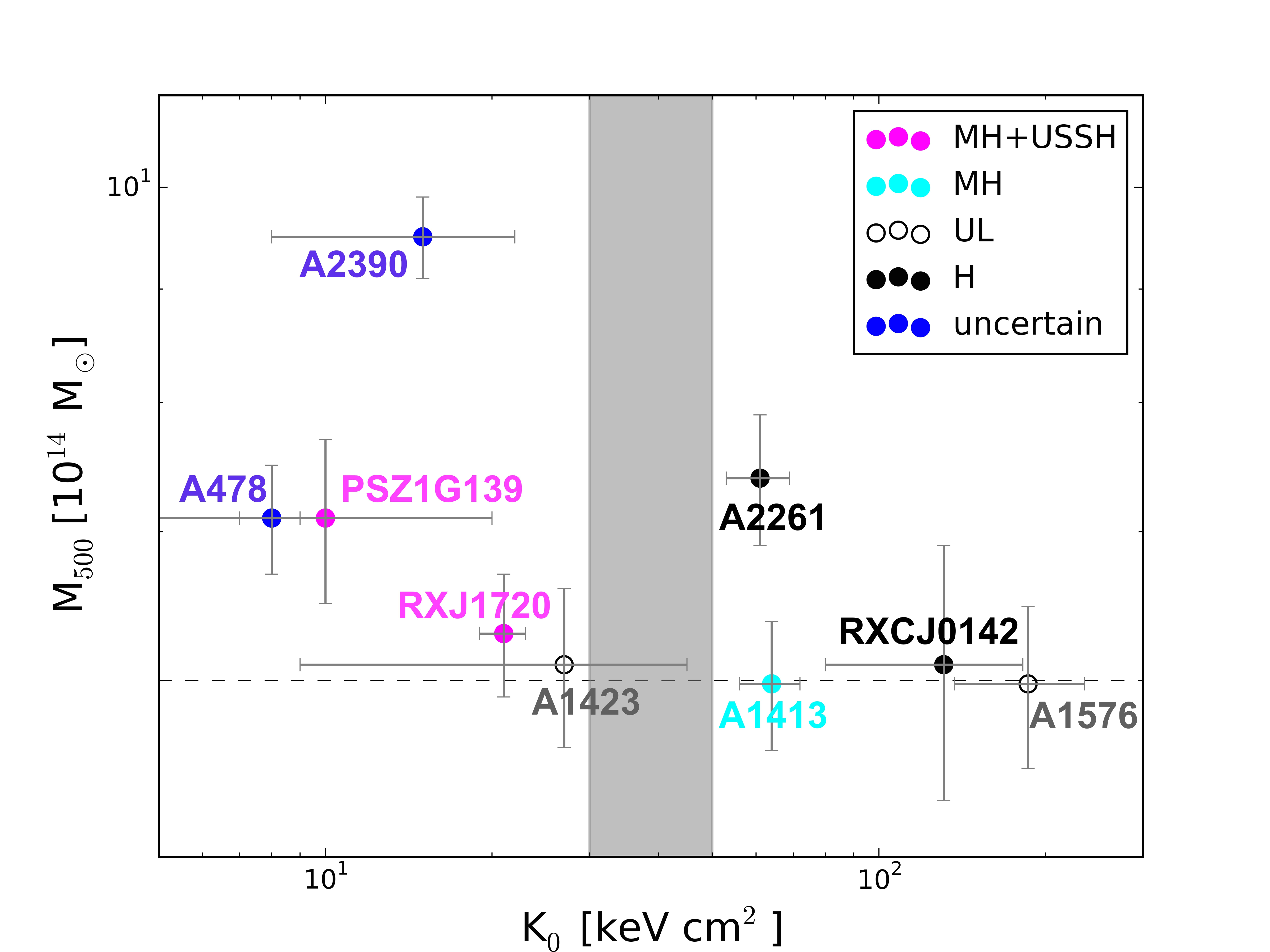}
         \includegraphics[width=\columnwidth]{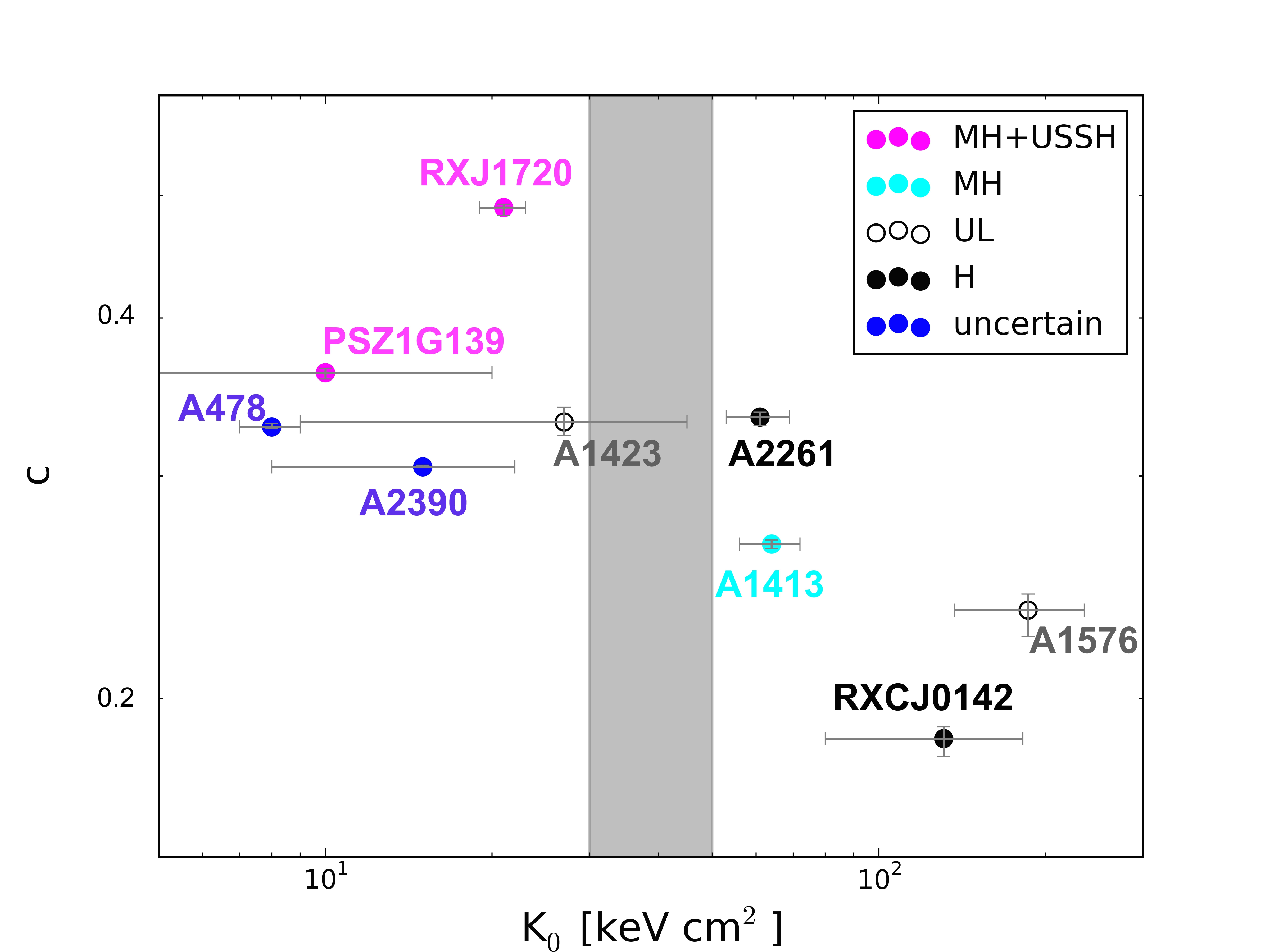}
         \includegraphics[width=\columnwidth]{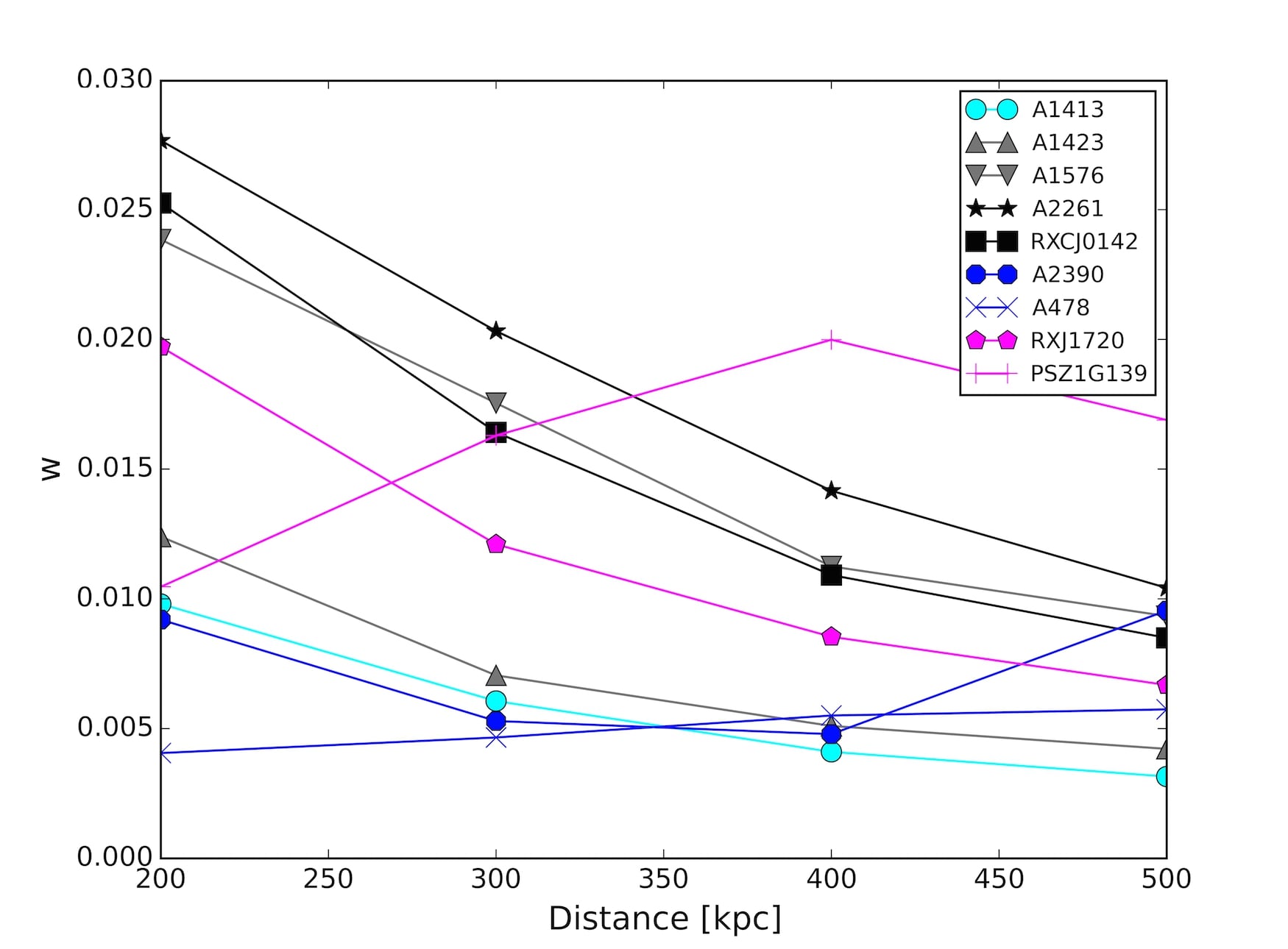}
         \includegraphics[width=\columnwidth]{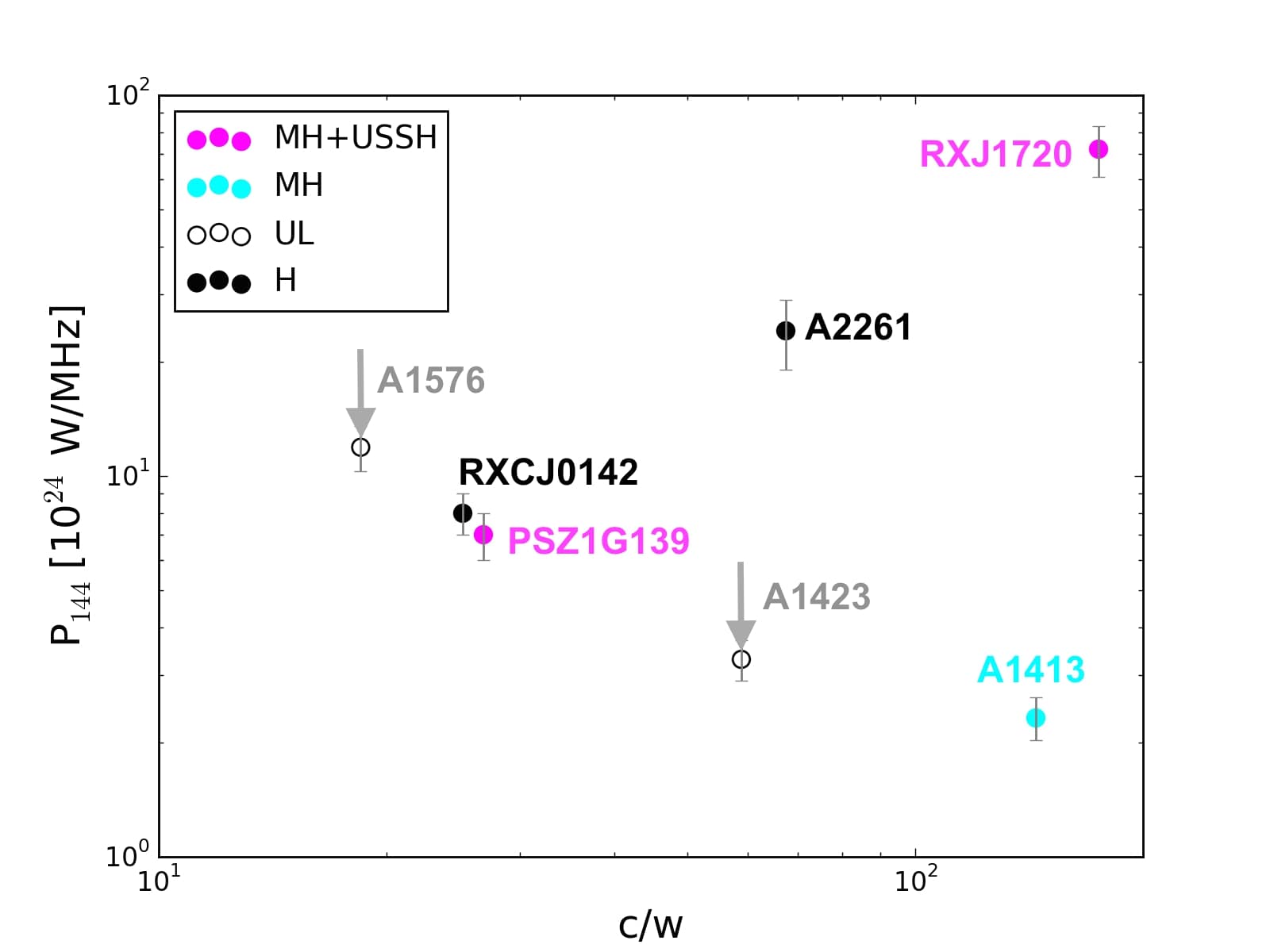}
  \caption{{\bf Top left panel:} Central entropy $K_0$ vs. cluster mass $M_{500}$ of our selected sample of clusters. The selection in mass ($M_{500} \ge 6 \times 10^{14} M_\odot$) is indicated by a dashed line. Clusters with low central entropies ($K_0 < 30 - 50$ keV cm$^2$) are expected to host a cool core, whilst high central entropy ($K_0 > 50$ keV cm$^2$) indicates a non-cool-core cluster \citep{Giaci2017}. The sources are coloured according to the new LOFAR findings: clusters with mini haloes and USS haloes (MH+USSH) are indicated by magenta circles, with candidate mini haloes (cMH) by cyan circles and radio haloes (H) by black circles. Clusters with no detected central diffuse radio emission, for which new upper limits (UL) were derived, are denoted by empty circles. {\bf Top right panel:} Central entropy $K_0$ vs. concentration parameter $c$. High values of $c$ and low values of $K_0$ indicate that the cluster host a cool core. In particular,  clusters with low central entropies ($K_0 < 30 - 50$ keV cm$^2$) host a cool core, whilst high central entropy ($K_0 > 50$ keV cm$^2$) indicates a non-cool-core cluster \citep{Giaci2017}. The only warm-core cluster hosting a mini halo is A1413. {\bf Bottom left panel:} Emission centroid shift $w$ vs. distance from the cluster centre at which $w$ is computed. The value $w$ was computed at a scale of 200, 300, 400 and 500 kpc. We note that the value of $w$ gets smaller going to outer distances for all the clusters but PSZ1G139.61+24. {\bf Bottom right panel:} Total radio power at 144 MHz of the diffuse emission vs. the ratio between the concentration parameter and emission centroid shift derived at 500 kpc.}
  \label{plot}
\end{figure*}

On the basis of the X-ray morphology, the nine clusters presented in this work are not currently undergoing a major merger. As shown in the top panels of Fig. \ref{plot}, five of these clusters host a cool core, whilst the remaining host a warm core, according to the classification based on the central gas entropy value \citep{Giaci2017}. The overall picture as seen in the low-frequency radio band by LOFAR is very diverse with the presence of radio diffuse emission in the form of two radio haloes, three mini haloes, and two uncertain cases, while two clusters do not host diffuse emission at all. Even though the sample we studied is not large enough to derive statistical results, we note that this is the largest sample of galaxy clusters studied within LoTSS, and we can draw a number of conclusions that can also be indicative for future low-frequency observations. In our sample, the diffuse radio emission appears to be uncorrelated with the dynamical state as indicated by the centroid shifts, $w$, of the X-ray emission computed at $500$ kpc. Even looking at $w$ on smaller scales, from 200 to 500 kpc, as shown in the bottom left panel of Fig.~\ref{plot}, no correlation between the radio emission and $w$ is found. Clusters that possess similar dynamical properties do not show the same radio properties, for example the non-cool-core clusters RXC0142.0+2131 and A1576, which have similar X-ray properties (relaxed morphology, comparable $c$ and $K_0$) and also comparable cluster masses, host a radio halo and no diffuse emission, respectively. Also plotting the power at 144 MHz versus the ratio of the X-ray concentration parameter, $c$, to $w$ (see bottom right panel of Fig. \ref{plot}) does not reveal a clear connection between the radio emission and the dynamical state of the cluster. Haloes are not necessary found in clusters with low $c$ and high $w$; see for example A2261.\\

In the top right panel of Fig.~\ref{plot}, we plot $c$ versus $K_0$ for each cluster. Two-component radio haloes (MH+USSH) are found in clusters with high $c$ and low $K_0$, while giant haloes are detected in clusters with low $c$ and higher $K_0$. Clusters for which no radio emission is found at the sensitivity level of these observations are found in both sides of the plot. Finally, a mini halo, which is the only mini halo found in a non-cool-core cluster that is also in the sample studied in \citet{Giaci2017}, is found in a cluster with low $c$ and high $K_0$.\\
%\textbf{We find cluster-scale radio emission in clusters with higher $c$ and low central entropy, $K_0$, that are known to host mini halos. The presence of a dense, compact, cool core appears to be indicative, hence $c$ and $K_0$, are more reliable predictors for diffuse radio emission. 

%The reason for this may be that the centroid shift is affected by the BCG or the presence of substructures, whereas $c$ and $K_0$ are more immediate measures of the thermal and, indirectly, the turbulent state of the central ICM.\\

Our observations show that the two clusters, PSZ1G139.61+24 and RXJ1720.1+2638, that host a mini halo and have the largest value of $c$,  also show larger-scale ultra-steep diffuse emission that extends beyond the cold fronts. In our sample, the only mini halo that is confined to the central region of a cluster is hosted by A1413, which is not a cool-core cluster. This might indicate that the presence of a dense cool core is required to initiate the re-acceleration of CRe in the regions surrounding the core through gas sloshing triggered by a minor merger, as suggested for PSZ1G139.61+24 by \citet{Savini2018b}. The observation that mini haloes can have a flatter core that is surrounded by a faint, steep corona is helpful in resolving their origin. A contribution from hadronic collisions cannot be excluded for the central region of cool-core clusters, but this contribution cannot explain the large-scale ultra-steep emission (e.g. \citealp{Bru2008}, \citealp{Maca2010}). Hence, different scenarios for the cosmic-ray origin might co-exist in these clusters that show hints of gas sloshing, with a distinction between the region within and outside the core.\\

%However, in a turbulent re-acceleration model \citep{ZuHone2013} it is expected that the halos become steeper at larger distances from the centre of the mini halo.\\
 
%It is unlikely that the steep part of the mini halos in PSZ1G139.61+24 and RXJ1720.1+2638 is caused by hadronic processes. 
%As pointed out in ZuHone et al. (2015), hadronic models for mini halos require that the emission follows a power law, except for regions where the magnetic field has been amplified substantially. 

%magnetic field amplification as this would imply that the magnetic fields have been amplified in a spherical shell. 
%

In our sample, no giant radio haloes in cool-core clusters are found. A2261 and RXC0142.0+2131, where haloes are found, do not host a cool core and the extended emission in A2390 is likely to  merely originate from the central radio galaxy, whose very diffuse lobes could have accidentally been classified as a halo at 1.4 GHz in \citet{Sommer2017}. We also note that both A2261 and RXC0142.0+2131 show traces of the occurrence of minor mergers. However, the non-cool-core cluster A1576, which presents hints of a minor merger, does not host diffuse emission.\\

%\textbf{The radio halo in RXC0142.0+2131 is smaller than a classical giant radio halo, such that in A2261, while both have an indication of steep spectrum. They might originate from the occurrence of minor mergers...}

%Interestingly, the discovery of giant radio halos in these two clusters, as claimed by \citet{Sommer2017}, was a little surprising given that giant halos are generally associated with merging clusters and not with cool-core clusters. However, our results suggest that giant halos in cool-core clusters, if any, are yet to be observed.\\
Many head-tail radio galaxies have been observed in the cluster environment, however a clear connection with the central diffuse emission is only seen in the case of RXJ1720.1+2638, where the tail might provide seed electrons for re-acceleration.

%The head-tail radio galaxy in A478 blends with the central source but no centrally-located diffuse emission is seen at the sensitivity level of our LOFAR observations.\\

%\begin{itemize}
%\item Coma magnetic field model
%\item assumptions on the gas density and temperature profiles
%\item using the upper limit on A1576 and A1423, we constrain the percentage of cosmic ray with respect to the gas thermal energy in these two clusters
%\end{itemize}
%I modelli prevedono che merger meno energetici diano origine ad aloni che si vedono sono a bassa frequenza (e con uno spettro ripido) e che sono piu' piccoli degli aloni normali, e quetso potrebbe essere proprio uno di quei casi. D'altra parte, ci sono aloni "piccoli" tipo i 2 trovati da Virginia (2017) e uno trovato da me (CIZAJ1938, nel 2015) che sono piccoli, stanno sotto la correlazione MA non hanno evidenza di avere uno spettro ripido, per cui in questo quadro diventerebbe molto importante avere una stima precisa dello spettro di questo alone, che non e' tanto piccolo, ma sta sicuramente sotto la correlazione.

\subsection{Limits on cosmic-ray protons}
%AB: ti ho cambiato questo incipt perche; iniziare dicendo che abbiam fatto cose gia' fatte (che non e' vero) non mi piaceva ;-p
In this section, we aim to use some of our results to constrain the energy density of relativistic protons in the ICM by computing the maximum possible radio emission that can originate from hadronic collisions \citep[][]{bl99,de00}. 
We use a similar albeit more complicated procedure as in \cite{Bru2007}, which we describe in the following. 

The total admissible energy density of cosmic rays is given by requiring that this radio emission does not exceed our upper limit, given a model for the thermal gas pressure and the magnetic fields.\\

In this work, we place new upper limits on the radio power for the clusters A1576 and A1423. While the new upper limit for A1576 is comparable to the value previously derived in literature, for A1423 we were able to obtain a limit that is almost a factor of two deeper. Hence, we created a model for the gas density and temperature of A1423 following  \citet{Ghi2018}, who presented best-fit formulas for the thermodynamical properties of cool-core and non-cool-core clusters based on the analysis of a sample of clusters observed with XMM-Newton. We rescaled the values to the cluster mass and temperature\footnote{We used the core-excised temperature given by \citet{Giaci2017}.} of A1423, $M_{\rm 500}=6.04 \times 10^{14} M_{\odot}$ and $T_{\rm 500}=6.4$ keV.\\ 

We assumed that the magnetic field in A1423 is similar to the best-fit 3D model of the magnetic field in the Coma cluster presented in \citet{Bona2010} owing to the similarity in the cluster total mass. We generated a power-law distribution of the magnetic vector potential $A$ in Fourier space for a $(256)^3$ grid with a fixed resolution of $\Delta x=10$ kpc that is randomly drawn from the Rayleigh distribution. The magnetic field in real space follows $B=\nabla \times A$, which ensures that $\nabla \cdot B=0$ by construction.
As in Coma, we assumed that the maximum coherence scale of the field is $\sim$ 40 kpc, the power law of fluctuations exhibits a Kolmogorov spectrum and the average magnetic field strength in the centre is $B_0 \sim 4 \, \mu$G. The volume-averaged magnetic field within a radius equal to $R_{\rm H}$ (i.e. the radius used for the upper limit, $436$
kpc for A1423) is $\langle B \rangle = 1.2 \, \mu$G.\\

To be self-consistent with the profile of the mock halo that we injected, we assumed that the ratio of the spatial distribution of the cosmic-ray energy density, $E_{\rm CR}$, over the gas energy density profile, $E_{\rm gas}$, is

\begin{equation*}
\frac{E_{\rm CR}}{E_{\rm gas}} = \Big( \frac{E_{\rm CR}}{E_{\rm gas}} \Big)_0 \, \Big(\frac{r}{\Delta x}\Big)^{\alpha_{\rm CR}}, 
\end{equation*}
%E_{\rm CR} = E_{\rm CR}^0 \, \Big( \frac{ E_{\rm gas}} {E_{\rm gas}^{\rm max} }\Big)^{\alpha_{\rm CR}},

where the index $0$ indicates the ratio computed at the cluster centre, and $\alpha_{\rm CR}$ is a shape parameter that allows for the non-linear scaling between cosmic rays and gas matter, as in \citet{Donn2010} and \citet{Brunetti2012}.
We fixed $\alpha_{\rm CR} = 1$ based on the requirement that the profile of the simulated radio emission within 1$\sigma$ (see below) matches the input halo model used in Sec. \ref{a1423-sec}. 
We also imposed that the ratio $E_{CR}/E_{gas}$ is no larger than 0.1.

The radio emission from hadronic collisions was derived following \citet{Pf2004}, i.e. assuming a Dermer model to compute the cross section of the proton-proton interaction, and integrating the CRp population from $E_{\rm min}=0.1$ GeV to $E_{\rm max}=100$ GeV for a particle spectrum of $\alpha_p = 2.6$, corresponding to a radio spectral index of $\alpha \sim$ -1.3. Finally, we projected the emission and measured the flux density within a radius $R_{\rm H}$ from the cluster centre, and compared it to the upper limit we derived from the LOFAR observation of A1423.\\

Through this procedure, we find that $\Big( \frac{E_{\rm CR}}{E_{\rm gas}} \Big)_0 \sim 0.1\%$ yields $\Big( \frac{E_{\rm CR}}{E_{\rm gas}} \Big) \sim 3\%$, averaged on the volume of a sphere of radius $R_{\rm H}$.\\ 

%This limit is in agreement with the limit derived from the lack of hadronic $\gamma$-ray emission reported by the \textit{Fermi} satellite for a sample of  galaxy clusters (e.g. \citealp{Ack2014}). In particular, the most stringent limit derived for the Coma cluster, even though using a bigger volume than the one we use here, is 1\% - 2\%. 
A comparison with the limit derived from the lack of hadronic $\gamma$-ray emission reported by the \textit{Fermi} satellite for a sample of  galaxy clusters (e.g. \citealp{Ack2014}) is not straightforward, and would require, for instance, a more detailed treatment of the cosmic-ray energy density distribution. However, we note that the limit we derived is of the same order of magnitude. Our constrain demonstrates the great potential of future, deeper LOFAR observations to constrain the energy budget of CRp in the ICM.

%Following \citet{Donn2010}, we  fix the maximum ratio $E_{\rm CR}/E_{\rm gas}$ implied by the above scaling to $0.1$ \textbf{MB: WHY IS THE LATTER CONSTRAINT NECESSARY?}, which is the maximum value found in state-of-the-art cosmological simulations of cosmic rays in galaxy clusters \citep{Vaz2016}.

% \textbf{MB: SAY WHAT THE FERMI LIMIT ON $X_0$ IS}.

\section{Summary}

In this paper, we have presented the results of the largest campaign of LOFAR observations targeting galaxy clusters so far. The mass-selected sample consists of clusters with no sign of major mergers observed in the frequency range between 120 MHz and 168 MHz. Data reduction was performed following the facet calibration scheme. Below we summarise our main results:

\begin{enumerate}

\item We find central, diffuse emission in the form of mini haloes surrounded by USS haloes in the cool-core clusters RXJ1720.1+2638 and PSZ1G139.6+24. Hence, we argue that the sloshing of the dense core after a minor merger can be responsible for the formation, not only of a central mini halo, but also of larger-scale emission that is visible at low radio frequencies. Moreover, the presence of a cool core, indicated by high values of the concentration parameter, $c$, and low values of the central entropy, $K_0$, might be significant for the formation of radio diffuse emission on scales larger than the cluster core that hosts a mini halo.

\item We confirm the presence of a mini halo in the non-cool-core cluster A1413 as proposed by \citet{Govo2009}. 

\item We discover a radio halo in the non-cool-core cluster RXCJ0142.0+2131 with a scale of $570$ kpc and a spectral index of $\alpha^{610}_{144} < -1.3$.

\item We confirm the presence of a radio halo in the non-cool-core cluster A2261 as proposed by \citet{Sommer2017}.

\item The central radio galaxy discovered in the massive cool-core cluster A2390 might account for most or even all the radio flux that was attributed to the giant radio halo proposed by \citet{Sommer2017}; high-resolution observations in the frequency range between 144 GHz and 1.4 GHz are needed to confirm the morphology of the two radio jets and lobes and exclude the presence of the giant radio halo.

\item At LOFAR frequencies no centrally-located diffuse emission is observed in the cool-core cluster A478. We injected a mock mini halo, and placed a limit on the spectrum, i.e. $\alpha > -1$. 

\item No cluster-scale diffuse radio emission is found in the cluster A1576 and in the cool-core cluster A1423 at the sensitivity of the observations, hence we derive new upper limits on the total radio power. 

\item We use the limit on the radio power of A1423 to constrain the energy budget of CRp in the ICM and compare the result with the constraints derived from the lack of hadronic $\gamma$-ray emission reported by the FERMI satellite. We find that our LOFAR observations are competitive with the deepest limits derived by FERMI for the Coma cluster.

\item We discover head-tail radio galaxies in the clusters A1423 and A1413. We note also the presence of head-tail radio galaxies in A478, and  in RXJ1720.1+2638, where the tail appears to be connected to the central source, and it might be a possible source of seed particles that can be re-accelerated.

\item No giant radio haloes in the cool-core clusters of our sample are found.

\end{enumerate}

Low-frequency radio observations are ideal for discovering diffuse emission, in particular steep-spectrum emission from low-energy CRe in galaxy clusters. Future studies performed on a larger sample of clusters will provide statistical information and help to further investigate the connection between the formation of radio emission from the ICM and its dynamical state.

%We focus on clusters whose dynamical status has been classified as non-merging from X-ray observations. \textbf{Key questions: is there a connection between mini halos and halos? Are mini halos produced by cluster-scale phenomena such as minor mergers or are they related to the central AGN in the cluster core?} Our goals:\\
%, and therefore permit placement of upper limits on the spectral index values
%to search for (limits on) cluster-wide steep radio emission. Such sources might indicate a transitional phase between mini halos and giant radio halos. 
%Higher frequency observations are available for all the clusters of the sample, therefore a spectral index analysis will be performed for each target.\\
%Performing spectral index measurements on a sample of non-merging clusters will help to discriminate between re-acceleration or hadronic models. 

%For re-acceleration models we expect a break in the observed spectrum while for hadronic models the spectrum extends to very high energies if CRp spectrum extend to very high energies. 

\begin{acknowledgements}

This paper is based (in part) on data obtained with the International LOFAR
Telescope (ILT) under the project codes listed in Tab.\ref{info1}. LOFAR (van Haarlem et al. 2013) is the LOw Frequency ARray designed and constructed by ASTRON. It has observing, data
processing, and data storage facilities in several countries, which are owned by
various parties (each with their own funding sources) and are collectively
operated by the ILT foundation under a joint scientific policy. The ILT resources
have benefitted from the following recent major funding sources: CNRS-INSU,
Observatoire de Paris and Université d'Orléans, France; BMBF, MIWF-NRW, MPG,
Germany; Science Foundation Ireland (SFI), Department of Business, Enterprise and
Innovation (DBEI), Ireland; NWO, The Netherlands; The Science and Technology
Facilities Council, UK; Ministry of Science and Higher Education, Poland.

%LOFAR, the Low Frequency Array designed and constructed by ASTRON, has facilities owned by various parties (each with their own funding sources), and that are collectively operated by the International LOFAR Telescope (ILT) foundation under a joint scientific policy.\\ 
Part of this work was carried out on the Dutch national e-infrastructure with the support of the SURF Cooperative through grant e-infra 160022 \& 160152.  
The LOFAR software and dedicated reduction packages on https://github.com/apmechev/GRID\_LRT were deployed on the e-infrastructure by the LOFAR e-infragroup, consisting of J. B. R. Oonk (ASTRON \& Leiden Observatory), A. P. Mechev (Leiden Observatory) and T. Shimwell (Leiden Observatory) with support from N. Danezi (SURFsara) and C. Schrijvers (SURFsara).\\
This research made use of the NASA/IPAC Extragalactic Database (NED), which is operated by the Jet Propulsion Laboratory, California Institute of Technology, under contract with the National Aeronautics and Space Administration.\\
A. Bonafede acknowledges support from the ERC-StG grant DRANOEL, n. 714245. F. Vazza acknowledges financial support from the ERC Starting Grant MAGCOW, no.714196. H. Rottgering and R. van Weeren acknowledge support from the ERC Advanced Investigator programme NewClusters 321271 and the VIDI research programme with project number 639.042.729, which is financed by the Netherlands Organisation for Scientific Research (NWO). F. de Gasperin is supported by the VENI research programme with project number 1808, which is financed by the Netherlands Organisation for Scientific Research (NWO). A. Drabent acknowledges support by the BMBF Verbundforschung under the grant 05A17STA.
    
\end{acknowledgements}

%-------------------------------------------------------------------
\bibliographystyle{aa} % style aa.bst
\bibliography{SAMPLE} % your references Yourfile.bib

\newpage
%\section{Supplementary Material}

\begin{figure*}
\centering
 \includegraphics[width=\textwidth]
  {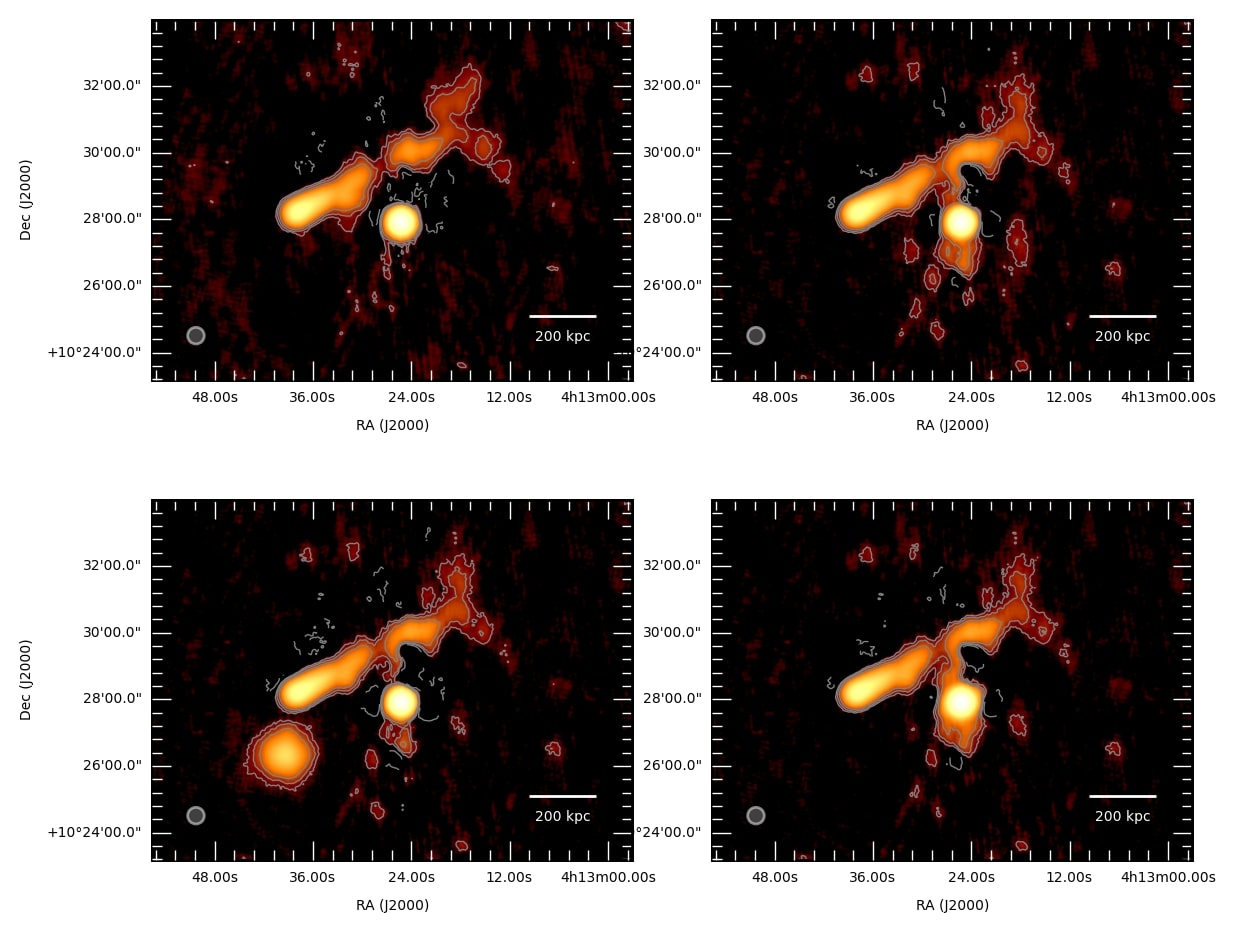}
   \caption{Low-resolution ($30'' \times 30''$) LOFAR image of the cluster A478 at 144 MHz. The contour levels are $(1, 2, 4)\, \times \, 3\sigma$, where $\sigma$ = 620 $\mu$Jy\,beam$^{-1}$. No mock mini halo is injected in the data set in the top left panel. A mock mini halo with $I_0 = 1$ $\mu$Jy\,arcsec$^{-2}$ is injected at the cluster centre with $r_e = 100$ kpc in the top right panel. A mock mini halo with $I_0 = 13$ $\mu$Jy\,arcsec$^{-2}$ is injected at the cluster centre with $r_e = 25$ kpc at the cluster centre in the bottom right panel, and in a close-by void region in the bottom left panel. The values for $r_e$ and $I_0$ are chosen accordingly to the top panel of Fig. 5 in \citet{Murgia2009} and are referred to measurements at 1.4 GHz.}
  \label{inj}
\end{figure*}

\begin{figure*}
\centering
   \includegraphics[width=0.5\textwidth]
  {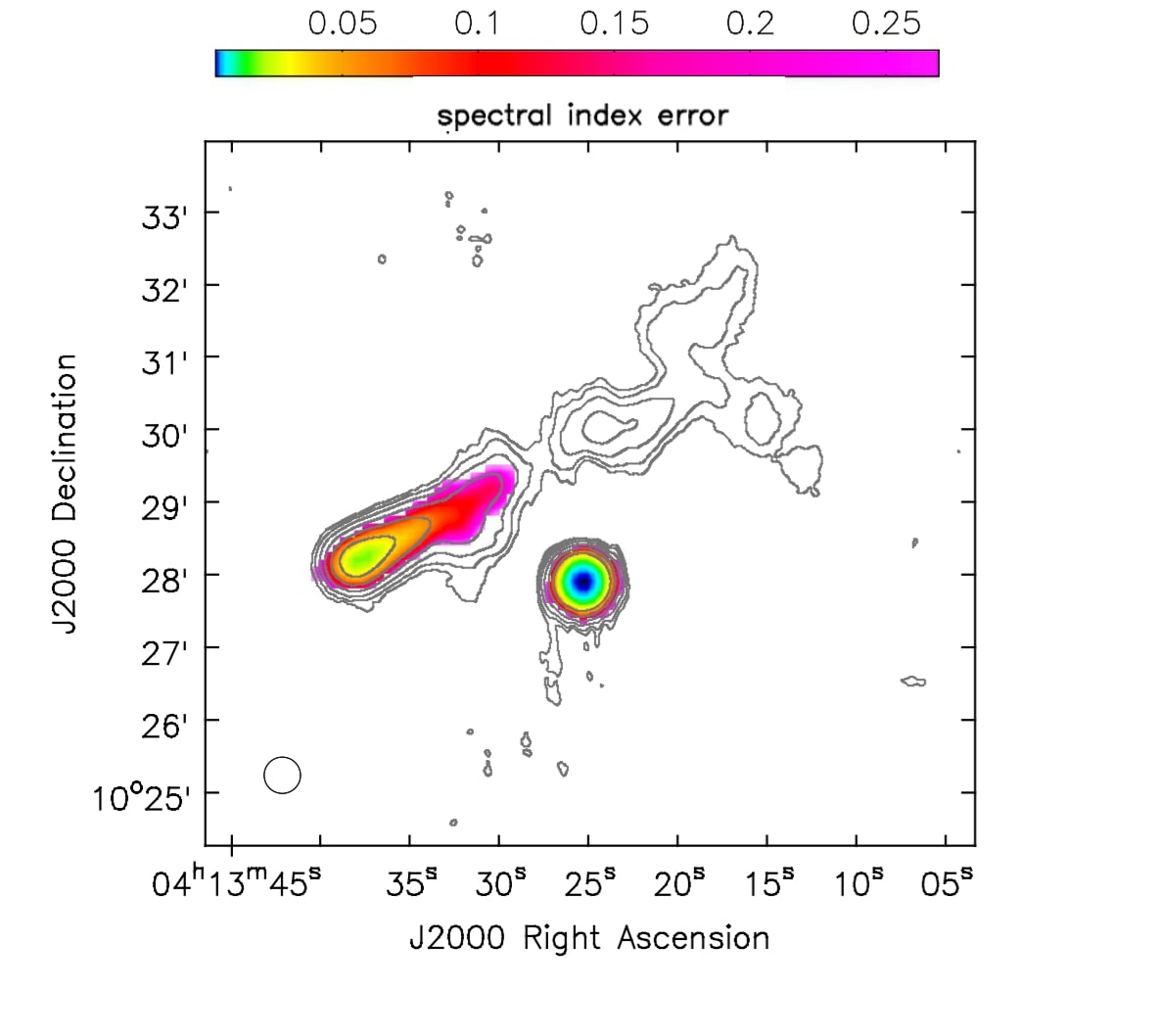}
   \caption{Spectral index error map of A478 ($\alpha^{610}_{144}$) related to Fig. \ref{a478-spix}. Pixels below 3$\sigma$ are blanked.}
  \label{spix-err1}
\end{figure*}

\begin{figure*}
\centering
 \includegraphics[width=0.45\textwidth]
  {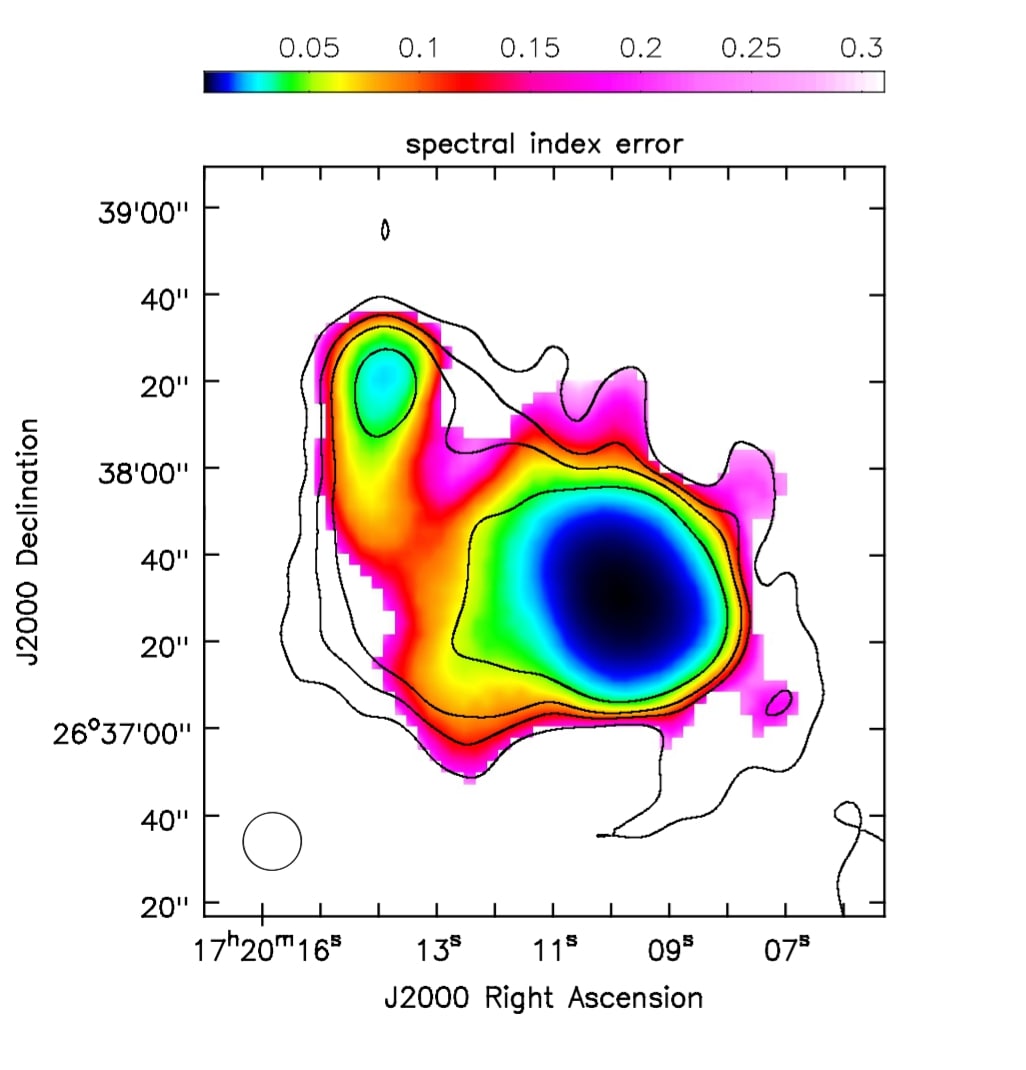}
  \caption{Spectral index error map of RXJ1720.1+2638 ($\alpha^{610}_{144}$) related to Fig. \ref{rxj-spix}. Pixels below 3$\sigma$ are blanked.}
  \label{spix-err2}
\end{figure*}

\begin{figure*}
\centering
 \includegraphics[width=0.5\textwidth]
  {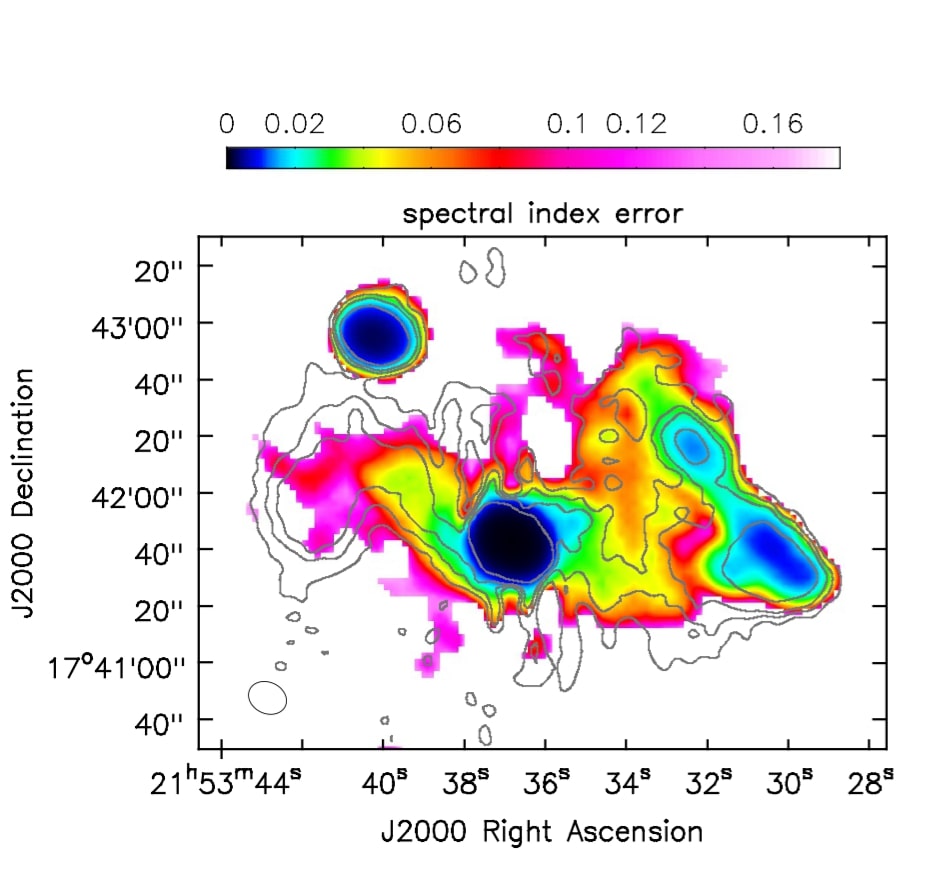}
  \caption{Spectral index error map of A2390 ($\alpha^{1500}_{144}$) related to Fig. \ref{a2390-spix}. Pixels below 3$\sigma$ are blanked.}
  \label{spix-err3}
\end{figure*}

\begin{figure*}
\centering
 \includegraphics[width=\textwidth]
  {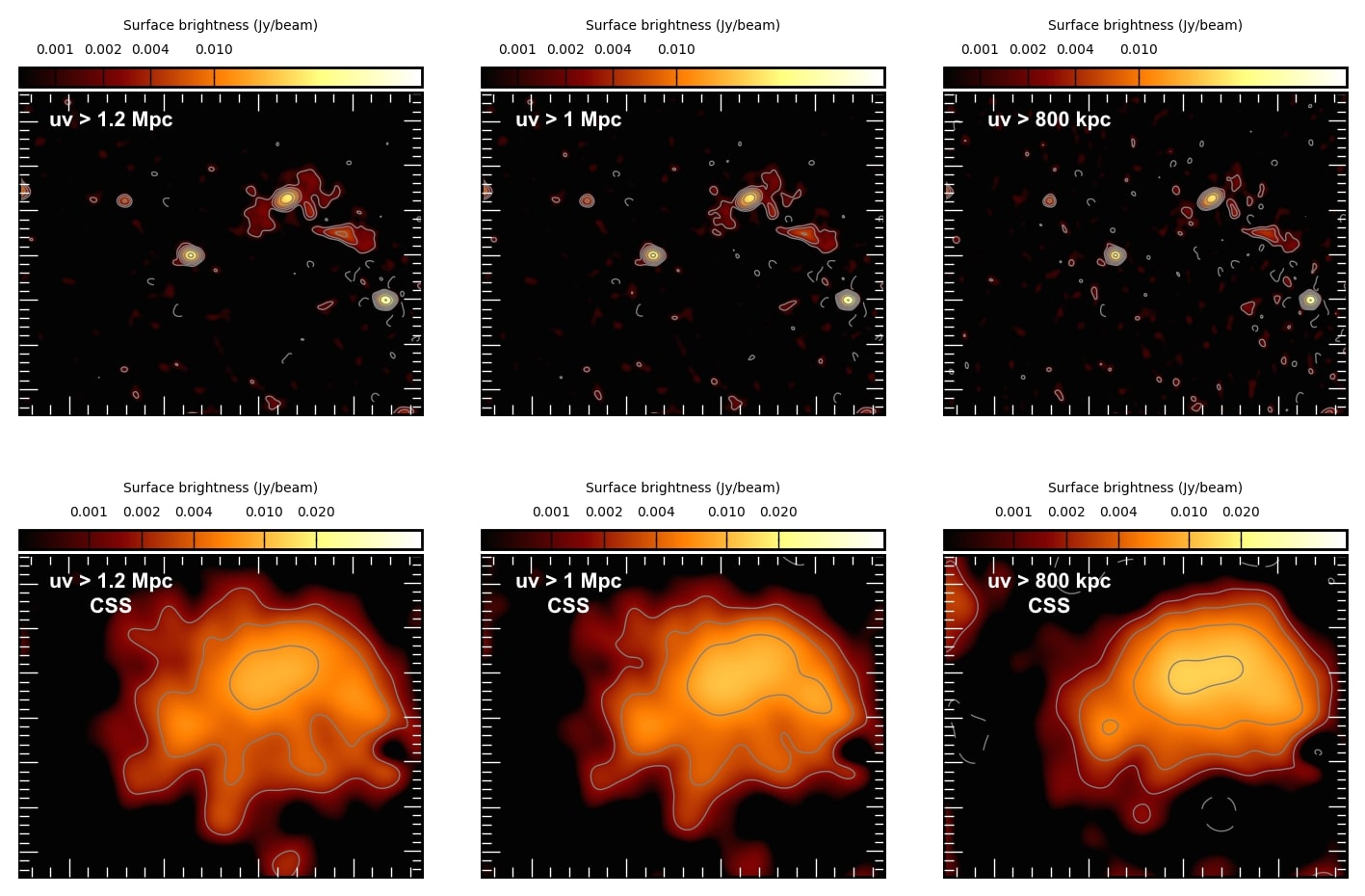}
   \caption{\textbf{Top row}: Images obtained by changing the \textit{uv}-range of the LOFAR observation of A2261 with different cuts. Contours start at 3$\sigma$, where $\sigma = 450$ $\mu$Jy beam$^{-1}$, and are spaced by a factor of 2. \textbf{Bottom row}: Corresponding \textit{uv}-subtracted images. Different models of the discrete sources do not affect the detection of large-scale diffuse emission. Contours start at 3$\sigma$, where $\sigma = 600$ $\mu$Jy beam$^{-1}$, and are spaced by a factor of 2.}
  \label{uvsub}
\end{figure*}

\begin{figure*}
 \includegraphics[width=0.9\textwidth]
  {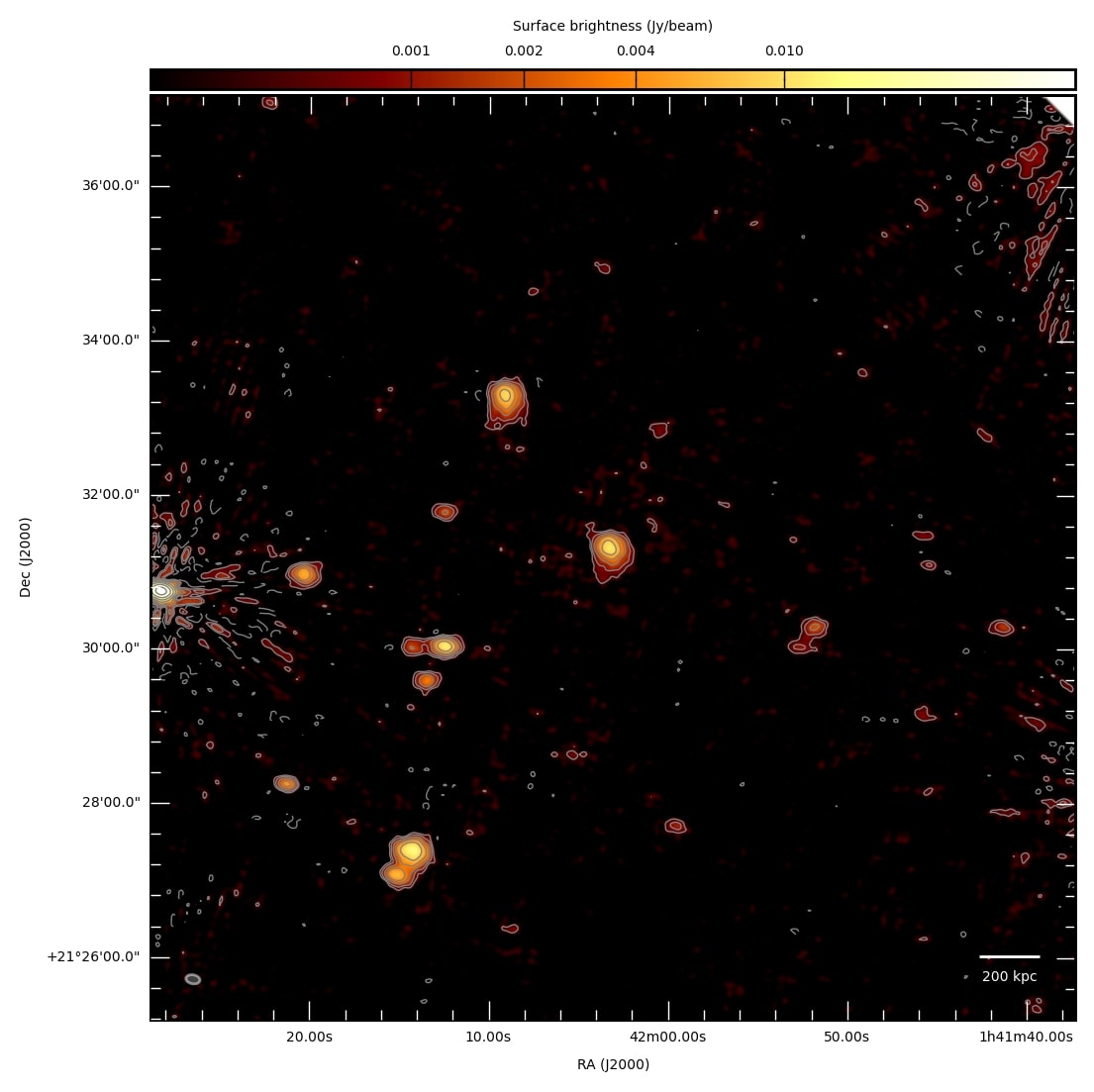}
   \caption{
   High-resolution 144 MHz LOFAR image of RXCJ0142.0+2131. The contour levels start at $3\sigma$, where $\sigma$ = 150 $\mu$Jy\,beam$^{-1}$, and are spaced by a factor of two. The negative contour level at $-3\sigma$ is overlaid with a dashed line. The beam is $11'' \times 7''$ and is shown in grey in the bottom left corner of the image. }
  \label{supp-rxc}
\end{figure*}

\begin{figure*}
 \includegraphics[width=0.9\textwidth]
  {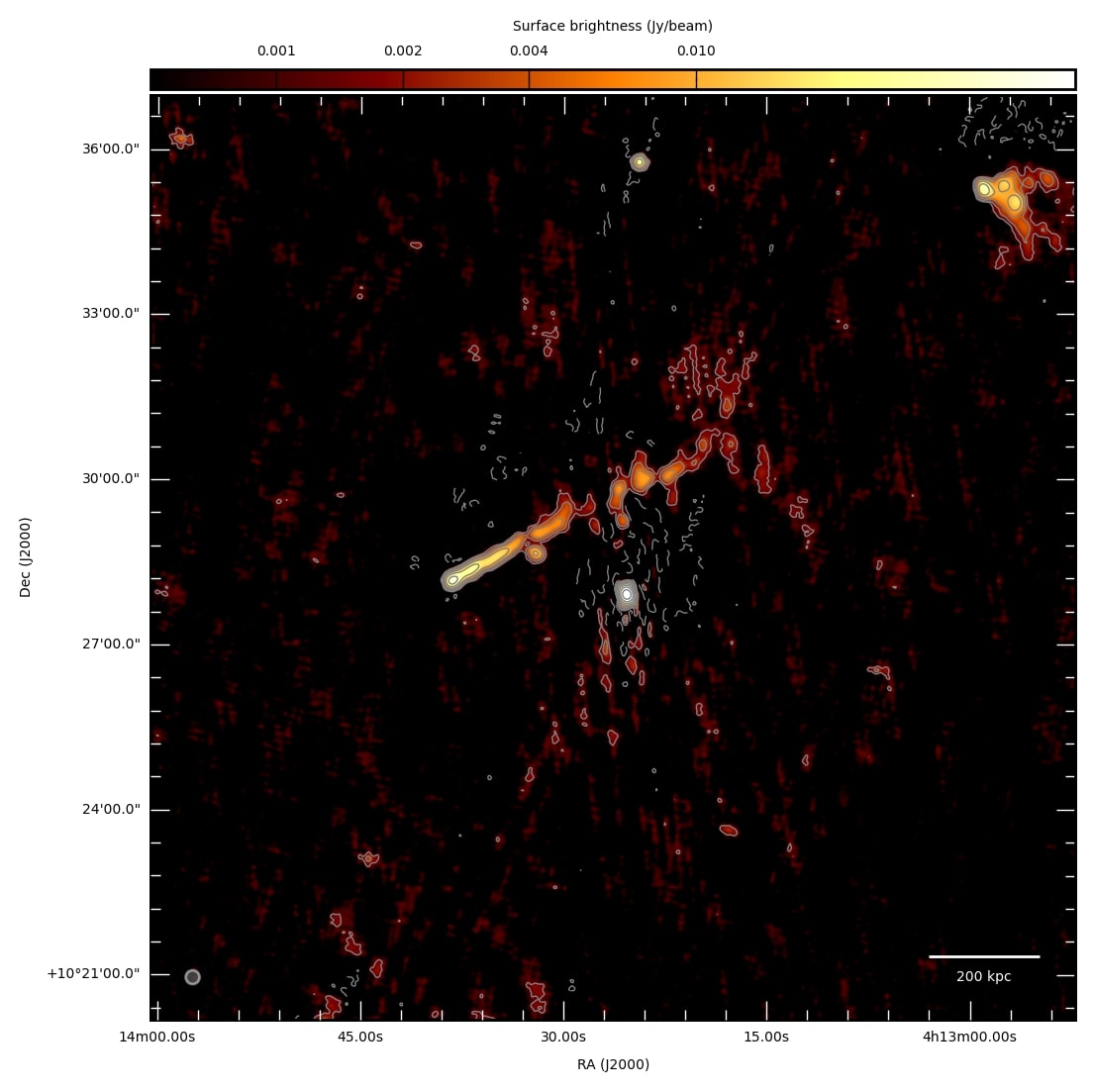}
   \caption{
   High-resolution 144 MHz LOFAR image of A478. The contour levels start at $3\sigma$, where $\sigma$ = 450 $\mu$Jy\,beam$^{-1}$, and are spaced by a factor of two.  The negative contour level at $-3\sigma$ is overlaid with a dashed line. The beam is $15'' \times 15''$ and is shown in grey in the bottom left corner of the image. We note the presence in the field of an head-tail radio galaxy located to the north-west of the cluster centre, which has an optical counterpart but no redshift information.}
  \label{supp-a478}
\end{figure*}

\begin{figure*}
 \includegraphics[width=0.9\textwidth]
  {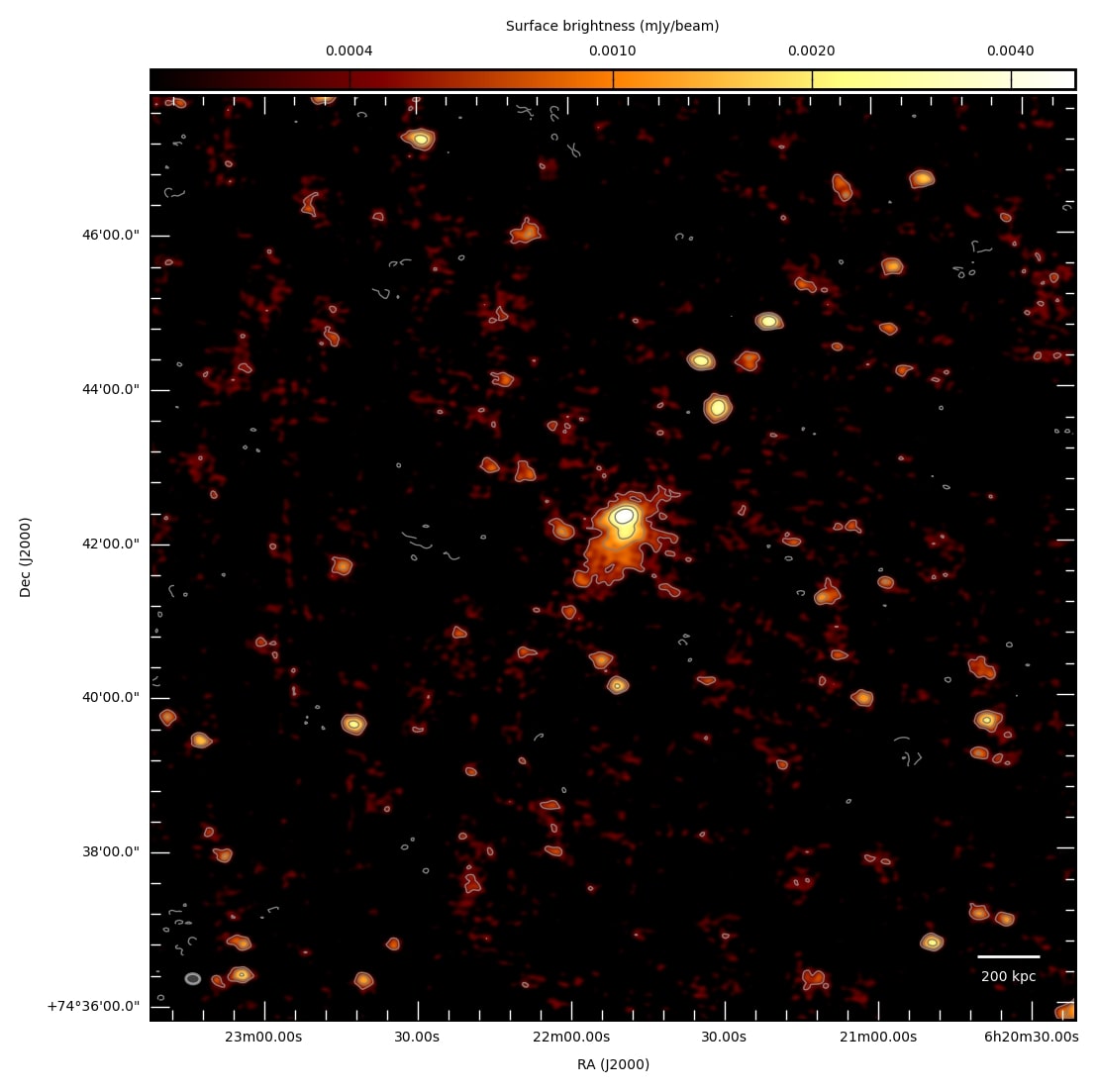}
   \caption{
 High-resolution 144 MHz LOFAR image of PSZ1G139.61+24. The contour levels start at $3\sigma$, where $\sigma$ = 150 $\mu$Jy\,beam$^{-1}$, and are spaced by a factor of two.  The negative contour level at $-3\sigma$ is overlaid with a dashed line. The beam is $11'' \times 8''$ and is shown in grey in the bottom left corner of the image. }
  \label{supp-psz}
\end{figure*}

\begin{figure*}
 \includegraphics[width=0.9\textwidth]
  {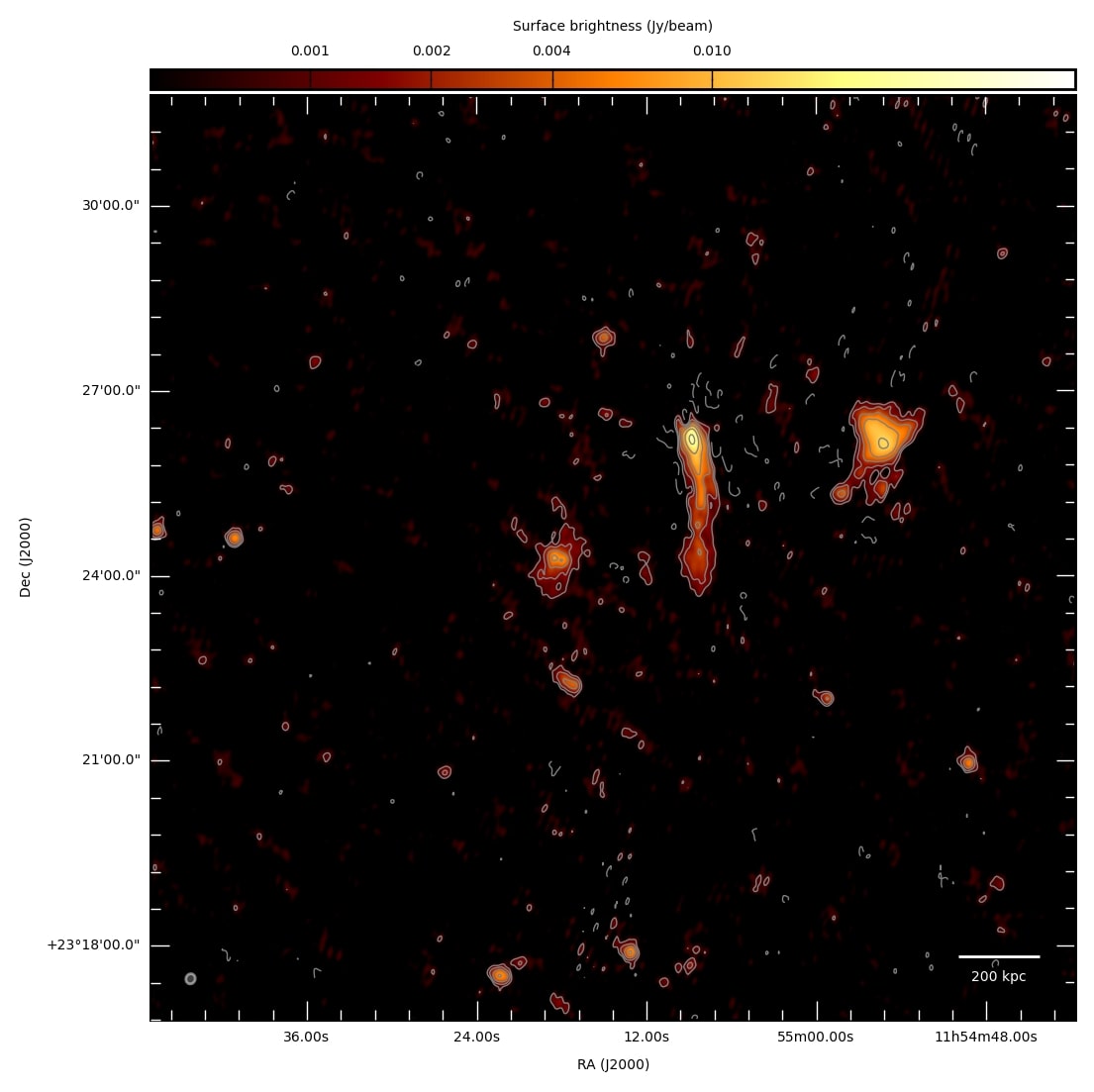}
   \caption{
 High-resolution 144 MHz LOFAR image of A1413. The contour levels start at $3\sigma$, where $\sigma$ = 270 $\mu$Jy\,beam$^{-1}$, and are spaced by a factor of two.  The negative contour level at $-3\sigma$ is overlaid with a dashed line. The beam is $10'' \times 9''$ and is shown in grey in the bottom left corner of the image. We note the presence in the field of a patch of diffuse emission located to the north-west of the cluster centre, which has no optical counter part.}
  \label{supp-a1413}
\end{figure*}

\begin{figure*}
 \includegraphics[width=0.9\textwidth]
  {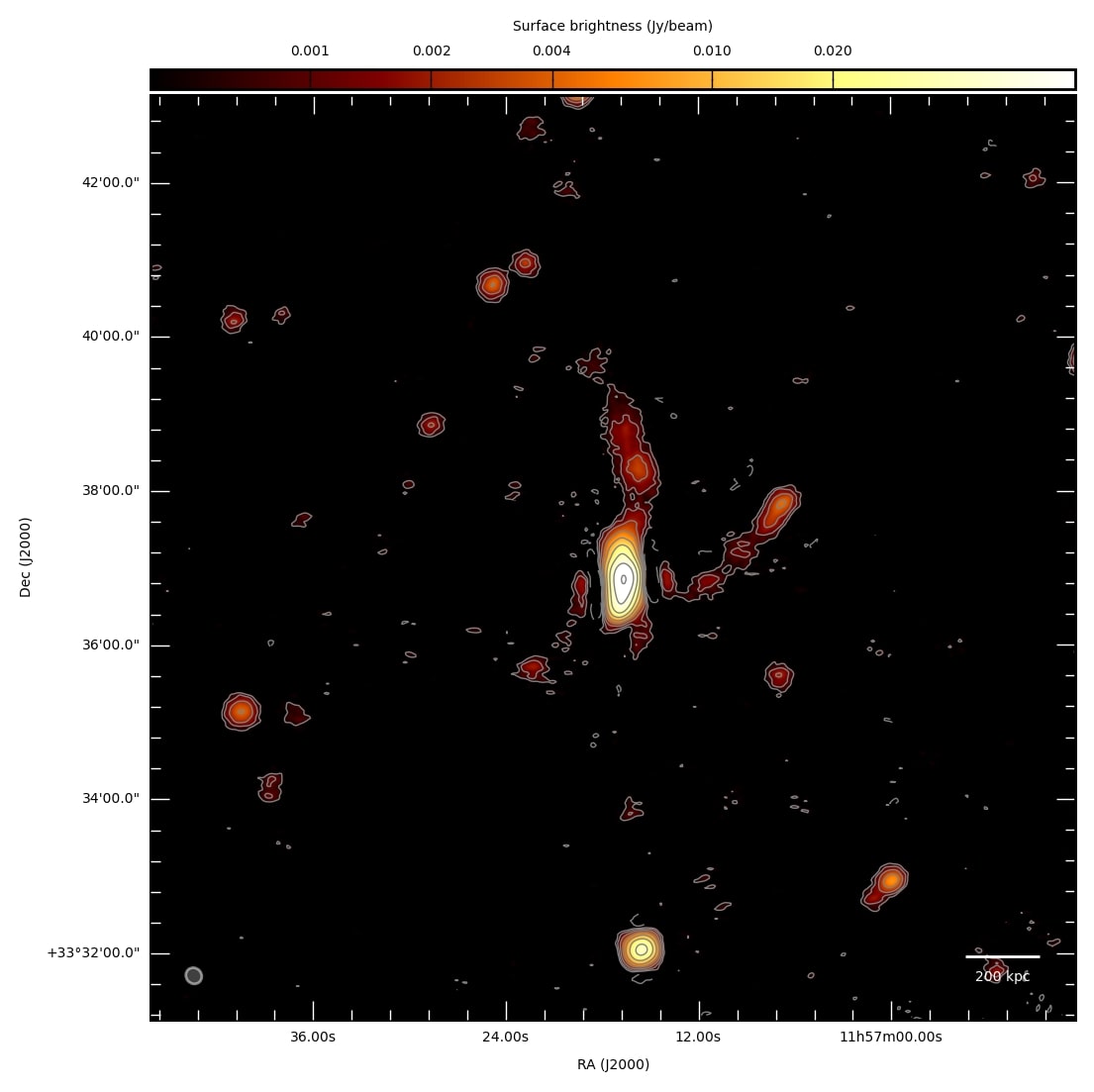}
   \caption{
 High-resolution 144 MHz LOFAR image of A1423. The contour levels start at $3\sigma$, where $\sigma$ = 170 $\mu$Jy\,beam$^{-1}$, and are spaced by a factor of two.  The negative contour level at $-3\sigma$ is overlaid with a dashed line. The beam is $13'' \times 12''$ and is shown in grey in the bottom left corner of the image. }
  \label{supp-a1423}
\end{figure*}

\begin{figure*}
 \includegraphics[width=0.9\textwidth]
  {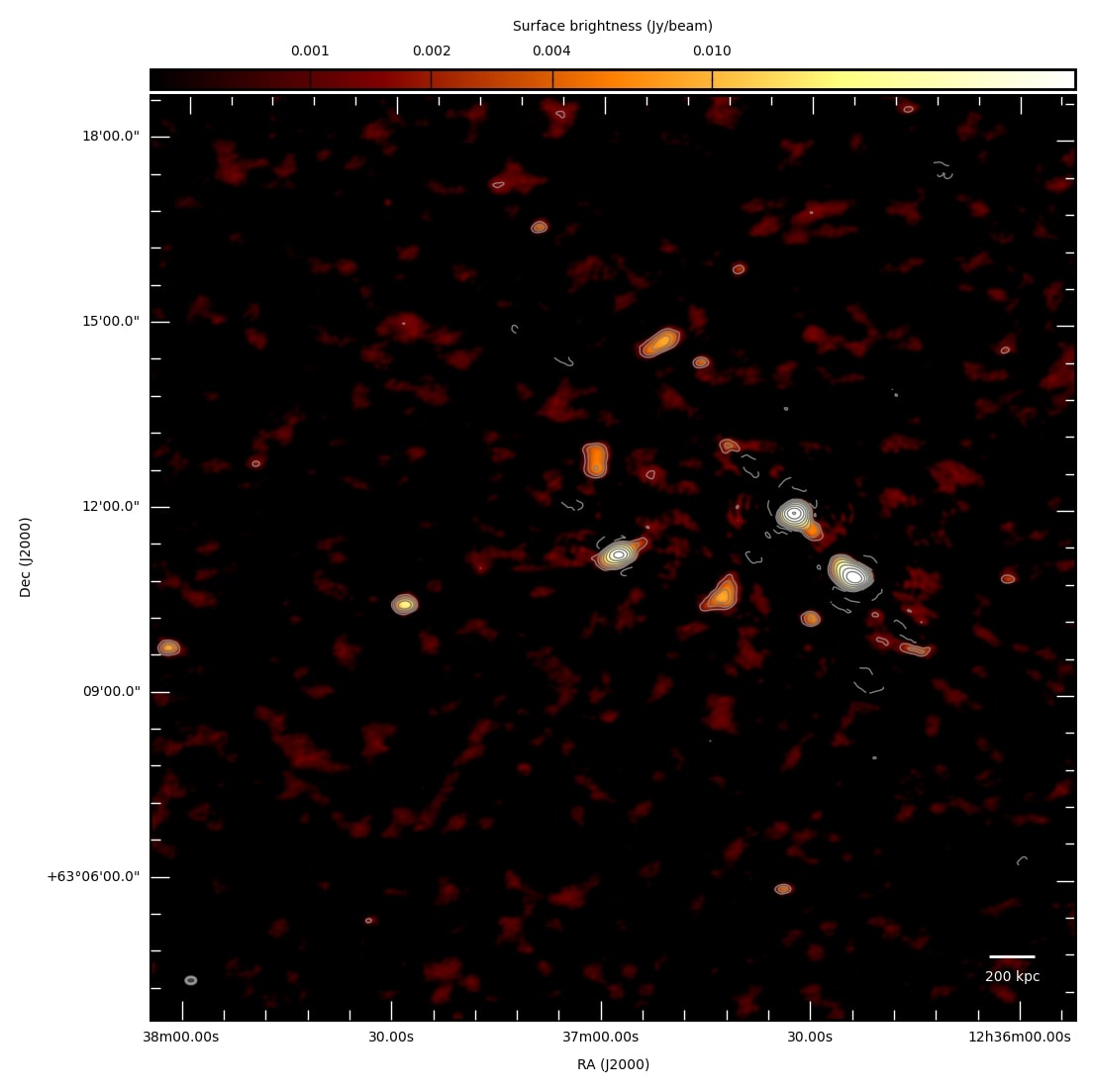}
   \caption{
   High-resolution 144 MHz LOFAR image of A1576. The contour levels start at $3\sigma$, where $\sigma$ = 500 $\mu$Jy\,beam$^{-1}$, and are spaced by a factor of two.  The negative contour level at $-3\sigma$ is overlaid with a dashed line. The beam is $10'' \times 7''$ and is shown in grey in the bottom left corner of the image. 
}
  \label{supp-a1576}
\end{figure*}

\begin{figure*}
 \includegraphics[width=0.9\textwidth]
  {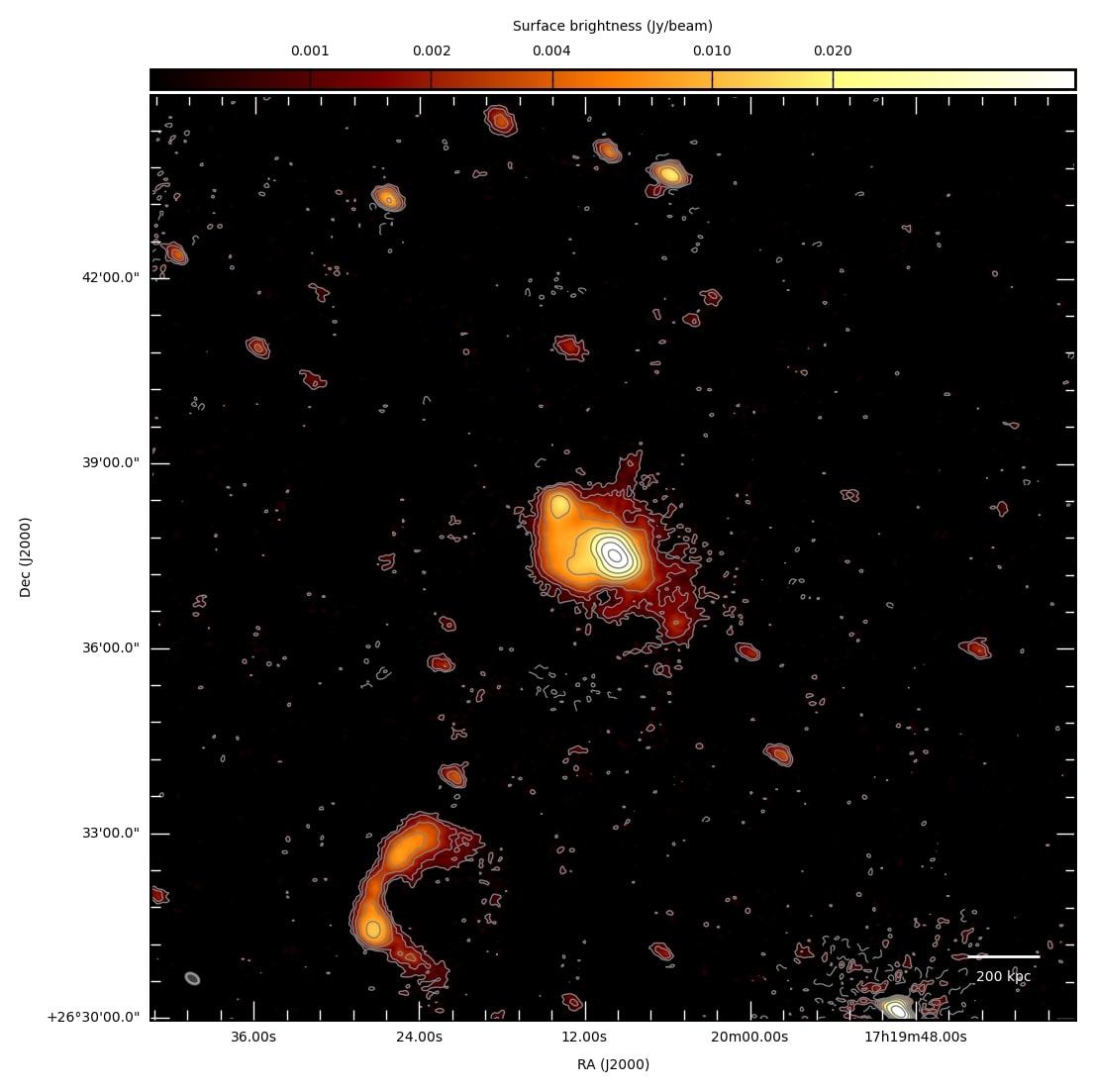}
   \caption{
   High-resolution 144 MHz LOFAR image of RXJ1720.1+2638. The contour levels start at $3\sigma$, where $\sigma$ = 200 $\mu$Jy\,beam$^{-1}$, and are spaced by a factor of two. The negative contour level at $-3\sigma$ is overlaid with a dashed line. The beam is $14'' \times 9''$ and is shown in grey in the bottom left corner of the image. We note the presence in the field of a wide-angle tail radio galaxy located to the south-east of the cluster centre, which is associated with a cluster member galaxy at $z = 0.159$.}
  \label{supp-rxj1720}
\end{figure*}

\begin{figure*}
 \includegraphics[width=0.9\textwidth]
  {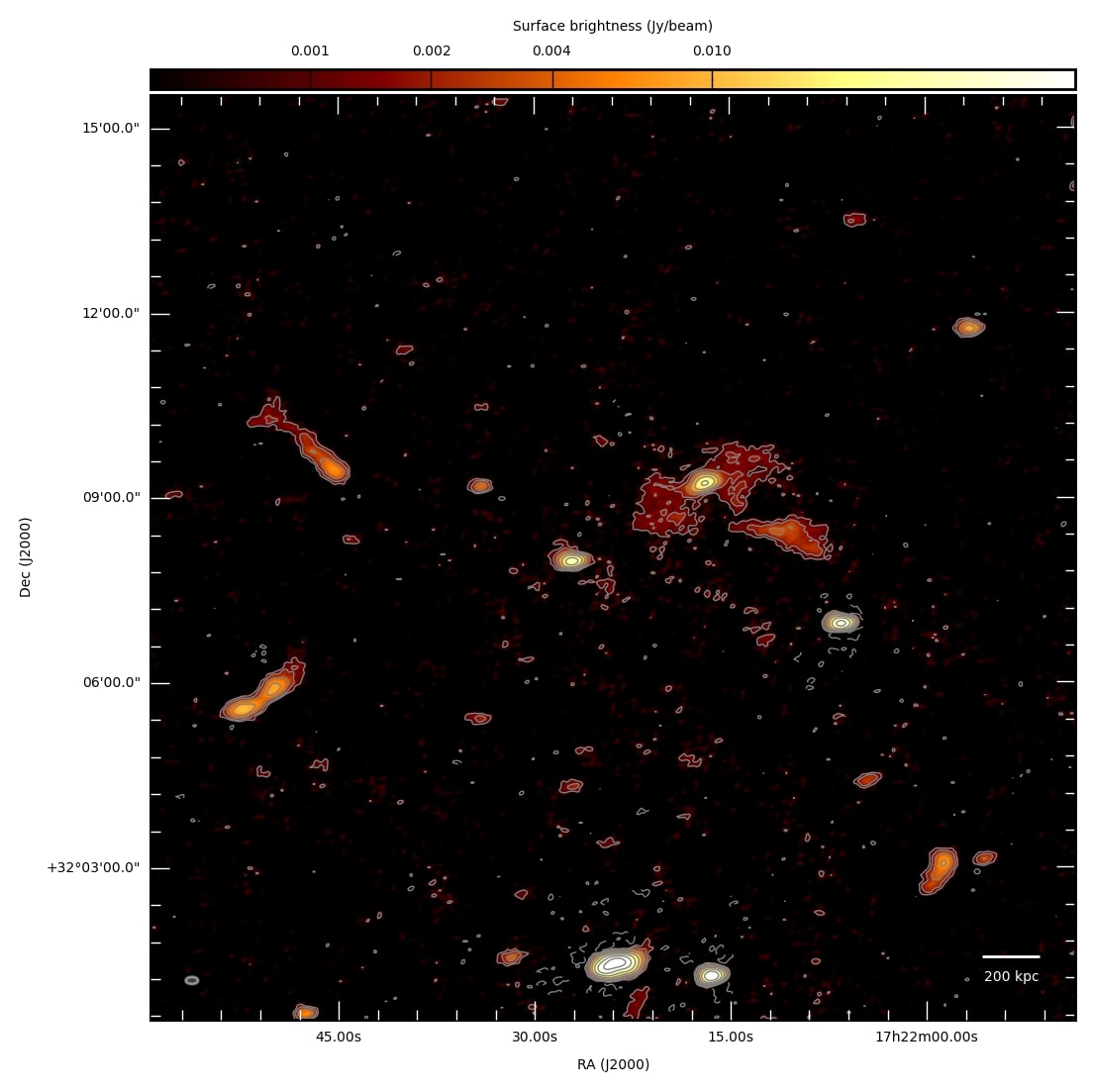}
   \caption{
   High-resolution 144 MHz LOFAR image of A2261. The contour levels start at $3\sigma$, where $\sigma$ = 270 $\mu$Jy\,beam$^{-1}$, and are spaced by a factor of two.  The negative contour level at $-3\sigma$ is overlaid with a dashed line. The beam is $12'' \times 7''$ and is shown in grey in the bottom left corner of the image. We note the presence in the field of a tail located to the north-east of the cluster centre, which has an optical counterpart but no redshift information.}
  \label{supp-a2261}
\end{figure*}

\begin{figure*}
 \includegraphics[width=0.9\textwidth]
  {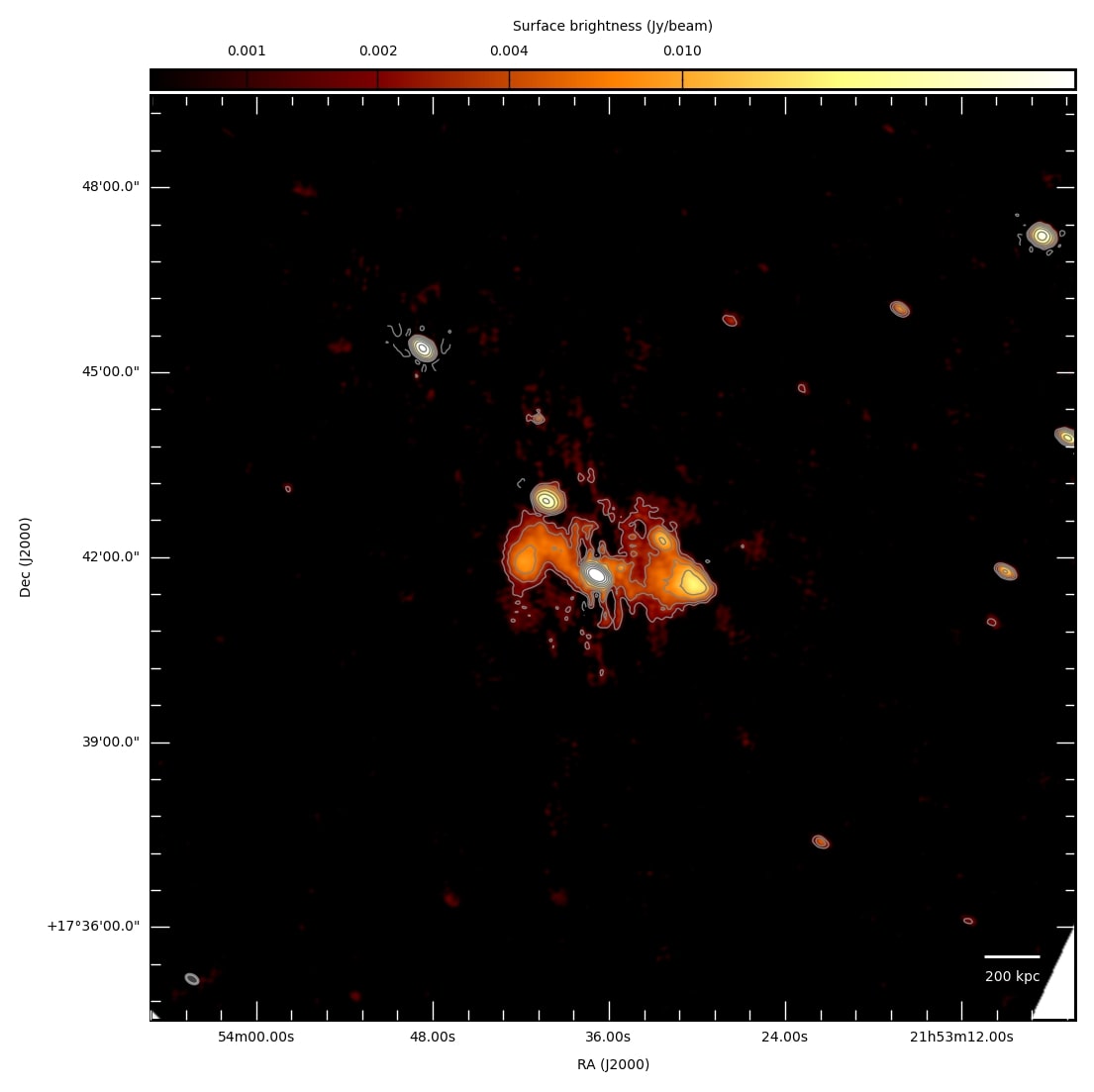}
   \caption{
   High-resolution 144 MHz LOFAR image of A2390. The contour levels start at $3\sigma$, where $\sigma$ = 400 $\mu$Jy\,beam$^{-1}$, and are spaced by a factor of two.  The negative contour level at $-3\sigma$ is overlaid with a dashed line. The beam is $13'' \times 8''$ and is shown in grey in the bottom left corner of the image.}
  \label{supp-a2390}
\end{figure*}

\end{document}